\documentclass[12pt,document,nofootinbib,superscriptaddress,onecolumn,preprintnumbers,balancelastpage]{revtex4}
\pdfoutput=1
\usepackage{latexsym}
\usepackage{amssymb}
\usepackage{epsfig,amsmath,graphics}
\usepackage{listings}
\usepackage{color}
\usepackage{dcolumn}
\usepackage{slashed}
\usepackage{cancel}
\usepackage{comment,latexsym} 
\usepackage{bm}
\usepackage{verbatim}
\usepackage{tabularx}
\usepackage{dcolumn}
\definecolor{Mahogany}{rgb}{0.62,0.24,0.15}
\definecolor{colorLink}{rgb}{0.6,0,0}
\definecolor{colorCite}{rgb}{0,.6,0}
\definecolor{colorURL}{rgb}{0,0.6,0.0}
\definecolor{colorTC}{rgb}{.2,.7,.2}
\definecolor{colorML}{rgb}{.7,.7,.2}
\usepackage[colorlinks=true,linktocpage=true,linkcolor=black,citecolor=colorCite,urlcolor=colorURL]{hyperref}
\usepackage{setspace}
\usepackage{xspace}
\usepackage{multirow}
\usepackage{titlesec}
\usepackage[bbgreekl]{mathbbol}
\usepackage{booktabs}
\usepackage{makecell}
\usepackage{array}

\newcolumntype{L}[1]{>{\raggedright\let\newline\\\arraybackslash\hspace{0pt}}m{#1}}
\newcolumntype{C}[1]{>{\centering\let\newline\\\arraybackslash\hspace{0pt}}m{#1}}
\newcolumntype{R}[1]{>{\raggedleft\let\newline\\\arraybackslash\hspace{0pt}}m{#1}}

\def\be{\begin{equation}}
\def\ee{\end{equation}}
\newcommand{\beq}{\begin{equation}}
\newcommand{\eeq}{\end{equation}}

\newcommand{\lsim}{\!\mathrel{\hbox{\rlap{\lower.55ex \hbox{$\sim$}} \kern-.34em \raise.4ex \hbox{$<$}}}}
\newcommand{\gsim}{\!\mathrel{\hbox{\rlap{\lower.55ex \hbox{$\sim$}} \kern-.34em \raise.4ex \hbox{$>$}}}}

\newcommand{\tr}{{\rm tr}\,}

\newcommand{\dark}{{\rm dark}}

\newcommand{\MET}{\diagup \!\!\!\!\! E_{T}}

\newcommand{\inv}{\textrm{inv}}

\expandafter\def\expandafter\normalsize\expandafter{%
    \normalsize
    \setlength\abovedisplayskip{8pt}
    \setlength\belowdisplayskip{8pt}
    \setlength\abovedisplayshortskip{8pt}
    \setlength\belowdisplayshortskip{8pt}
}

\usepackage{feynmp-auto}
\titleformat{\section}{\center\normalfont\fontsize{14}{15}\bfseries}{\thesection.}{1em}{}
\titleformat{\subsubsection}{\center\normalfont\fontsize{12}{15}}{\thesubsubsection.}{1em}{}

\begin{document}
$\quad$
\vskip 60 pt

\preprint{PUPT 2529}
 
\title{
LHC Searches for Dark Sector Showers
} 

\author{Timothy Cohen}
\affiliation{Institute of Theoretical Science, University of Oregon, Eugene, OR, 97403
\vspace{-4pt}}
\author{Mariangela Lisanti}
\affiliation{Department of Physics, Princeton University, Princeton, NJ 08544
\vspace{-4pt}}
\author{Hou Keong Lou}
\affiliation{Department of Physics, University of California, Berkeley, California 94720
\vspace{-10pt}\\
Theoretical Physics Group, Lawrence Berkeley National Laboratory, Berkeley, California 94720}
\vspace{-4pt}
\author{Siddharth Mishra-Sharma}
\affiliation{Department of Physics, Princeton University, Princeton, NJ 08544
\vspace{-4pt}}

\begin{abstract}
\vskip 1 pt
\begin{center}
{\bf Abstract}
\end{center}
\vskip -30 pt
$\quad$
\begin{spacing}{1.05}\noindent
This paper proposes a new search program for dark sector parton showers at the Large Hadron Collider (LHC).  
These signatures arise in theories characterized by strong dynamics in a hidden sector, such as Hidden Valley models.  A dark parton shower can be composed of both invisible dark matter particles as well as dark sector states that decay to Standard Model particles via a portal.  The focus here is on the specific case of `semi-visible jets,' jet-like collider objects where the visible states in the shower are Standard Model hadrons. We present a Simplified Model-like parametrization for the LHC observables and propose targeted inclusive search strategies for regions of parameter space that are not covered by existing analyses.  Following the `mono-$X$' literature, the portal is modeled using either an effective field theoretic contact operator approach or with one of two ultraviolet completions; sensitivity projections are provided for all three cases.  We additionally highlight that the LHC has a unique advantage over direct detection experiments in the search for this class of dark matter theories.
\end{spacing}
\end{abstract}

\maketitle
\newpage
\begin{spacing}{1.3}
\pagebreak

\section{Introduction}
The Large Hadron Collider (LHC) provides a unique opportunity to discover dark matter (DM) and study its properties.  To date, LHC DM searches have largely been focused on Weakly Interacting Massive Particles (WIMPs), neutral particles with weak-scale mass and interactions.  The signature of WIMPs at the LHC is relatively clean: 
 they simply leave the detector and their presence is inferred by enforcing transverse momentum conservation in each collision.  In contrast, non-WIMP scenarios can lead to very different collider signatures that require their own dedicated analyses.  An additional challenge lies in organizing the enormous variety of self-consistent theories into a finite number of inclusive searches.   To this end, we focus on a broad class of models characterized by  strong dynamics in a hidden dark sector.  We present a proposal for a new analysis framework that builds upon the existing DM program at the LHC and targets the distinctive phenomenology of these models. 
 
In many theories, the DM resides within a `dark sector'~\cite{Goldberg:1986nk, Strassler:2006im, Pospelov:2007mp, Feng:2008ya, ArkaniHamed:2008qn, Hochberg:2014kqa, Buen-Abad:2015ova}, defined by some internal set of new particles, forces, and interactions.  This hidden sector can communicate with the visible, \emph{i.e.}, Standard Model, sector through portal interactions---the renormalizable examples are the Higgs, photon, and lepton portals.  Strongly interacting dark sectors arise in a wide variety of new physics scenarios~\cite{Nussinov:1985xr, Barr:1990ca, Khlopov:2005ew, Gudnason:2006ug, Gudnason:2006yj, Khlopov:2008ty, Ryttov:2008xe, Foadi:2008qv, Alves:2009nf, Mardon:2009gw, Kribs:2009fy, Frandsen:2009mi, Lisanti:2009am, Belyaev:2010kp, Alves:2010dd, Lewis:2011zb, Buckley:2012ky, Hietanen:2012qd, Blinov:2012hq, Cline:2013zca, Bai:2013xga, Appelquist:2014jch, Detmold:2014qqa, Detmold:2014kba, Hochberg:2014dra, Hochberg:2014kqa, Appelquist:2015zfa, Appelquist:2015yfa}; the canonical example is the Hidden Valley~\cite{Strassler:2006im,Han:2007ae,Strassler:2006ri}.  Because the dynamics in the hidden sector can be arbitrarily complicated, these models tend to yield LHC signatures characterized by high-multiplicity final states, displaced vertices, and novel collider objects such as lepton, photon, or emerging jets~\cite{Strassler:2006im,Harnik:2008ax,Falkowski:2010cm,Falkowski:2010gv,Co:2015pka,Izaguirre:2015zva,Garcia:2015toa,Zhang:2016sll,Freytsis:2016dgf,Daci:2015hca,Hochberg:2015vrg,Schwaller:2015gea,Knapen:2016hky,Curtin:2015fna}.  This paper establishes a systematic study of yet another exotic possibility, semi-visible jets~\cite{Cohen:2015toa}.

We will assume that the strongly coupled hidden sector includes some families of dark quarks that bind into dark hadrons at energies below a dark confinement scale $\Lambda_d$.  While the dark hadrons interact strongly with each other, they interact only weakly with visible states through the portal.  If a dark quark is produced with transverse momentum $p_T \gg \Lambda_d$ in an LHC collision, it will shower and ultimately hadronize, producing collimated sprays of dark hadrons.  These states are invisible at colliders unless they can decay to the Standard Model.  Depending on the symmetries of the theory, some fraction of these states are likely to be stable, providing good DM candidates.  However, many of the hadrons should decay back to the visible sector through the portal coupling, which is required to produce the dark quarks in the first place.  Their decays may lead to a hadronic shower with DM interspersed amongst the visible states.  

\begin{figure}[t] 
   \centering
   \includegraphics[width=0.6\textwidth]{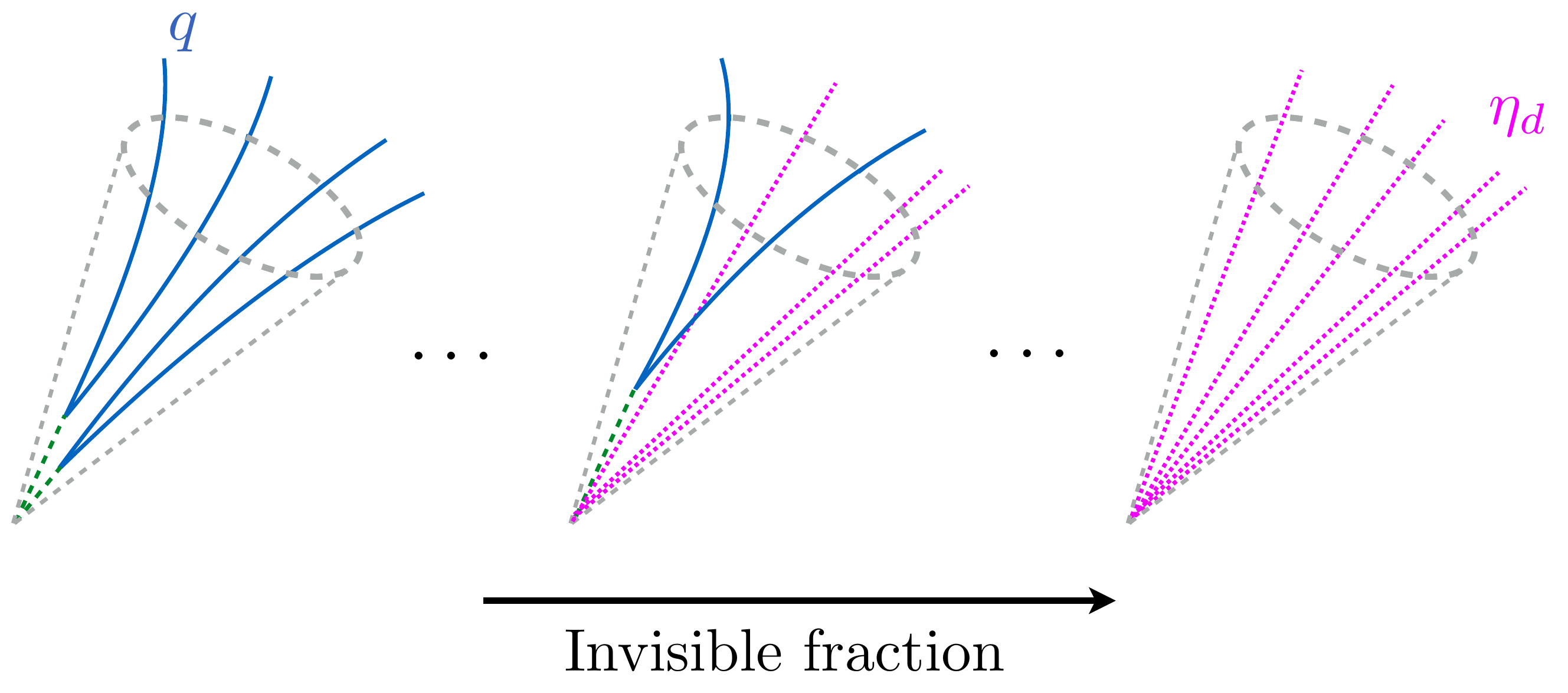} 
   \caption{The anatomy of a dark sector parton shower.  Unstable dark hadrons (green dashed) decay to Standard Model quarks, $q$ (solid blue).  If all the dark hadrons are unstable, then the jet is easily mistaken for an ordinary QCD jet (left panel).  However, some fraction of the dark hadrons, $\eta_d$, can be collider stable (pink dotted).  If both stable and unstable hadrons are produced in a collision, the end result is a semi-visible jet (middle panel).  In this case, the missing energy can be aligned along the direction of one of the jets.  If all the dark hadrons are stable, then only missing energy is inferred (right panel).  The LHC search strategy depends on the invisible fraction of the jet. } 
\label{fig:anatomy}
\end{figure}

Characterizing the individual shower constituents is difficult because of the large number of nearly collinear, low-$p_T$ states.  Greater success can be achieved by clustering the final states into jets and focusing on generic properties of the shower as a whole.  Figure~\ref{fig:anatomy} illustrates a range of allowed final states that can result, depending on the detailed particle content and parameter choices of the dark sector.  In the left-most diagram, all the hadrons are unstable and decay to light quarks.  The result looks very much like an ordinary QCD jet, although differences exist at the substructure level.  In the right-most diagram, all the dark hadrons are collider stable\footnote{The DM candidate proposed here is not necessarily assumed to constitute all of the observed relic density.}  and do not result in any direct visible signatures---in fact, these would be nearly indistinguishable from WIMP signatures, as we emphasize below.  The central diagram illustrates what happens when some fraction of the dark hadrons decay to quarks.  The result is a cluster of visible hadronic states that would be constructed as a jet, albeit an unusual one.  Because this jet has dark hadrons interspersed throughout, we refer to it as a `semi-visible' jet~\cite{Cohen:2015toa}.  Figure~\ref{fig:anatomy} illustrates the case for hadronic decay modes, but the same holds for any decay scenario.  One can, for example, consider dark hadron decays to heavy quarks, leptons, or photons.   

In the following, we present a search program for strongly interacting dark sectors that yield semi-visible jets. We will see that semi-visible jets generally lead to a new collider signal topology where the total momentum of the DM is correlated with the momentum of the visible states. In Sec.~\ref{sec:simplifiedmodels}, we introduce a simplified parametrization that covers the phase space realized by these theories.  Motivated by the standard LHC WIMP searches (referred to as `mono-$X$' searches, where $X$ can be a jet(s), a weak gauge boson, etc.), we focus on several different production channels.  
To begin, we remain agnostic about the new states that connect the dark sector to the Standard Model and rely on an effective theory framework where the interaction is modeled by a contact operator; this is discussed in Sec.~\ref{sec:contact}.  In Sec.~\ref{sec:resolved}, we consider dedicated searches for two ultraviolet (UV) completions of this contact operator.  Throughout, we emphasize the complementarity with standard LHC searches, indicating regions of parameter space where current analyses already have coverage, and other regions where new dedicated analyses are required.  In Sec.~\ref{sec:directdetection}, we show that direct detection experiments have limited sensitivity to these DM models, thereby emphasizing the critical role played by a dedicated LHC program.  We conclude in Sec.~\ref{sec:conclusions} with a discussion of additional final states, as well as control regions.  Two Appendices are included.  The first demonstrates the convergence of the separate UV models in the contact-operator limit, and the second shows the insensitivity of our search to variations in the dark sector parameters.  For the reader that would like to simulate the signal Monte Carlo used here, we provide all generation files at \url{https://github.com/smsharma/SemivisibleJets}.

\section{Signatures of Dark Sector Parton Showers}
\label{sec:simplifiedmodels}

Building an experimental program that systematically searches for all strongly coupled dark sectors is not feasible due to the large number of possible models.  This motivates inclusive searches with non-trivial signal efficiency to a wide range of scenarios.  The key is to realize that not all elements of a spectrum of new particle states and their ensuing interactions affect observable signatures at a collider.  This is why Simplified Models are now broadly used for supersymmetry~\cite{Alves:2011wf} and WIMP searches~\cite{Abdallah:2014hon, Abdallah:2015ter, Abercrombie:2015wmb, Kahlhoefer:2017dnp}.  The complicated dynamics of a dark sector have a limited number of effects on collider observables, primarily impacting the multiplicity of the final state, the fraction of invisible final-state particles, and the average $p_T$ of these states.  A search that targets these variables yields inclusive bounds in parameter space that can later be recast for any particular theoretical model. 

The remainder of this section provides concrete details on how to map an example dark sector Lagrangian onto a simplified parametrization,\footnote{By definition, a Simplified Model is written in terms of physical observables that are directly related to Lagrangian parameters.  It is not possible to do so for the dark sectors we consider here, as some of the observables depend on non-perturbative physics.  For this reason, we refer to our proposal as a `simplified parametrization,' even though it shares the same guiding principles as a Simplified Model.} and then translate it into Monte Carlo events.  The discussion is naturally divided into three parts.  Sec.~\ref{sec:dynamics} describes the hidden-sector dynamics, Sec.~\ref{sec:portals} focuses on the portal, and Sec.~\ref{sec:generation} details the signal and background generation, and describes the limit-setting procedure.

\subsection{Dark Sector Dynamics} 
\label{sec:dynamics}

This section elucidates the dark sector physics.  For illustration, we consider a toy scenario where the dark sector is an $SU(2)_d$ gauge theory with coupling $\alpha_d = g_d^2/(4\pi)$, containing two fermionic states $\chi_a = \chi_{1,2}$ in the fundamental representation:
\begin{align}
  \mathcal{L}_\dark \supset -\frac{1}{2}\tr G^d_{\mu\nu}G^{d\mu\nu} -\overline\chi_a \left(i\slashed D - M_{d,a}\right) \chi_a \, ,
  \label{eq:darkL}
\end{align}
where $G^d_{\mu\nu}$ is the dark gluon field strength, and $M_{d,a}$ is the mass for the $\chi_a$; we assume that the dark quarks have a common mass $M_d$.  Similar to QCD, the fermions act as dark quarks that interact strongly with coupling strength $\alpha_d$.  The dark quarks form bound states at the confinement scale $\Lambda_d$, where $\alpha_d$ becomes non-perturbative. It is technically natural for there to be large hierarchies between $\Lambda_d$ and $M_d$ due to an approximate chiral symmetry, however we focus on the case where  $M_d\sim\Lambda_d$ and take $\Lambda_d = M_d/2$ for concreteness.  The general spectrum of these dark hadronic states depends on non-perturbative physics and is difficult to calculate, though some attempts have been made for specific examples in the literature~\cite{Lewis:2011zb,Hietanen:2013fya,Detmold:2014qqa,Detmold:2014kba}.  Fortunately, most of the details concerning the spectrum are irrelevant for collider observables; we focus on three aspects that do have an effect.\footnote{For a study showing what can be learned about the mass spectrum at a future $e^+\,e^-$ collider, see~\cite{Hochberg:2015vrg, Hochberg:2017khi}.}  

It is important to distinguish between bound states that do or do not decay back into Standard Model particles. Clearly, a stable state (or one that decays only within the hidden sector) leaves the detector without a trace and results in missing energy. If it does decay to the Standard Model, the decay products can be observed directly.  Basic symmetry arguments can be used to determine the stability of the hadrons formed from pairs of the $\chi_a$. For generic masses, the dark-isospin number $U(1)_{1-2}$ and dark-baryon number $U(1)_{1+2}$ (where ``1'' and ``2'' refer to the flavor indices) are accidental symmetries of the theory.   For instance, the mesons $\chi_1\chi_2^\dagger$ and $\chi_1^\dagger \chi_2$ are charged under dark-isospin, and can be stabilized if these symmetries are preserved.  Similarly, the baryons $\chi_1\chi_2$ and $\chi_1^\dagger \chi_2^\dagger$ can be stable because they are charged under dark-baryon number. By contrast, the mesons $\chi_1^\dagger \chi_1$ and $\chi_2^\dagger \chi_2$ are not charged under either symmetry and are thus expected to decay.

Additionally, different spin and CP configurations of the bound states are possible.  For example, the dark hadrons can form scalar, pseudoscalar, vector and/or higher spin combinations. In what follows, we assume that the DM is the lightest stable scalar dark hadron, $\eta_d$.  The spin quantum numbers can determine aspects of the decay parametrics. For example,  vector mesons could decay promptly if coupled to the Standard Model through a vector portal, while the decay of an unstable (pseudo)scalar would be suppressed by additional mass insertions. The decay of higher spin states may also be suppressed by loop factors if they cannot decay within the hidden sector. This implies that generically some displaced vertices are expected, which could provide additional handles for improving signal discrimination.  However, we choose to design searches that are insensitive to the presence of displaced vertices, which can be strongly model-dependent.  In practice, we treat all decays as prompt throughout the rest of the paper.

The relative number of stable and unstable states in the dark sector can vary significantly depending on the details of the theory. For example, one can generalize this toy model to an arbitrary confining sector with any number of colors, $N_c$, or flavors, $N_f$. Extending the flavor symmetry to $U(1)^{N_f}$ would naively result in $N_f$ ``uncharged" mesons and $N_f(N_f-1)/2$ ``charged" mesons.  This impacts the fraction of possible stable to unstable states in the hidden sector, thereby changing the amount of observed missing energy. In addition, there should also be baryons, although their production in the shower will tend to be suppressed.

Introducing a mass splitting between the various mesons can also alter the multiplicity of the final state and the relative fraction of stable and unstable states.  Following the Lund string model for fragmentation, the yield of a meson is exponentially sensitive to the meson mass; heavier mesons are exponentially less likely to be produced during hadronization~\cite{Andersson:1983ia}.  This is captured by the suppression factor for estimating the ratio of $\chi_2$ to $\chi_1$ production:
\begin{align}
T_{21} = \text{exp}\Bigg[-\frac{4\,\pi \, (M_{d,2}^2-M_{d,1}^2)}{\Lambda_d^2}\Bigg]\,.
\end{align}
When the mass splitting between $\chi_1$ and $\chi_2$ is large compared to the dark confinement scale, the production of stable dark mesons (\emph{e.g.}, $\chi_1 \chi_2^\dagger$ and $\chi_1^\dagger \chi_2$) is suppressed.  This in turn reduces the number of invisible states in the dark parton shower.  

To capture the variation in the number of stable to unstable states in dark sector models, we introduce the following parameter:
\begin{align}
r_\inv \equiv \left \langle \frac{\# \text{ of stable hadrons}}{\# \text{ of hadrons}} \right \rangle \,.
\end{align}
If the dark hadrons decay entirely to visible states, then $r_\inv \rightarrow 0$.  The opposite limit arises when none of the dark hadrons decay back to the Standard Model (on collider timescales).  In this limit, $r_\inv \rightarrow 1$, and this scenario would be indistinguishable from WIMPs.  

The two important parameters that remain are (1) the characteristic mass scale for the dark hadrons, $M_d$, and (2) the dark strong coupling, $\alpha_d$.\footnote{While these are both physical parameters, we prefer to think of them in the same spirit as $r_\inv$.  This is justified since the mapping between the real Lagrangian parameters and what is actually computed by the simulation is an unsolved problem and is certainly not captured using current state-of-the-art tools.  Furthermore, due to the inclusive nature of the search, different dark sector Lagrangians can be mapped onto the same collider signatures.}  Both affect the number of dark hadrons that are produced during the dark shower, which subsequently impacts the multiplicity of the dark jet.  These parameters also alter the relative $p_T$ of the states produced in the shower, which manifests in the detector as the amount of missing and visible energy of the final states.  For simplicity, we assume that none of the dark hadron resonance structure is relevant, such that the collider observables are insensitive to any mass splittings in the dark sector spectrum.  This assumption only applies if the hard interaction scale of the new-physics event is much larger than the confining scale $\Lambda_d$.  Furthermore, we assume that $\sqrt{\hat{s}} \gg \Lambda_d$ for the events that populate the signal region such that the perturbative shower is a good approximation.  This is true for the searches described below due to the strong kinematic cuts.  

\begin{figure}[t!] 
\hspace{-0.9 cm} 
\includegraphics[width=0.5185\textwidth]{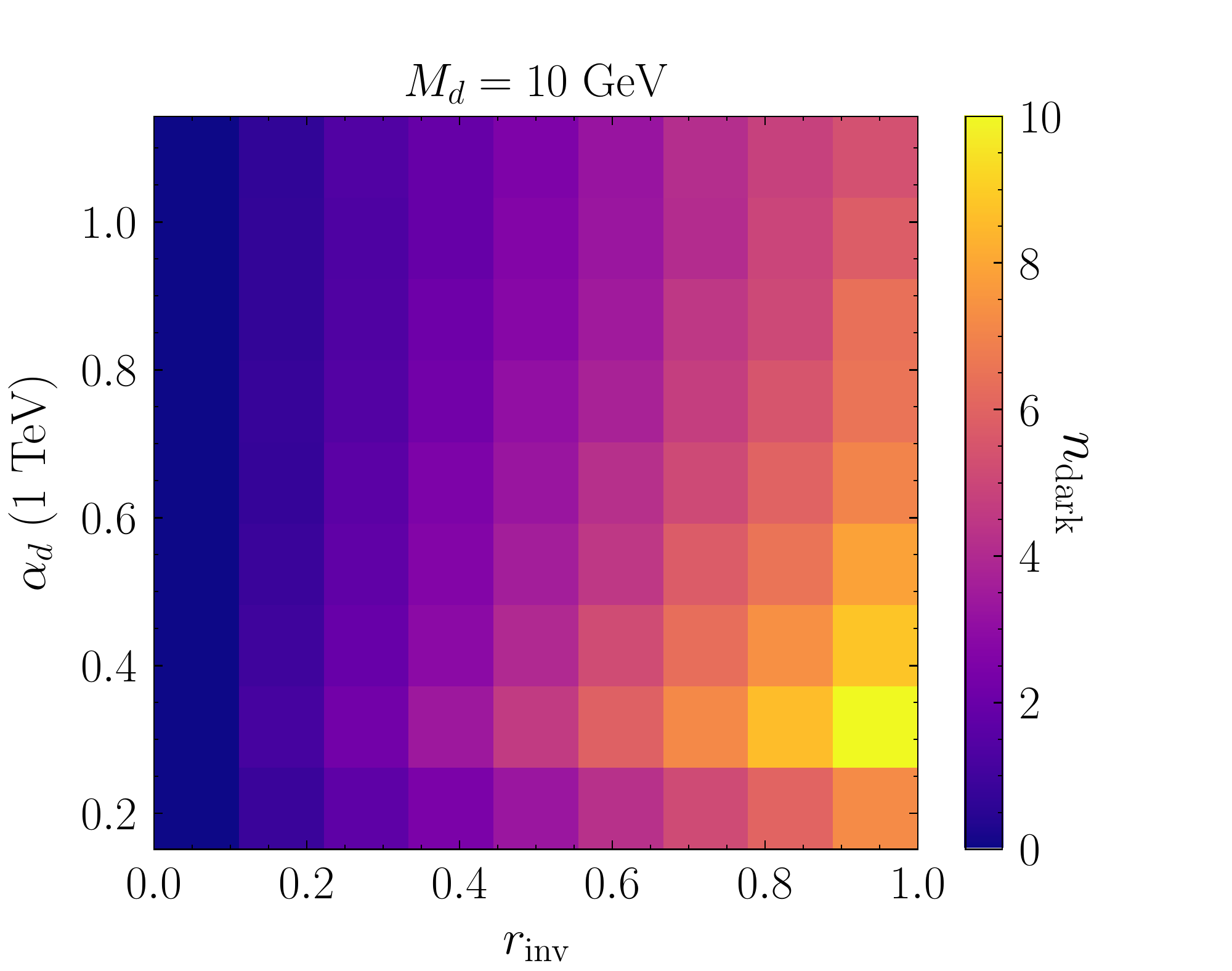} \hspace{-0.29 cm}~ 
 \includegraphics[width=0.528\textwidth]{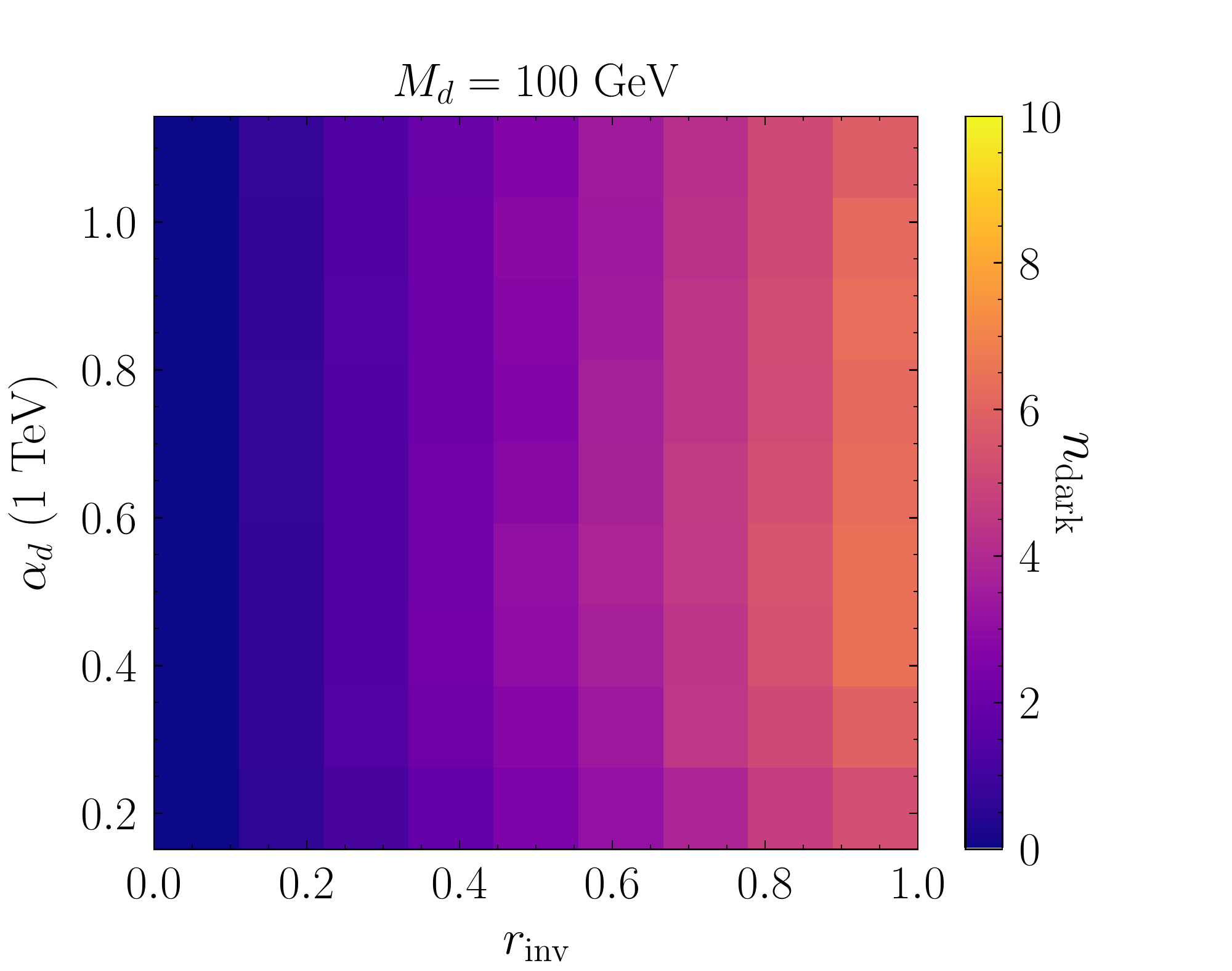}  \hspace{-0.8cm}
\caption{The number of dark hadrons that are produced per event for different values of $\alpha_d$ and $r_\mathrm{inv}$.  The left panel corresponds to the dark hadron mass scale $M_d = 10$~GeV, while the right panel corresponds to $M_d = 100$~GeV.  Here, $\alpha_d = 2\pi / (b\log(\frac{1~\mathrm{TeV}}{\Lambda_d}))$ where $\Lambda_d$ is the confinement scale and $b = \frac{11}{3}N_c - \frac{2}{3}N_f$. Note that $\alpha_d(1\,\text{TeV}) = 0.23\,(0.45)$ approximately corresponds to $\Lambda_d= 10\,(100) \text{ GeV}$.  The simulation used to generate this figure is described in Sec.~\ref{sec:generation}.}  
\label{fig:multiplicity}
\end{figure}

Figure~\ref{fig:multiplicity} demonstrates how the multiplicity in the shower changes with $\alpha_d$ and $r_\inv$ for two choices  of $M_d$.  The particle multiplicity is smaller than in Standard Model QCD showers due to the absence of light pion-like states with mass below the confinement scale.  The number of dark hadrons produced in the shower $n_\dark$ increases with $r_\inv$.  Additionally, $n_\dark$ generally gets larger as $\Lambda_d$ decreases, due to the growing hierarchy between the confinement and hard interaction scales. This enhancement stops when $\Lambda_d\lesssim M_d$, where the dark shower is cut-off by the dark quark mass.  For $M_d = 10 \,(100)$~GeV, this occurs for $\alpha_d(1~\text{TeV}) \sim 0.23\,(0.45)$.  Additionally, as $M_d$ increases from 10 to 100~GeV, the overall number of dark hadrons in the shower decreases.  Note that for $M_d\gg \Lambda_d$, the fragmentation should be dominated by dark glueballs---this effect is irrelevant for the parameter space explored in this paper where $M_d \sim \Lambda_d$.  Modeling the production of glueballs within \texttt{Pythia} is outside the scope of this work.

Armed with this parametrization of the dark sector physics, we next turn to the details of the portal that connects it to the Standard Model.

\subsection{Portal to the Dark Sector}
\label{sec:portals}

\begin{figure}[h!]
   \centering
   \includegraphics[width=0.8\textwidth]{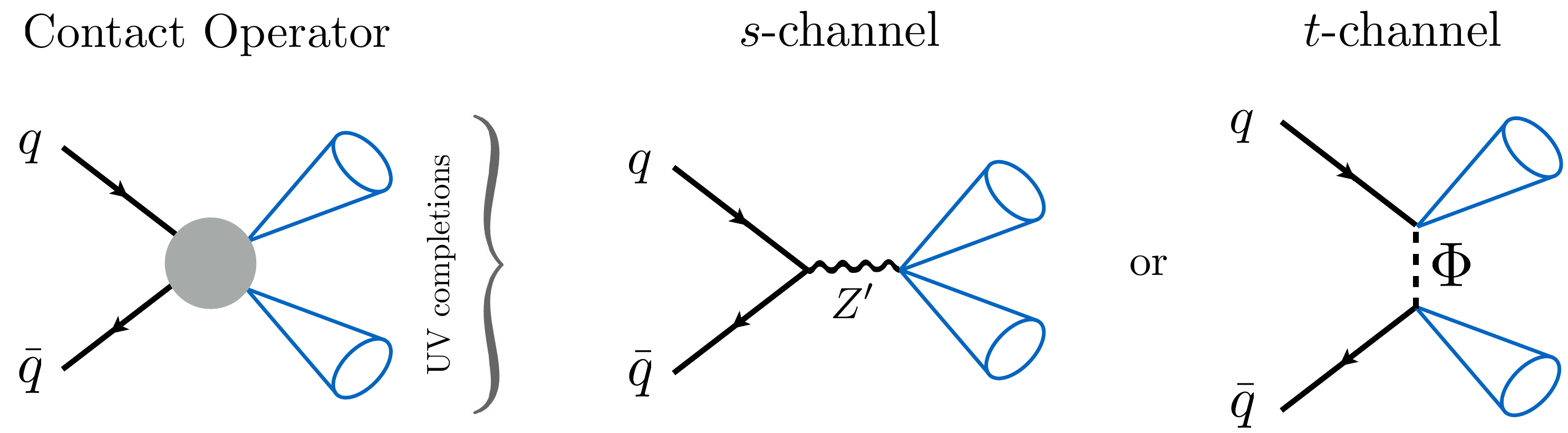} 
   \caption{The portals considered in this work.}
   \label{fig:portals}
\end{figure}

The portal describes how the hidden sector communicates with the visible Standard Model states.  This determines the production channels at the LHC and implies a particular set of decay modes.  Following the mono-$X$ literature, we study the three portals illustrated in Fig.~\ref{fig:portals}.  Specifically, we consider the contact operator limit~\cite{Goodman:2010yf, Beltran:2010ww, Fox:2011pm} where the mediator is integrated out, as well as two UV completions of this operator~\cite{Frandsen:2012rk, An:2012va, An:2012ue, An:2013xka, Buchmueller:2013dya, Chang:2013oia, Bai:2013iqa,  Dreiner:2013vla, DiFranzo:2013vra, Papucci:2014iwa, Buchmueller:2014yoa, Hamaguchi:2014pja, Garny:2014waa, Harris:2014hga, Jacques:2015zha, Liew:2016oon, Englert:2016joy}.  

To summarize, a strongly interacting hidden sector can be described by three dark sector parameters ($\alpha_d$, $M_d$, and $r_\inv$) and a portal parameter ($\Lambda$).  While we simulate an $SU(2)_d$ sector to derive the results that follow for concreteness, this same  approach can be applied to any strongly-interacting hidden sector that decays back to Standard Model quarks.  This provides a powerful framework in which the collider results can be presented in terms of generic parameters that can be mapped onto a range of strongly interacting dark sector theories.  

\subsection{Event Generation and Sensitivity Estimation}
\label{sec:generation}

Signal and background events are generated using~\texttt{MadGraph5\_aMC@NLO}~\cite{Alwall:2014hca} with parton distribution functions \texttt{NN23LO1}~\cite{Ball:2013hta} and are showered using \texttt{Pythia8}~\cite{Sjostrand:2007gs}.  To simulate the dark sector shower and hadronization, we use the Hidden~Valley module~\cite{Carloni:2011kk,Carloni:2010tw} in \texttt{Pythia8}, where we have implemented the running of the dark coupling $\alpha_d$ as in~\cite{Schwaller:2015gea}.  All events are then passed through the \texttt{DELPHES3}~\cite{deFavereau:2013fsa} detector simulator with CMS settings.  Jets are initially clustered using the anti-k$_T$~\cite{Cacciari:2008gp} algorithm with $R=0.5$~\cite{Nachman:2014kla}.  

To perform the searches described in this paper, we must implement $r_\text{inv}$ within our simulation framework.  First, we shower and hadronize in the dark sector, producing dark mesons.  Next, we decay all the dark mesons either to a quark pair or to invisible DM particles.  The invisible branching ratio is equal to $r_\text{inv}$.  

We generate 20,000 signal events, unless otherwise specified, for each parameter point at 13~TeV center-of-mass energy using the MLM~\cite{Alwall:2007fs} matching procedure implemented in \texttt{MadGraph}, with the \texttt{xqcut} parameter set to 100~GeV and matched up to 2 jets.  To model the background and estimate the sensitivity reach of the searches, we also generate 5~million $W^\pm/Z+\text{jets}$ events, matched up to two jets (\texttt{xqcut} = 40~GeV), and 10~million QCD events matched up to 4~jets (\texttt{xqcut} = 40~GeV).  Matched $t\,\overline{t}+\text{jets}$ backgrounds are generated in the semi-leptonic and di-leptonic channels with 5~million events each, including emission of up to one extra jet (\texttt{xqcut} = 80~GeV). We weigh the parton-level background events using the bias module implemented in \texttt{MadGraph} and set a leading parton jet $p_T$ cut of 200 GeV.  Both of these choices improve the background statistics in the high missing energy ($\MET$) tail. We validate our electroweak and $t\,\overline t$ background samples by comparing against Monte Carlo in~\cite{Aaboud:2016tnv}. We use the $\MET > 250$~GeV signal region in that study to calibrate the $K$-factors (accounting for NLO corrections to the overall cross section) for our backgrounds, finding values of 1.0, 1.1 and 1.7 for the $W^\pm+\text{jets}$, $Z+\text{jets}$ and $t\,\overline t+\text{jets}$ samples, respectively.  We obtain a $K$-factor $\sim$1.0 for QCD by matching to the di-jet distributions in~\cite{Aaboud:2017yvp}, and make the conservative choice to not implement a $K$-factor for the signal. 

There are two kinds of searches described in the following sections. For the cut-and-count approaches, we treat the background as an Asimov dataset to obtain the expected exclusion reach, following~\cite{Cowan:2010js}. Given the number of expected signal(background) events, $s(b)$, we then compute the Poisson log-likelihood ratio, $L(s+b, b)$, of the signal hypothesis to the background-only hypothesis. A 95\% confidence limit is set by varying the number of signal events such that $L(s+b,b)=2^2$. In the large background limit, $L(s+b,b)\rightarrow s^2/(s+b) $, and a standard $2\sigma$ Gaussian limit is recovered. To compute the expected exclusion reach for the shape analysis in Sec.~\ref{sec:sChan}, we treat the background as an Asimov dataset and the final Poisson log-likelihood ratio is computed by summing over the contribution from each bin.  Because we are primarily interested in comparing different search strategies, as opposed to the precise numbers provided by the projections themselves, this simple treatment of the statistics suffices.  For simplicity, no systematic errors are included in the searches proposed here.  A detailed study of the relevant systematic uncertainties is beyond the scope of this paper and will require careful study in any experimental implementation of this proposal.  

Now we are equipped with all the necessary technology to develop a semi-visible jet search strategy and provide an estimate of the mass reach that could be derived using the current LHC data set.

\section{Dark Sector Showers from Contact Operators}
\label{sec:contact}

In this section, we consider the case where the portal is modeled as a contact operator, and show that it leads to semi-visible jets.  We focus on the following dimension-six operator:  
\begin{align}
  \mathcal{L}_\text{contact}\supset \frac{c_{ijab}}{\Lambda^2}\,\big(\overline q_i \gamma^\mu q_j \big)\big (\overline \chi_a \gamma_\mu \chi_b\big) \, ,
  \label{eq:contact}
\end{align}
where $\Lambda$ is the characteristic dimensionful scale for the operator, and the $c_{ijab}$ are $\mathcal{O}(1)$ couplings that encode the possible flavor structures.  As discussed in Sec.~\ref{sec:dynamics} above, the DM $\eta_d$ is a scalar bound state comprised of the $\chi$'s.  Of course, a variety of operators can be written that span a range of effective interactions and spin states of $\chi_a$.  While the following analysis can be repeated for these different scenarios, we focus on the vector contact operator as an illustrative example.  We also restrict ourselves to the production mode $u\,\overline{u}, d\,\overline{d} \rightarrow \chi\,\overline{\chi}$, which corresponds to the flavor structure $c_{ij ab} = c\, \delta_{ij}\delta_{ab}$.   Flavor constraints generally allow a richer flavor structure, \emph{e.g.} one could apply the Minimal Flavor Violation (MFV) assumption to $c_{ijab}$.  Assuming MFV, heavy-flavor production channels dominate, leading to final states rich in bottom and top quarks.  In contrast, the diagonal flavor structure assumed here leads to dominantly light-flavor jets.  

\begin{figure}[t!] 
   \centering
   \includegraphics[width=0.8\textwidth]{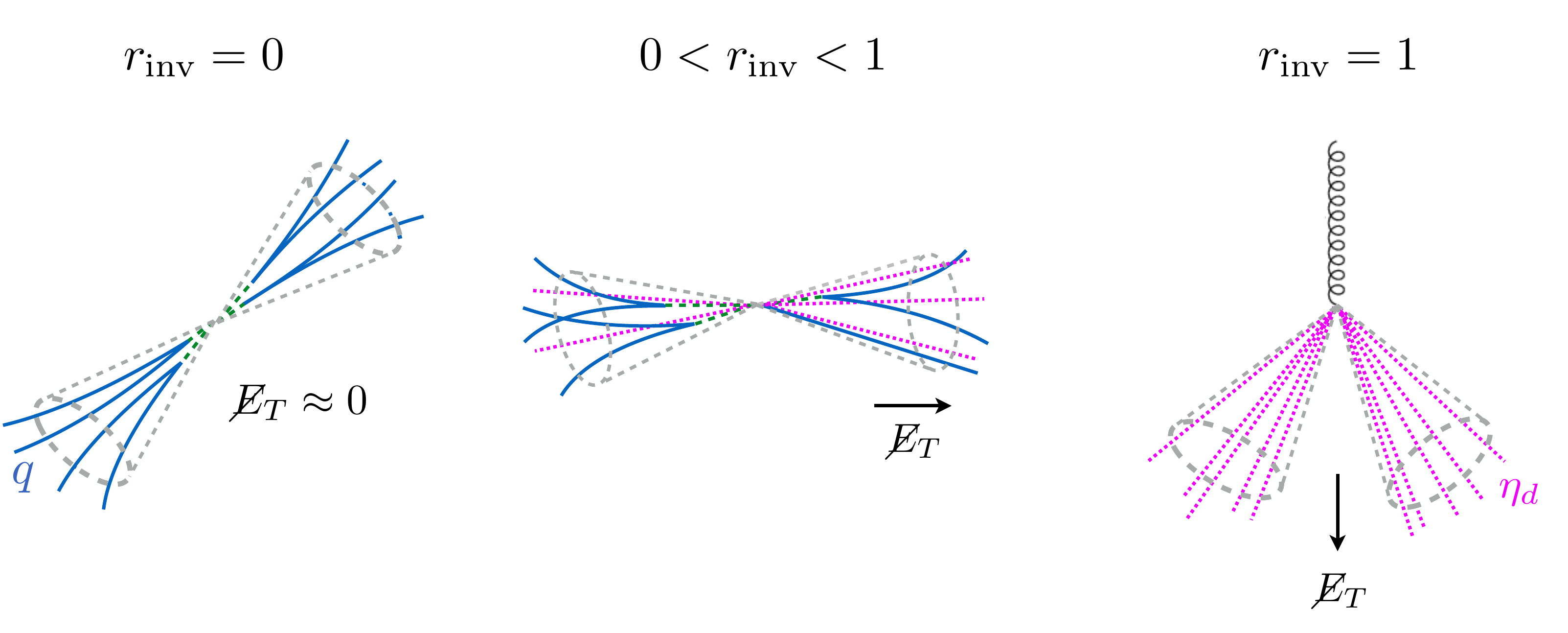} 
   \caption{Illustration of the typical missing energy direction for several different $r_\inv$ scenarios.}
   \label{fig:metfigure}
\end{figure}

When dark quarks are pair-produced at the LHC, they shower and hadronize in the hidden sector.  The magnitude and orientation of the missing energy in each event depends sensitively on the relative fraction of stable to unstable dark mesons that are produced in the shower.  The possibilities are illustrated in Fig.~\ref{fig:metfigure}.  When $r_\inv = 0$, all the dark hadrons decay to quarks and thus there is no parton-level missing energy (neglecting neutrinos that are produced from heavy-flavor quark decays).  When $r_\inv =1$, all the dark hadrons are collider stable.  Initial-state-radiation (ISR) is required to observe such events, as in the standard WIMP case.  The ISR jet boosts the dark hadrons in the antipodal direction, leading to non-vanishing missing energy that is oriented opposite the jet.\footnote{The ISR spectrum for $r_\inv=1$ is not identical to that for a WIMP.  While the number of WIMPs produced in each event is constant, the number of dark hadrons produced in a shower varies from event to event, which can affect the $\MET$ spectrum~\cite{Englert:2016knz}.}  In the intermediate $r_\inv$ scenario, two back-to-back semi-visible jets are produced and the missing energy points in the direction of the jet that contains the most stable mesons.

\begin{figure}[t!] 
\hspace{-0.9 cm} 
 \includegraphics[width=0.5\textwidth]{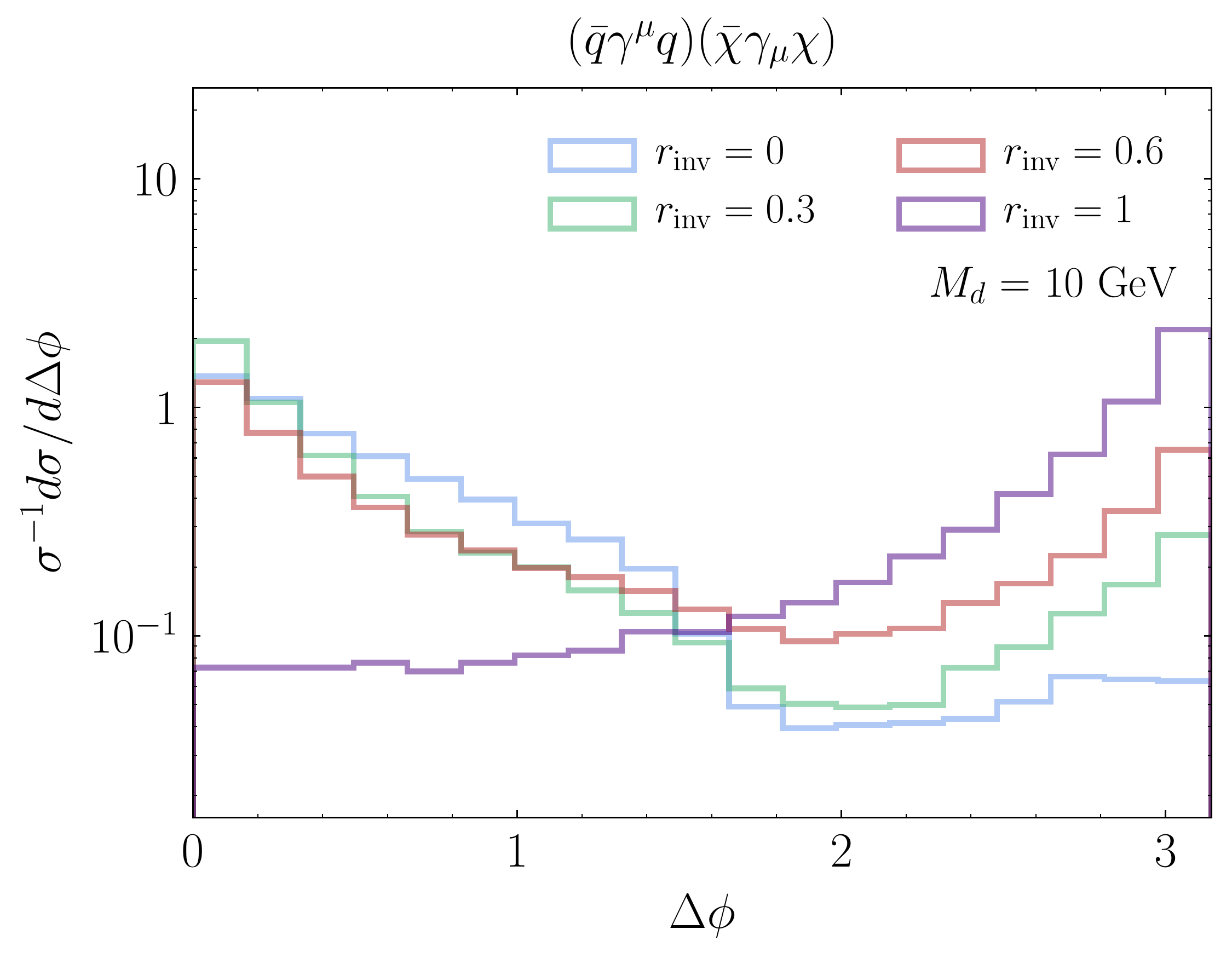}\hspace{-0.29 cm}~
\includegraphics[width=0.5\textwidth]{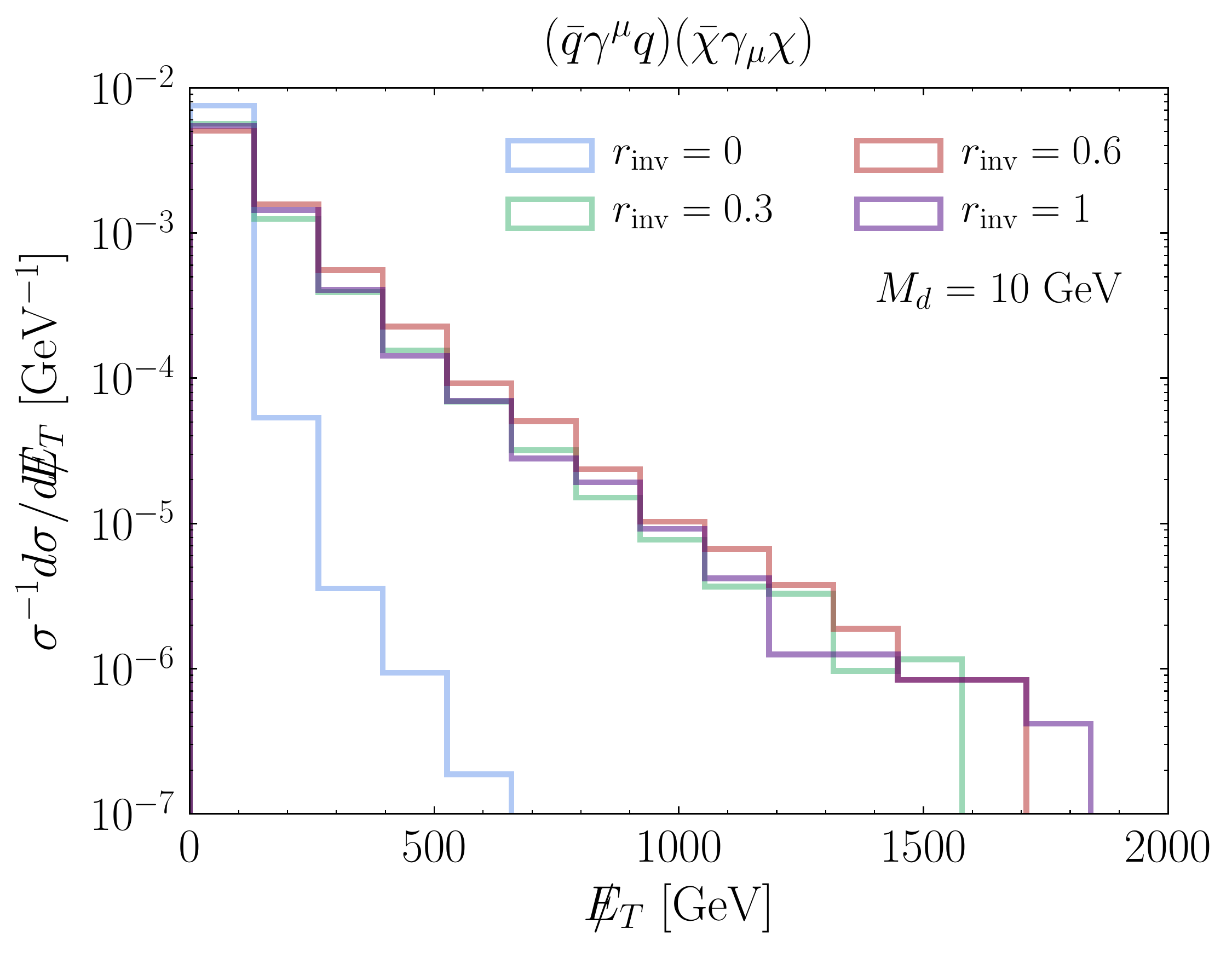}   \hspace{-0.8cm}
\caption{Kinematic distributions for $\Delta\phi$ (left) and missing energy (right) before trigger and preselection cuts are applied. The distributions correspond to the vector contact operator, with $M_d=10$~GeV and $r_\inv= 0, 0.3, 0.6, 1.0$ (blue, green, red and purple, respectively). }  
\label{fig: contact_distributions}
\end{figure}

To study this behavior quantitatively, we generate events for the vector contact operator by taking the large-mass limit for an $s$-channel mediator (see Appendix~\ref{sec:ApproachContactLimit} for further details), with  60,000 signal events produced over the range of $r_\mathrm{inv}$ values.  The mapping from cross section to $\Lambda$ is then evaluated for $c=1$.  The left panel of Fig.~\ref{fig: contact_distributions} shows the $\Delta \phi$ distributions for the signal, where 
\begin{equation}
\Delta \phi \equiv \underset{i\leq 4}{\text{min}} \Big\{ \Delta \phi_{j_i,\MET} \Big\} \, 
\end{equation}
and $ \Delta \phi_{j_i,\MET}$ is the angle in the rapidity-azimuthal angle plane between the $p_T$ of  the $i^\text{th}$ jet and the missing transverse momentum vector.  When $r_\inv = 1$, the missing energy is typically oriented opposite to the hardest jet in the event, as expected for the ISR regime.  As $r_\inv$ decreases, the distribution in $\Delta \phi$ becomes peaked towards zero, demonstrating that the missing energy becomes closely aligned along the direction of one of the jets in the event.    The right panel of Fig.~\ref{fig: contact_distributions} illustrates the $\MET$ distributions for $M_d = 10$~GeV and several values of $r_\inv$.  The amount of missing energy in the event increases as $r_\inv$ goes from $0$ to $1$.  

\begin{figure}[t!] 
\hspace{-0.9 cm} 
\includegraphics[width=0.5\textwidth]{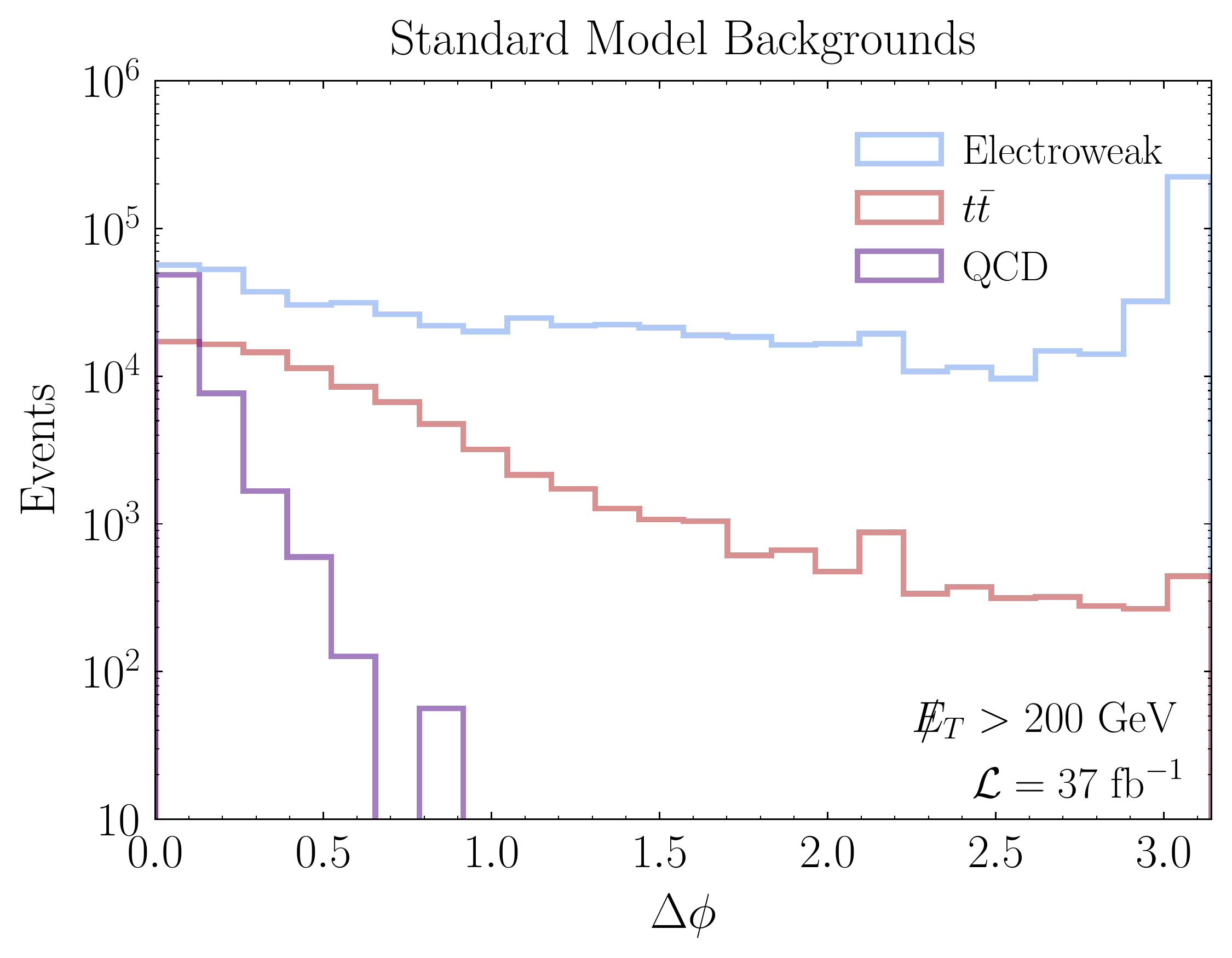} \hspace{-0.29 cm}~ 
\includegraphics[width=0.5\textwidth]{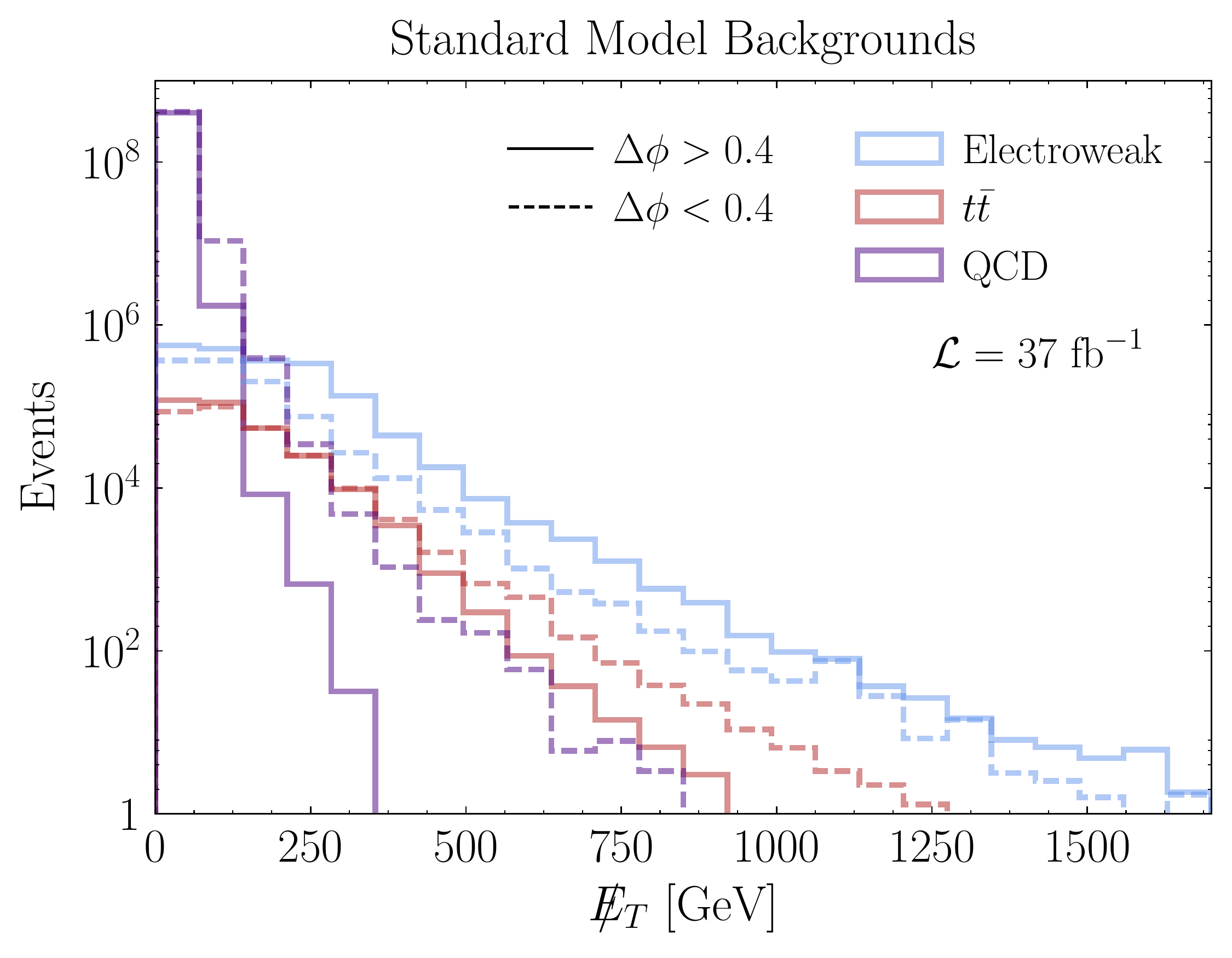} \hspace{-0.29 cm}~ 
\caption{(Left) $\Delta\phi$ distributions for the Standard Model backgrounds.   (Right)~Missing energy distributions for the Standard Model backgrounds with a cut of $\Delta\phi > 0.4$ (solid) and $\Delta\phi < 0.4$ (dashed).  No trigger or preselection cuts are applied, except for the requirement that $\MET > 200$~GeV in the left panel.}  
\label{fig: background_distributions}
\end{figure}

To study the projected sensitivity for the vector contact operator, we perform an optimized cut analysis on two separate signal regions---one with $\Delta \phi > 0.4$ and the other with $\Delta \phi < 0.4$.  The former is the standard requirement for most current searches at the LHC, and is implemented to minimize contamination from jet-energy mis-measurement.  This is exemplified by the left panel of Fig.~\ref{fig: background_distributions}, where the QCD background falls off steeply with $\Delta \phi$.  Requiring $\Delta \phi > 0.4$ removes a significant fraction of the high-$\MET$ QCD events, as demonstrated in the right panel of Fig.~\ref{fig: background_distributions}.  Even when $\Delta \phi < 0.4$, however, there is a negligible contribution from QCD above $\MET\sim800$~GeV.  In contrast, the top background is less steep and the  electroweak background is nearly isotropic such that cutting on $\Delta \phi$ has a less significant effect.  Note that the signal populates the control region currently utilized by standard searches when $\Delta \phi < 0.4$, which can significantly complicate the background determination in a data analysis.  We comment on this further in Sec.~\ref{sec:conclusions}. 

Considering two separate regions with $\Delta \phi$ greater/less than~0.4 allows us to study the complementarity between the two approaches.   At the trigger level, we require $\MET > 200$~GeV and a jet with $p_T > 250$~GeV and $|\eta| < 2.8$.  Additionally, events containing isolated electrons(muons) with $|\eta| < 2.5$ and $p_T > 20(10)$~GeV are vetoed.  We optimize the missing energy cut to maximize the signal sensitivity for a given $r_\inv$.  The cut is chosen from the list $\MET > [400, 600, 800, 1000, 1200]$~GeV; however, in cases where $\Delta\phi < 0.4$, the minimum $\MET$ requirement is not allowed to go below 800~GeV to avoid contamination from the QCD background.  An example cut-flow table for $\sigma(p\,p\rightarrow \overline{\chi}\,\chi$) = 1~pb is provided in Table~\ref{tab:cut-flowContactOp}.  

\begin{table}[t]
\footnotesize
\renewcommand{\arraystretch}{1.4}
\begin{tabular}{C{3cm}|C{1.9cm}C{1.9cm}C{1.9cm}|C{1.6cm}C{1.7cm}C{1.5cm}C{1.7cm}}
\multicolumn{8}{c}{\sc{Contact operator}}\\
\Xhline{3\arrayrulewidth}
 & \multicolumn{3}{c|}{\textbf{Signal} ($r_\inv$)} & \multicolumn{4}{c}{\textbf{Background}} \\
Cuts &    0.1 &   0.5 &   0.9 &       $Z+\text{jets}$ &       $W^\pm+\text{jets}$  &     $t\,\overline t+\text{jets}$ &      QCD  \\
\hline
Trigger and presel. &  2000[2.58] &  4920[6.34] &  2340[3.02] &  $2.3\times 10^5$ &  $2.5\times 10^5$ &  $6.9\times 10^4$ &  $5.7\times 10^4$ \\
$\MET > 800$ &    43[1.01] &   174[3.94] &   108[2.49] &    1160 &     536 &     80 &      0 \\
$\Delta\phi > 0.4$ &    0[0] &    31[0.89] &     73[2.0] &    1050 &     209 &      8 &      0  \\
or & & & & & \\  
$\Delta\phi < 0.4$ &    42[1.81] &   142[5.57] &    35[1.51] &     110 &     326 &     72 &      0  \\
\Xhline{3\arrayrulewidth}
\end{tabular}
\caption{Cut-flow table for the vector contact operator, assuming a production cross section $\sigma(p\,p\rightarrow \overline{\chi}\,\chi$) = 1 pb for $\mathcal L = 37$ fb$^{-1}$ at 13 TeV.  We show the number of signal and background events that remain after trigger/preselection cuts, as well as after the addition of a missing energy cut with either $\Delta\phi > 0.4$ or $<0.4$.   The numbers in brackets correspond to an estimate of the significance $s/\sqrt{s+b}$ at each stage of the cut-flow, where $s(b)$ is the number of signal(background) events. The $\MET$ cuts are optimized in each signal region; we only show the results for $\MET > 800$~GeV here as an example.}
\label{tab:cut-flowContactOp}
\end{table}

\begin{figure}[t!] 
\hspace{-0.9 cm} 
\includegraphics[width=0.5185\textwidth]{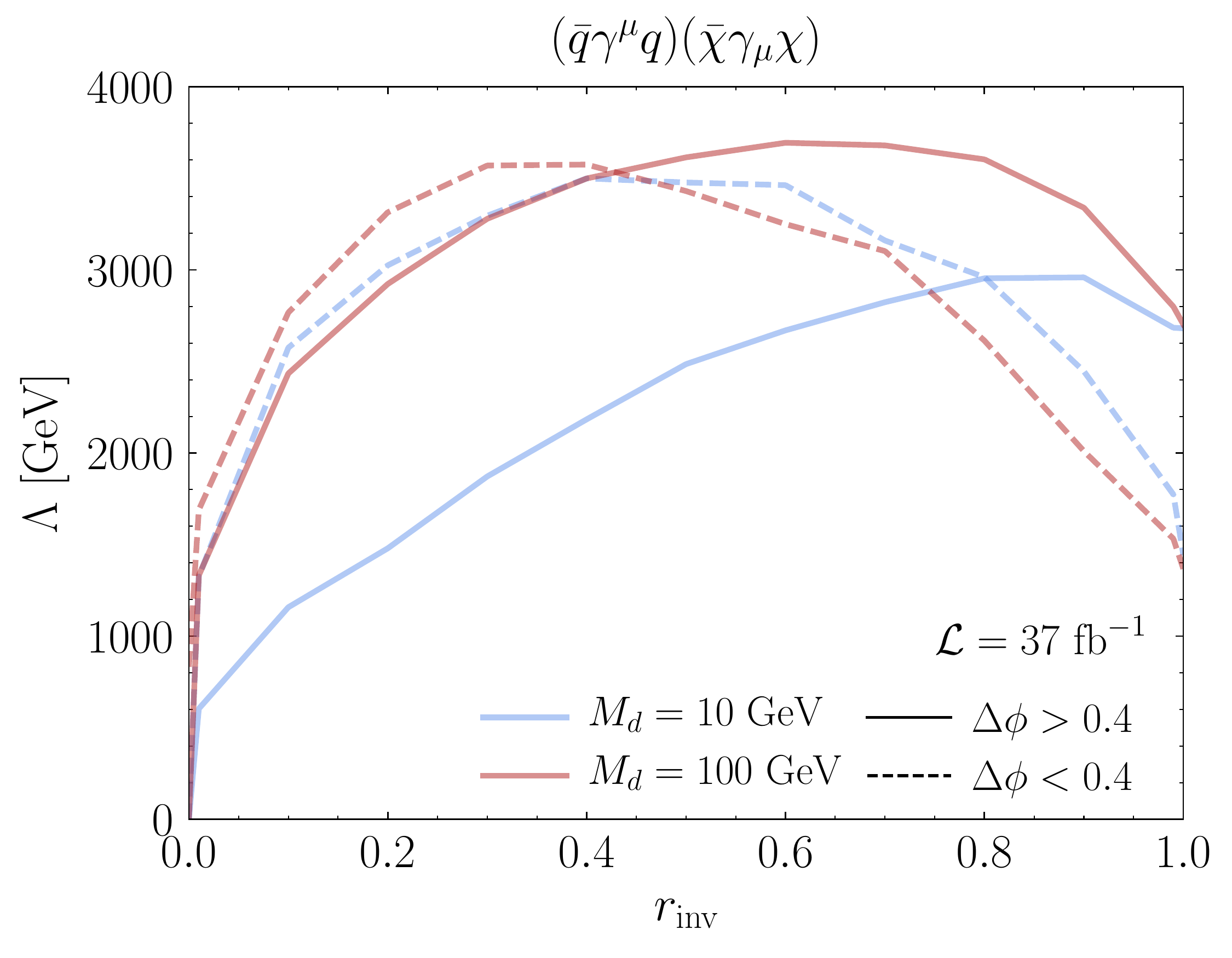}  \hspace{-0.29 cm}~ 
\includegraphics[width=0.5185\textwidth]{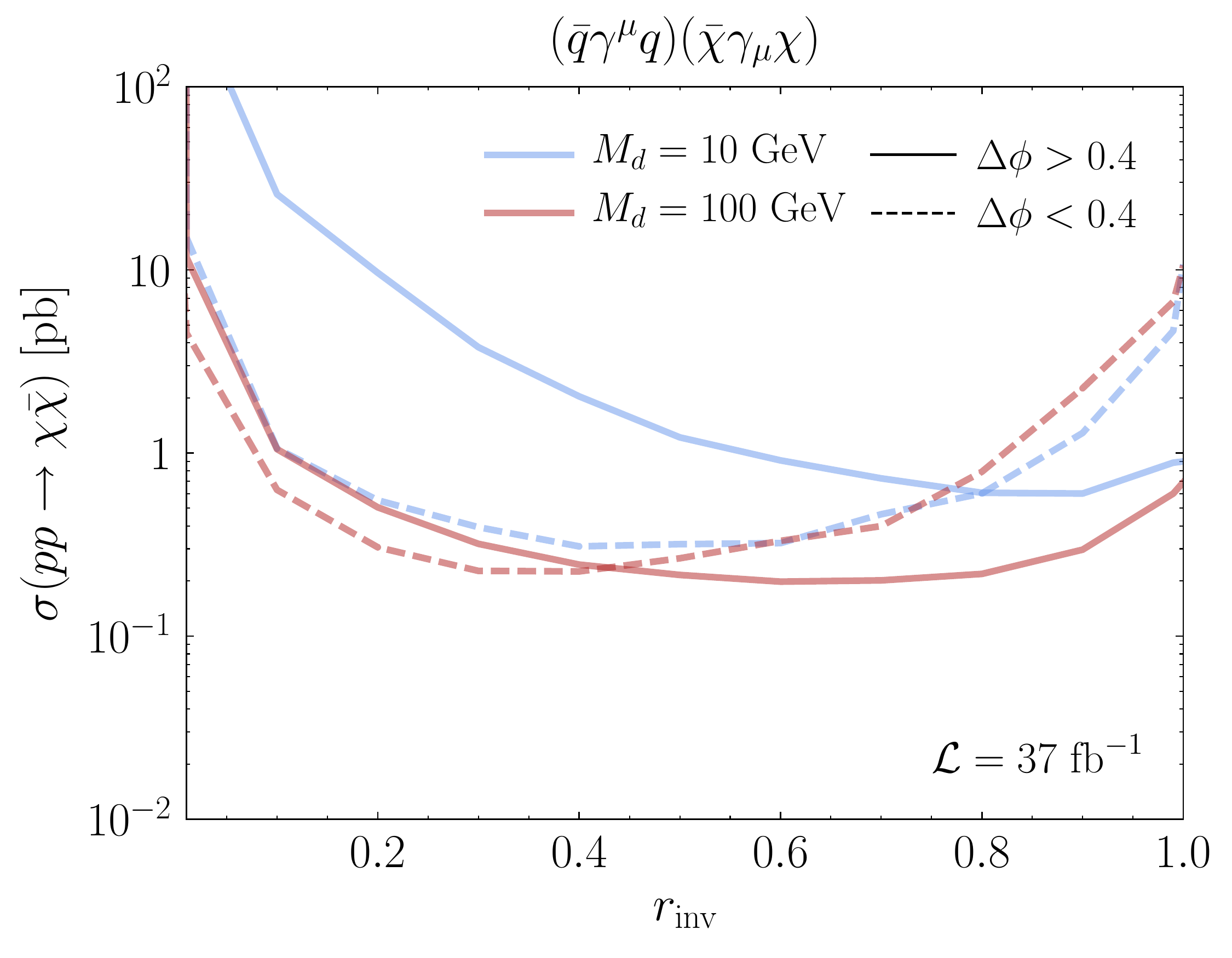}  \hspace{-0.8cm}
\caption{Projected sensitivity on the scale for the vector contact operator (left) and its associated production cross section (right) for dark hadron masses $M_d = 10$ and $100$~GeV (blue and red, respectively).  The limits are shown for $\Delta\phi<0.4$ GeV (dashed) and $\Delta\phi>0.4$ GeV (solid).  Note that for $r_\text{inv} \rightarrow 0$, a search strategy that does not have a minimum $\MET$ requirement should be investigated.   Appendix~\ref{sec:VaryParameters} demonstrates that there is minimal impact on the limits from varying additional dark sector parameters.} 
\label{fig:contact_limits}
\end{figure}

Figure~\ref{fig:contact_limits} highlights the complementarity between the two different search strategies in covering the full range of $r_\inv$.  The left panel shows the bounds on the effective contact operator scale, $\Lambda$, while the right panel shows the bounds on the production cross section $\sigma( p\,p \rightarrow \overline{\chi}\,\chi)$, as a function of $r_\inv$.  Solid lines show the results for a standard monojet search with $\Delta \phi > 0.4$, and the dashed lines show the corresponding limits placed by reversing this cut to $\Delta \phi < 0.4$.  Notice that the bounds using the standard search region improve as one moves to larger $r_\inv$, as expected,  because the jets are nearly invisible in this limit and ISR generates the non-trivial $\MET$.  In contrast, the bounds on the semi-visible search increase towards lower $r_\inv$.  We see, for example, that for $M_d = 100$~GeV, the monojet search takes over in sensitivity relative to the semi-visible search around $r_\inv\sim0.5$.  In comparison, this transition point is closer to $r_\inv\sim0.9$ when $M_d= 10$~GeV.   In general, the monojet limits are very sensitive to the dark hadron mass and become increasingly stronger as one moves from 10 to 100~GeV.  The limits from the semi-visible analysis are not as sensitive to the dark hadron mass.  We show that the strategies are robust to changes in other dark sector parameters in Appendix~\ref{sec:VaryParameters}.  Finally, it is worth noting that an additional search strategy to target the small $r_\text{inv}$ region could in principle be developed; we leave this investigation to future work.

Now that we have explored the basic search strategy for semi-visible jets, we will study how the searches change when the contact operator is resolved into the $s$-channel and $t$-channel UV completions.  As we will see, the $s$-channel model motivates a significantly different strategy, while the $t$-channel model is covered by the same simple $\MET$-driven approach that we used for the contact operator limit. 

\section{Dark Sector Showers from Resolved Contact Operators}
\label{sec:resolved}
Next, we resolve the contact operator at tree-level with two simple UV completions.  We characterize these two cases as $s$-channel and $t$-channel, which refer to the Feynman diagrams that dominate the production of $p\,p \rightarrow \overline{\chi}\,\chi$ at the LHC for the two models.

\subsection{\bf\emph{s}-channel}
\label{sec:sChan}

A pair of dark quarks can be produced through a new heavy resonance, $Z'$, that couples to the Standard Model baryon-number current and the DM flavor-number current via 
\begin{equation}
\mathcal{L}_{s\text{-channel}} \supset - Z'_\mu \sum_{i,a} \big( g_q \, \overline{q}_i\gamma^\mu q_i + g_\chi \, \overline{\chi}_a \gamma^\mu \chi_a  \big) \, ,
\label{eq:schannelL}
\end{equation}
where $g_{q,\chi}$ are coupling constants and $i,a$ are flavor indices.  The $Z'$ can potentially couple to other visible states, but we focus on the quark current here as we are interested in purely hadronic events.  It is worth emphasizing that Eq.~\eqref{eq:schannelL} is a simple phenomenological parametrization.  Specifically, we remain agnostic about the new particle content that is needed to appropriately cancel anomalies---see~\cite{Ismail:2017ulg} for a recent discussion---and do not model-build the mixing structure that is required to give $g_q \neq  g_\chi$.  We assume that the Higgs sector which gives the $Z'$ its mass does not impact the collider signatures, and thus do not specify it.     In this subsection, we revisit the analysis first proposed in~\cite{Cohen:2015toa} for this $s$-channel production mode to explore its complementarity with existing LHC searches, as well as the contact operator case. 
\begin{figure}[t] 
\hspace{-0.9 cm} 
 \includegraphics[width=0.5\textwidth]{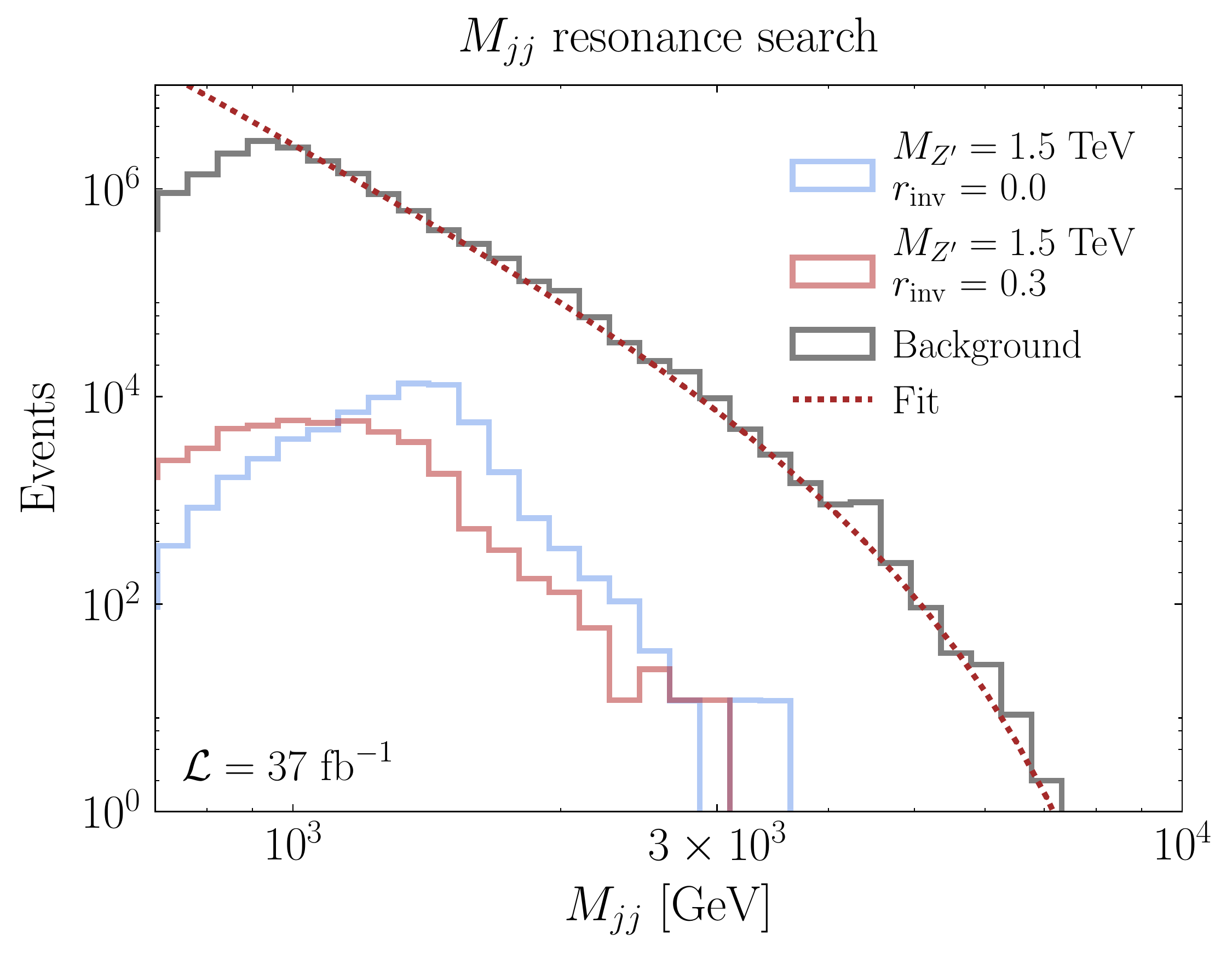} \hspace{-0.29 cm}
 \includegraphics[width=0.5\textwidth]{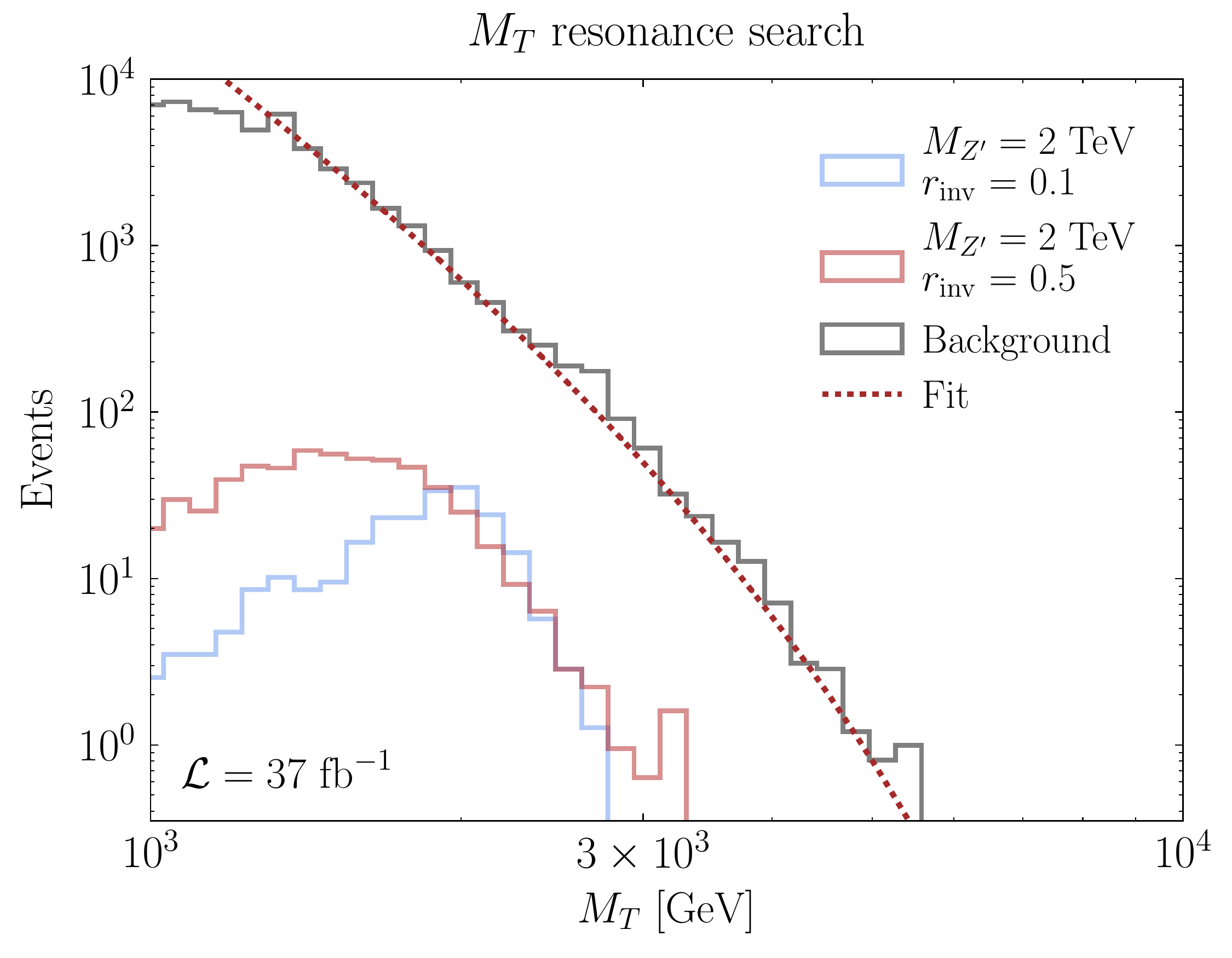} ~  \hspace{-0.8cm}
\caption{(Left) Invariant mass distribution for the dedicated $s$-channel search.  The background distribution is shown in black, while benchmark signal distributions are shown for $M_{Z'} = 1.5$~TeV and $r_\inv = 0.0$ and $0.3$ in blue and red, respectively. (Right)~The transverse mass distribution, this time for $M_{Z'} = 2$~TeV and $r_\inv = 0.1$ and $0.5$ in blue and red, respectively.  For each panel, the background fit used in that analysis is shown in dotted red.  In both cases, the signal is plotted assuming $g_q = 0.1$ and $g_\chi = 1$.}
\label{fig:Zprime_distributions}
\end{figure}

We generate events for the $s$-channel production in \texttt{MadGraph} using the \texttt{DMsimp}~\cite{Neubert:2015fka, Mattelaer:2015haa, Backovic:2015soa} model file implemented through \texttt{FeynRules}~\cite{Alloul:2013bka}, taking as fixed $g_q =0.1$ and $g_\chi = 1$; note that the $Z'$ width is calculated self-consistently in the generation.  When the $Z' $ decays predominantly to visible quarks, di-jet searches provide the best sensitivity regardless of the details of the dark sector.  In this case, $r_\inv\rightarrow 0$ and the final state resembles two QCD jets whose invariant mass ($M_{jj}$) reconstructs  the $Z'$ mass. Following the ATLAS di-jet analysis~\cite{Aaboud:2017yvp}, we require that the $p_T$ of the leading and sub-leading jets be at least 440 and 60~GeV, respectively, at the trigger and preselection level. We further require that $|\Delta y| < 1.2$ between the two leading jets.  The left panel of Fig.~\ref{fig:Zprime_distributions} shows the invariant mass distribution for $M_{Z'} = 1.5$~TeV, taking $r_\inv = 0$ and $0.3$.  As the invisible fraction increases, the width of the signal's invariant mass distribution broadens, reducing the sensitivity of a bump hunt.
\begin{table}[t]
\footnotesize
\renewcommand{\arraystretch}{1.4}
\begin{tabular}{C{4cm}|C{2cm}C{2cm}|C{1.6cm}C{1.7cm}C{1.5cm}C{1.8cm}C{1.8cm}}
\multicolumn{7}{c}{$s$-\sc{channel}}\\
\Xhline{3\arrayrulewidth}
 & \multicolumn{2}{c|}{\textbf{Signal} ($r_\inv$, $M_{Z'}$~[GeV])} & \multicolumn{4}{c}{\textbf{Background}}\\
 
$\bm{M_{jj}}$ \textbf{resonance} &   (0.1, 2000) &    (0.4, 1000) &       $Z+\text{jets}$ &       $W^\pm+\text{jets}$  &     $t\,\overline t+\text{jets}$ &      QCD  \\
\hline

Trigger and presel. & 9860[1.8] &  6770[1.23] &  70900 &  $1.4\times10^5$ &  54100 &  $3\times10^7$  \\
$|y^*| < 1.2$ &  6630[1.62] &  5060[1.24] &  41100 &   83200 &  36700 &  $1.7\times10^7$ \\

\hline 
\hline
\textbf{$\bm{M_T}$ resonance} &   (0.1, 2000) &    (0.5, 2000) &       $Z+\text{jets}$ &       $W^\pm+\text{jets}$  &     $t\,\overline t+\text{jets}$ &      QCD \\

\hline
Trigger and presel. &  634[1.03] &   1360[2.2] &  $1.1\times10^5$ &  $1.4\times10^5$  &  68100 &  64400 \\
$\MET > 0.15 \times M_T$ &  403[0.69] &  1250[2.13] &  $10^5$ &  $1.3\times10^5$  &  63700 &  46300   \\
$|\eta| < 1.1$ &  250[0.58] &   756[1.75] &   51700 &   71200 &  38900 &  24900 \\
$\Delta\phi < 0.4$ &  239[0.79] &   637[2.11] &   11100 &   33400 &  21800 &  24300 \\
\Xhline{3\arrayrulewidth}
\end{tabular}
\caption{Cut-flow table for $s$-channel production for $\mathcal L = 37$ fb$^{-1}$ at 13 TeV LHC.  The couplings $g_q = 0.1$ and $g_\chi = 1$ are assumed for the signal.  The numbers in brackets correspond to an estimate of the significance $s/\sqrt{s+b}$ at each stage of the cut-flow, where $s(b)$ is the number of signal(background) events.}
\label{tab:cut-flow_s-ch}
\end{table}

 In the limit of large $r_\inv$, a resonance search in the transverse mass, $M_T$, of the two final-state jets is more effective than one in $M_{jj}$ because the latter is considerably broadened due to the invisible states within the jet.  We choose a preselection cut requiring $\MET > 200$~GeV and at least two $R=0.5$ anti-k$_T$ jets, each with $p_T > 100$~GeV and $|\eta| < 2.4$.  For the selection cuts, the jets in the event are reclustered into $R=1.1$~Cambridge/Achen (CA) jets~\cite{Bentvelsen:1998ug} with $|\Delta \eta | < 1.1$.  Additionally, each event is required to have $\MET/M_T > 0.15$ and no electrons(muons) with $p_T  >10 (20)$~GeV and $|\eta|<2.4$.  Finally, we require that $\Delta\phi < 0.4$.  These cuts are designed to isolate events with significant missing energy aligned along one of the jets produced in the $Z'$ decay.  The right panel of Fig.~\ref{fig:Zprime_distributions} illustrates the shape of the $M_T$ distribution after selection cuts, for $M_{Z'} = 2$~TeV and several values of $r_\inv$.  In the case of a 2~TeV $Z'$, this search continues to have sensitivity even up to values of $r_\inv \simeq 0.9$, as we will show.  Table~\ref{tab:cut-flow_s-ch} summarizes the cut-flow for both the $M_{jj}$ and $M_T$ searches.  

A bump-hunt can be performed over the variable of interest after all the selection cuts are applied. The background distributions for both $M_{jj}$ and $M_T$ are well-approximated by the following fit function:
\begin{equation}
f(x) = p_0 \frac{(1-x)^{p_1 + p_2 \ln x}}{x^{p_3 + p_4 \ln x}},
\qquad x=\frac{M_{jj}}{\sqrt{s}} \quad{\rm or }\quad \frac{M_{T}}{\sqrt{s}}\, ,
\label{eq:app_fit}
\end{equation}
where the $p_i$ are free parameters.  The best-fit distributions are shown in Fig.~\ref{fig:Zprime_distributions}.  

\begin{figure}[tb] 
\hspace{-0.9 cm} 
 \includegraphics[width=0.555\textwidth]{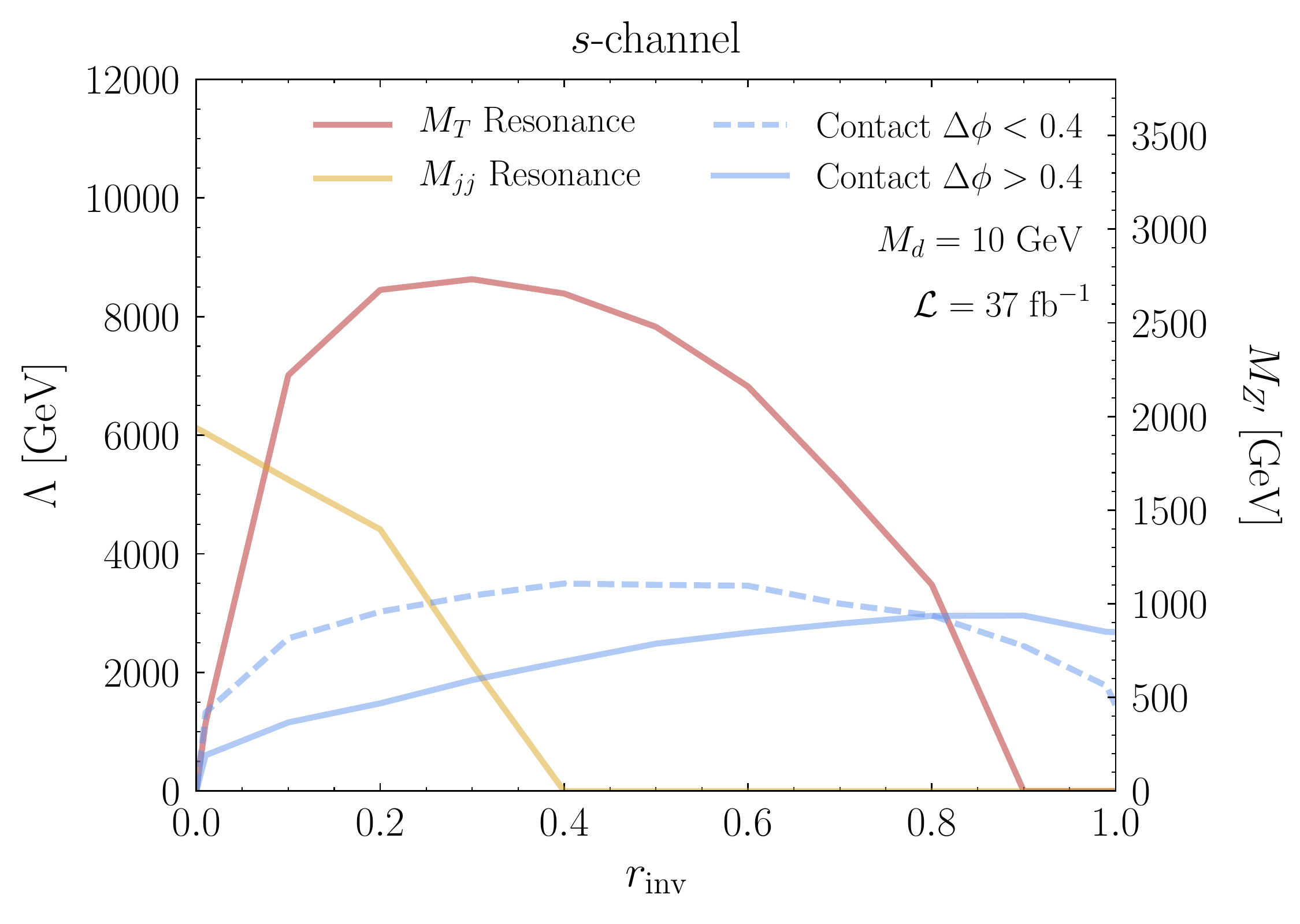} \hspace{-0.29 cm}
 \includegraphics[width=0.5\textwidth]{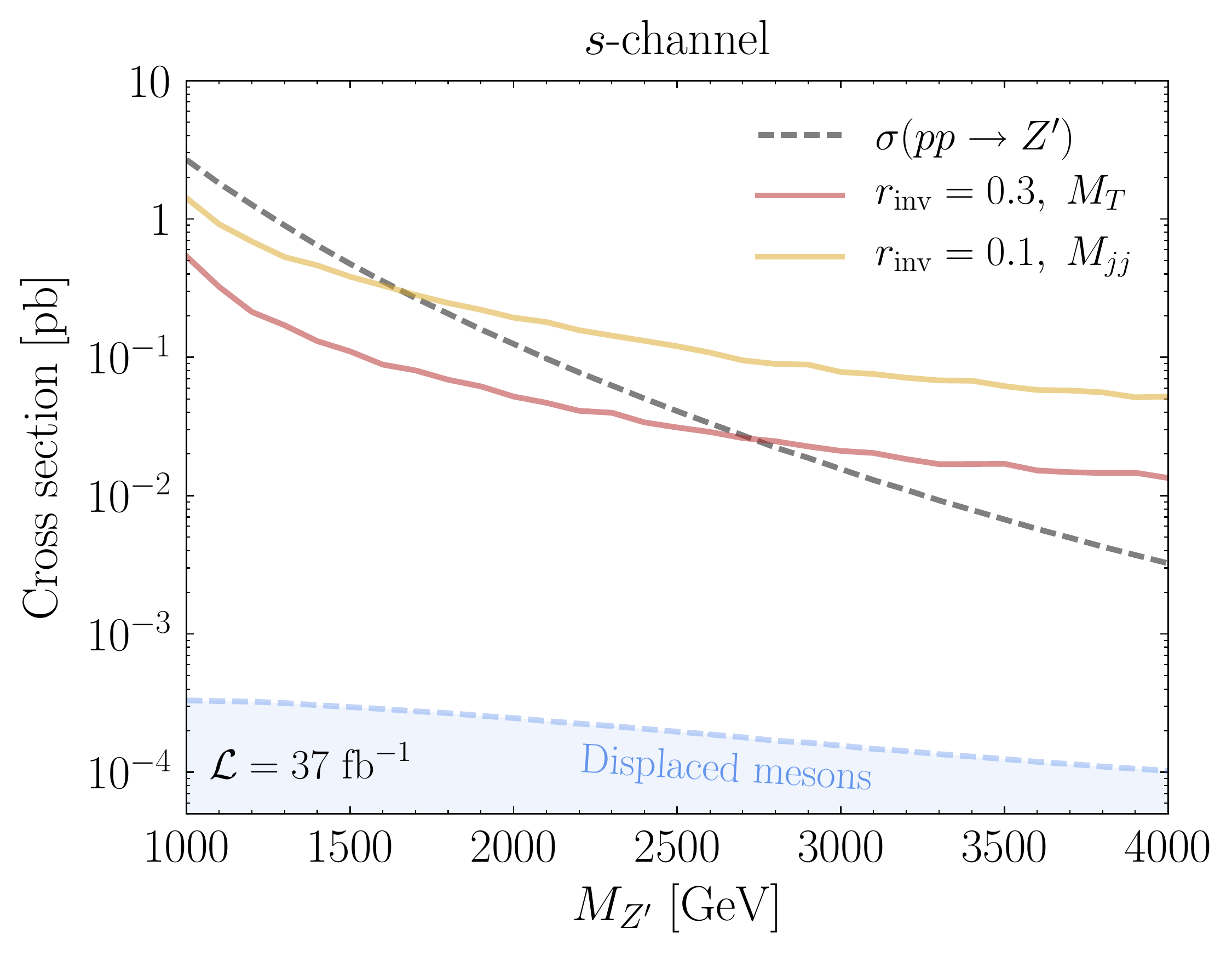} ~  \hspace{-0.8cm}
\caption{(Left) Projected sensitivity on the operator scale (or $Z'$ mass) for the $s$-channel model.  The result for the $M_{T}$($M_{jj}$) bump hunt is shown in red(yellow). The blue lines show the limits from the contact operator searches with $\Delta \phi > 0.4$ (solid) or $<0.4$ (dashed), as in Fig.~\ref{fig:contact_limits}. The mapping onto the contact operator limit is $\Lambda = M_{Z'}/\sqrt{g_q\,g_\chi}$.   (Right)~The 95\% exclusion limits on the production cross section as a function of $M_{Z'}$ for $r_{\inv}=0.1$ or $0.3$ using the $M_{jj}$ or $M_{T}$ search, respectively.  The dashed black line indicates the production cross section for the $Z'$ model, assuming $g_q = 0.1$ and $g_\chi = 1$.  The blue shaded region  indicates where a search targeting displaced vector mesons may improve the sensitivity reach.  Note that this region is a rough estimate and is quite sensitive to the vector meson mass, which we take to be  $m_{\rho_d} = 20 \text{ GeV}$ here.  }
\label{fig:Zprime_limits}
\end{figure}

The left panel of Fig.~\ref{fig:Zprime_limits} shows the limits on $\Lambda$ (or, correspondingly, $m_{Z'}$).  In order to compute this limit, we fix the couplings to be $g_q = 0.1$ and $g_\chi = 1$, and scan over the $Z'$ mass.  The bounds from the $M_T$ analysis (solid red) are strongest for $r_\inv\sim 0.3$.  The $M_T$ search loses sensitivity as $r_\inv \rightarrow 0$ because no stable hadrons are produced in the dark shower and the $\MET$ requirement is consequently too strong.  In this regime, however, the $M_{jj}$ analysis proves to be useful, with sensitivity peaking at $r_\inv = 0$ (solid yellow).  For comparison, we also show the limits from the contact operator analysis discussed in Sec.~\ref{sec:contact}.  For most values of $r_\inv$, either the $M_{jj}$ or $M_T$ analysis does considerably better.  However, the contact operator search provides the strongest bounds near $r_\inv\sim1$.  In this limit, the $M_T$ analysis loses sensitivity as events tend to fail the jet number and $p_T$ cut.

The right panel of Fig.~\ref{fig:Zprime_limits} shows the bounds on the production cross section, as a function of $Z'$ mass for $r_\inv = 0.3$ using the $M_T$ search (yellow) and for $r_\inv = 0.1$ using the $M_{jj}$ search (red).  When computing these limits, we fix the mass and vary the production cross section.  We assume a fixed signal shape and branching ratio derived with $g_q = 0.1$ and $g_\chi = 1$; this is a good approximation for the range of cross sections excluded by the two search strategies.  The production cross section for the mediator is shown in dashed black, for the benchmark case with $g_q = 0.1$.  When the $Z'$ becomes sufficiently heavy, the vector mesons in the shower can manifest displaced decays~\cite{Cohen:2015toa} for\footnote{For a discussion of the decay rate for the scalar mesons, see~\cite{Strassler:2006im}.}
\begin{align}
g_q \lesssim 10^{-2} \left(\frac{1}{g_\chi}\right) \sqrt{\frac{B}{10}} \left(\frac{M_{Z'}}{3\text{ TeV}}\right)^2 \left(\frac{20 \text{ GeV}}{m_{\rho_d}}\right)^{5/2}\,,
\label{eq:displaced}
\end{align}
where $B\sim 10$ is the average boost factor as computed by the simulation,\footnote{This choice for $B$ is conservative, since the majority of the mesons produced have a smaller boost, but the tail of this distribution is relatively broad.  For example, we find that $\sim 80\%$ of the mesons have $B < 10$.} $m_{\rho_d}$ is the mass of the vector meson, and the inequality is saturated for a lab-frame displacement of a millimeter.    This parameter range is indicated by the blue shaded region in Fig.~\ref{fig:Zprime_limits}.  If the cross section limit reaches this level of sensitivity, a search that relies on displaced signatures should be implemented, perhaps along the lines of the proposed strategies for emerging jets~\cite{Schwaller:2015gea}.  We stress that Eq.~(\ref{eq:displaced}) is a rough estimate and that the value of $g_q$ depends quite sensitively on the vector meson mass, which we simply take to be $m_{\rho_d} = 20$~GeV in the figure.

Now that we understand how the search strategy and sensitivity changes for a scenario described by an $s$-channel UV completion, we move on to the example of a model where the dark quark pair production occurs via a $t$-channel diagram.

\subsection{\bf \emph{t}-channel}
The collider physics for the $t$-channel UV completion is governed by the coupling
\begin{equation}
\mathcal{L}_{t\text{-channel}}  \supset \sum_{i,j,a,b} \lambda_{ijab} \, \overline{\chi}_a\, \Phi^*_{bi}\, q_{Rj}  \, ,
\end{equation}
 where $a,b$ are DM-flavor indices, $i,j$ are Standard Model-flavor indices, and $q_{Rj}$ represents both up- and down-type quarks. The dark and visible sectors communicate via the scalar bi-fundamental $\Phi_{bi}$, which is in the fundamental representation under both visible QCD and the dark non-Abelian gauge group.  For simplicity, we have only introduced a coupling to the right-handed quarks, which requires the $\Phi_{bi}$ to carry hypercharge. There is no obstruction to coupling with left-handed quarks $q_{Lj}$; this would require the $\Phi_{bi}$ to form electroweak doublets, which is not considered here. Additionally, we take all the flavor structure to be proportional to the identity $\lambda_{ijab} = \lambda\, \delta_{ij}\,\delta_{ab}$ and assume a common mass $M_\Phi$ for the scalar bi-fundamentals. 
 
\begin{figure}[tb] 
\hspace{-0.9 cm} 
 \includegraphics[width=0.48\textwidth]{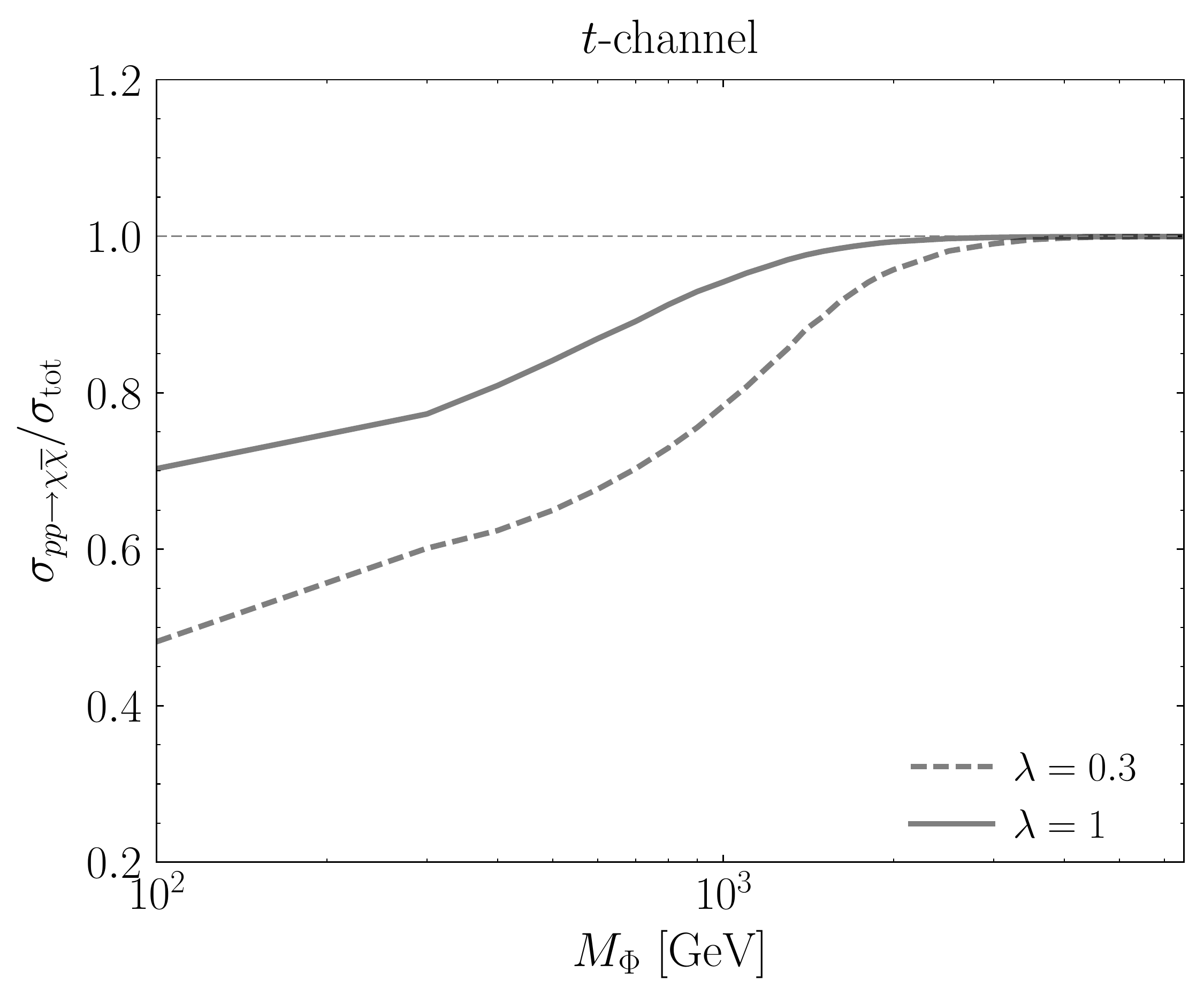} \hspace{-0.29 cm}
 \includegraphics[width=0.52\textwidth]{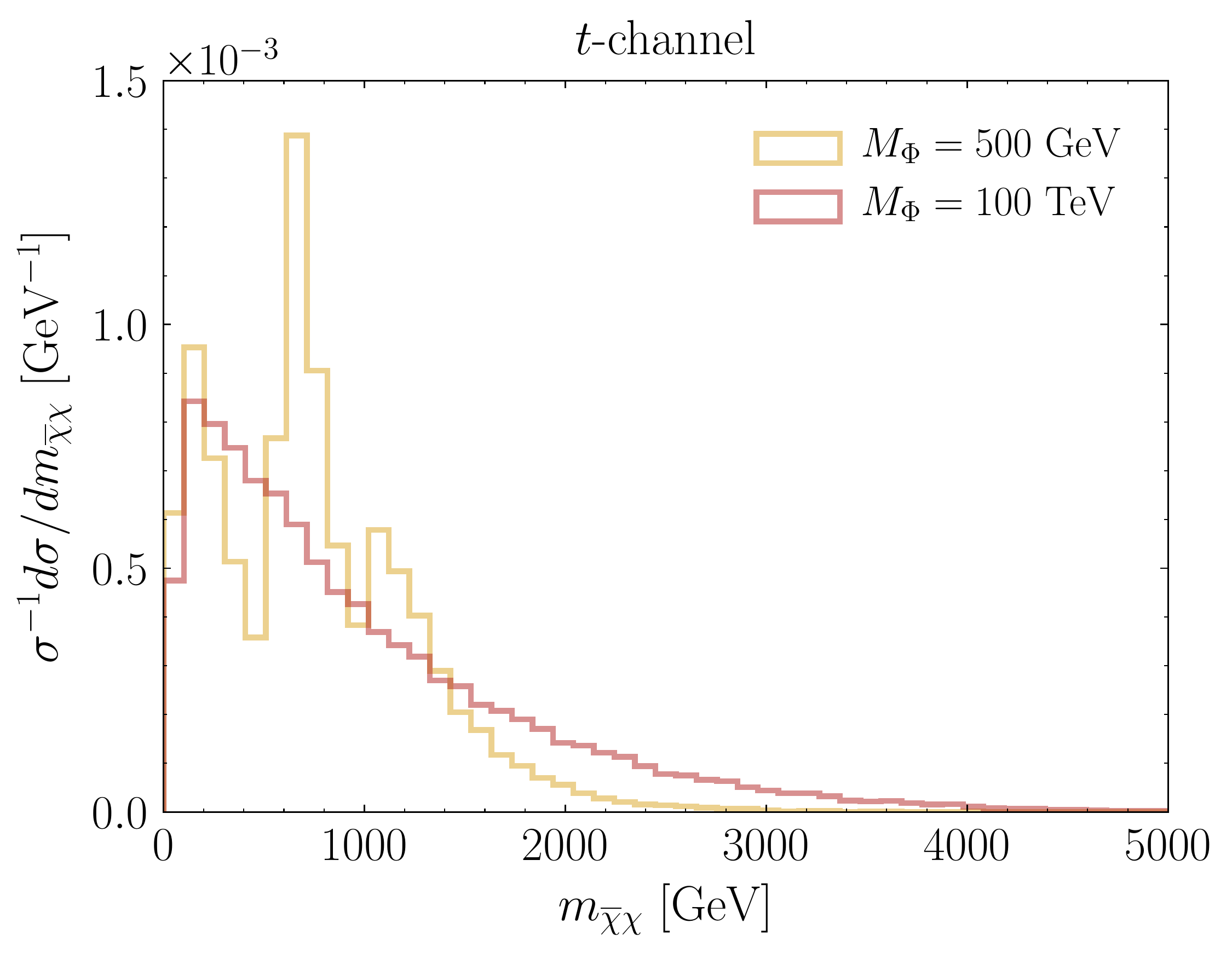} \hspace{-0.29 cm}
\caption{(Left) Ratio of the $t$-channel direct production cross section to the total cross section as a function of the bi-fundamental mass. The total cross section includes processes with additional quarks in the final states through $\Phi_{ai} \rightarrow \overline \chi_a \, q_{i} $. The $t$-channel production mode accounts for a larger proportion of the total cross section as the mediator mass is increased towards the contact operator limit. (Right) The parton-level invariant mass distribution for the $\overline{\chi}\,\chi$ particles, $m_{\overline{\chi}\chi}$, for $M_\Phi = 500 \text{ GeV}$ and $100 \text{ TeV}$ (the contact operator limit) and $\lambda = 1$.  One can clearly identify the threshold as each production channel turns on.  This shows that the high $\overline{\chi}\,\chi$ tail falls off more rapidly when $M_\Phi$ is light, which results in weaker limits.}
\label{fig:tchannel_production}
\end{figure}

 A variety of production modes are possible for this scenario.  In addition to direct pair-production of the dark quarks, the bi-fundamentals may also be directly produced if they are light enough.  For example, the $\Phi$ can be pair-produced via its coupling to visible gluons/quarks ($g \, g , q \, \bar{q} \rightarrow \Phi \, \Phi^*$) or associatively ($q \, g \rightarrow \Phi \, \overline\chi$).  The large number of production modes results in a complicated dependence of the production cross section on $M_\Phi$.  This behavior is demonstrated in the left panel of Fig.~\ref{fig:tchannel_production}, which plots the fractional contribution of the $t$-channel direct production process as a function of $M_\Phi$ for two choices of $\lambda$. 
In the Appendix, we show how large $M_\Phi$ must become such that the $t$-channel and $s$-channel distributions are identical, demonstrating that the contact operator limit is reached for masses of $\mathcal{O}(10\text{ TeV})$.

In practice, when generating events for this model, we produce matched samples of $p\,p\rightarrow \overline{\chi}\,\chi  + \text{jets}$ events with 0, 1, and 2 jets.  This implies that production modes involving one or two intermediate $\Phi$'s are generated and decayed within \texttt{Madgraph}.   Furthermore, the width of $\Phi$ is computed for each parameter point in the simulation, ensuring that finite-width effects are appropriately modeled.  A larger number of events are required in order to ensure stability of the cross section using our implementation of this model, and 200,000 events per parameter point were generated. The resulting parton-level $m_{\overline{\chi}\chi}$ distribution is shown in the right panel of Fig.~\ref{fig:tchannel_production} for $\lambda = 1$ and $M_\Phi = 500 \text{ GeV}$.  There are three clear contributions to this distribution:  $\overline{\chi}\, \chi$ production turns on at threshold, followed by the turn-on of the diagrams with one(two) $\Phi$ intermediate states at around 500 GeV(1 TeV).  This figure also shows the shape of the same distribution for $M_\Phi = 100 \text{ TeV}$, where the model is well-approximated by the contact operator.  The fall-off is more rapid for smaller $M_\Phi$ because the non-trivial momentum dependence in the propagator becomes important in this limit.  We choose $\lambda = 1$ as our benchmark for this model.  Note that the analysis presented here would be the only probe of the model since there are no competing final states as in the $s$-channel case.

To assess the reach for this model, we optimize a search with cuts that are motivated by standard jets + $\MET$ analyses, \emph{e.g.}~\cite{Aad:2015zva}.   After applying a trigger-level cut of $p_{T,1} > 250$ GeV and $\MET> 200$~GeV, we optimize the signal reach by scanning in $\MET > [600, 800,1000,1200]$~GeV.  We repeat this procedure for the case where $\Delta \phi < 0.4$ and $>0.4$.  As in the contact operator case, when $\Delta\phi < 0.4$ we restrict ourselves to $\MET \geq 800$ GeV.  This is identical to the search strategy for the contact operator limit, presented in Sec.~\ref{sec:contact} above.  We also investigated the impact of additional cuts on $H_T$, as well as the $p_T$ of the jets.  We find improved performance for smaller values of $r_\text{inv}$ when cuts on the $p_T$ of the third and fourth jets are imposed because they target the additional hard jets produced by the intermediate $\Phi$ states.  For example, at $r_\text{inv} = 0.2$, the $\Delta \phi < 0.4$ limit on $M_\Phi$ improves from $\sim 1000$ to $\sim 1500 \text{ GeV}$ with these additional cuts.  We only show the results for the optimized $\MET$ cuts (and not the additional jet $p_T$ cuts) so that the comparison with the contact operator search is transparent.  The cut-flow for a few benchmarks is provided in Table~\ref{tab:cut-flow_t-ch}.

\begin{table}[t]
\footnotesize
\renewcommand{\arraystretch}{1.4}
\begin{tabular}{C{3cm}|C{1.9cm}C{1.9cm}|C{1.6cm}C{1.7cm}C{1.5cm}C{1.7cm}}
\multicolumn{7}{c}{$t$-\sc{channel}}\\
\Xhline{3\arrayrulewidth}
 & \multicolumn{2}{c|}{\textbf{Signal} ($r_\inv$, $M_\Phi$ [GeV])} & \multicolumn{4}{c}{\textbf{Background}} \\
Cuts &    (0.5, 1500) &    (0.9, 2000) &       $Z+\text{jets}$ &       $W^\pm+\text{jets}$  &     $t\,\overline t+\text{jets}$ &      QCD  \\
\hline
Trigger and presel. &  2091[2.7] &  467[0.6] &  $2.3\times 10^5$ &  $2.5\times 10^5$ &  $6.9\times 10^4$ &  $5.7\times 10^4$ \\
$\MET > 800$ &    50[1.17] &  96[2.22]  &    1160 &     536 &     80 &      0 \\
$\Delta\phi > 0.4$ &    13[0.38] &  64[1.77] &     110 &     326 &     72 &      0  \\
or & & & & & \\  
$\Delta\phi < 0.4$ &    36[1.57] &  31[1.35] &    1050 &     209 &      8 &      0  \\
\Xhline{3\arrayrulewidth}
\end{tabular}
\caption{Cut-flow table for $t$-channel production for $\mathcal L = 37$ fb$^{-1}$ at 13 TeV LHC.  The coupling $\lambda = 1$ is taken for the signal.  The numbers in brackets correspond to an estimate of the significance $s/\sqrt{s+b}$ at each stage of the cut-flow, where $s(b)$ is the number of signal(background) events.}
\label{tab:cut-flow_t-ch}
\end{table}

\begin{figure}[tb] 
\hspace{-0.9 cm} 
 \includegraphics[width=0.555\textwidth]{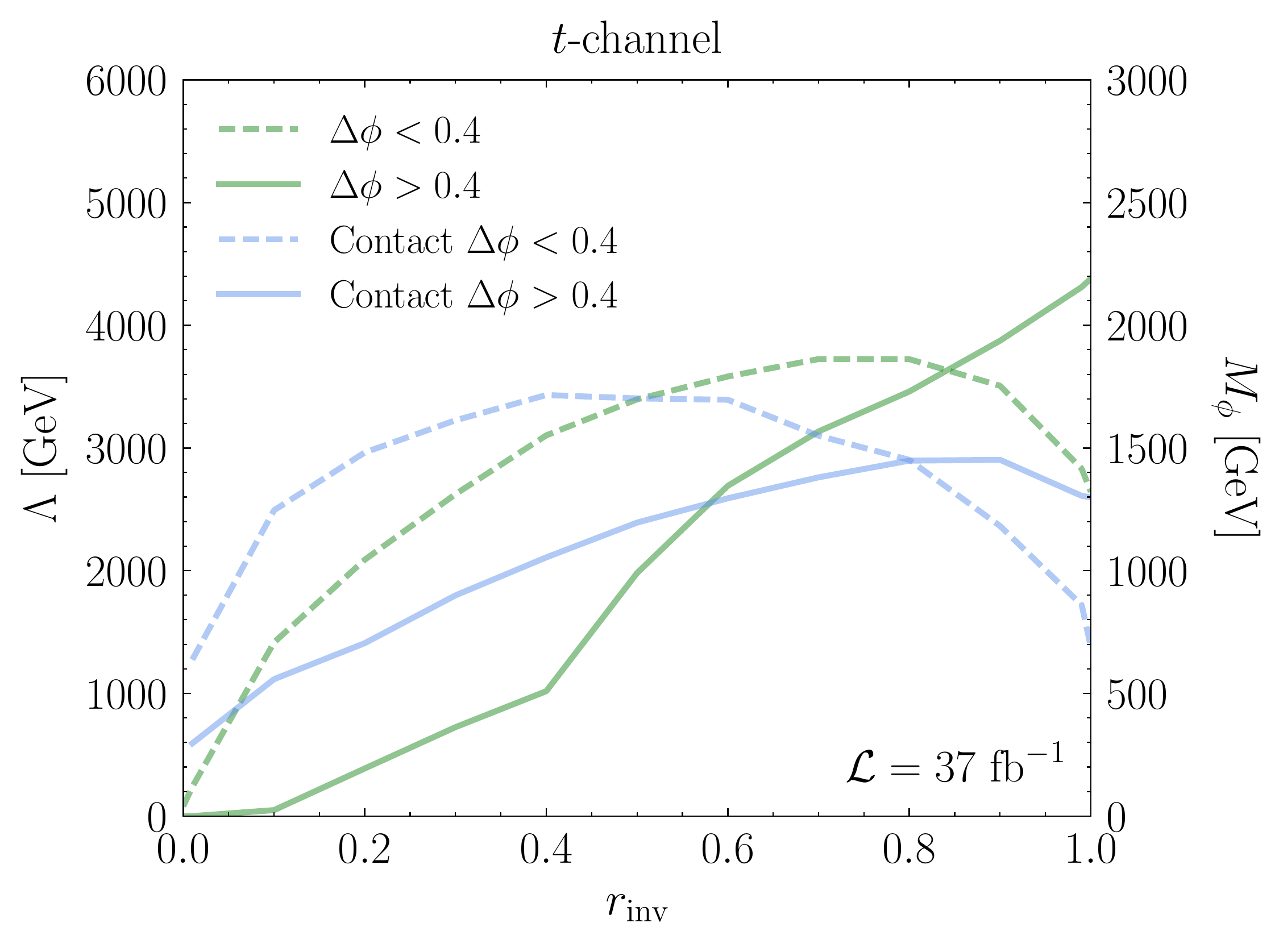} \hspace{-0.29 cm}
 \includegraphics[width=0.5\textwidth]{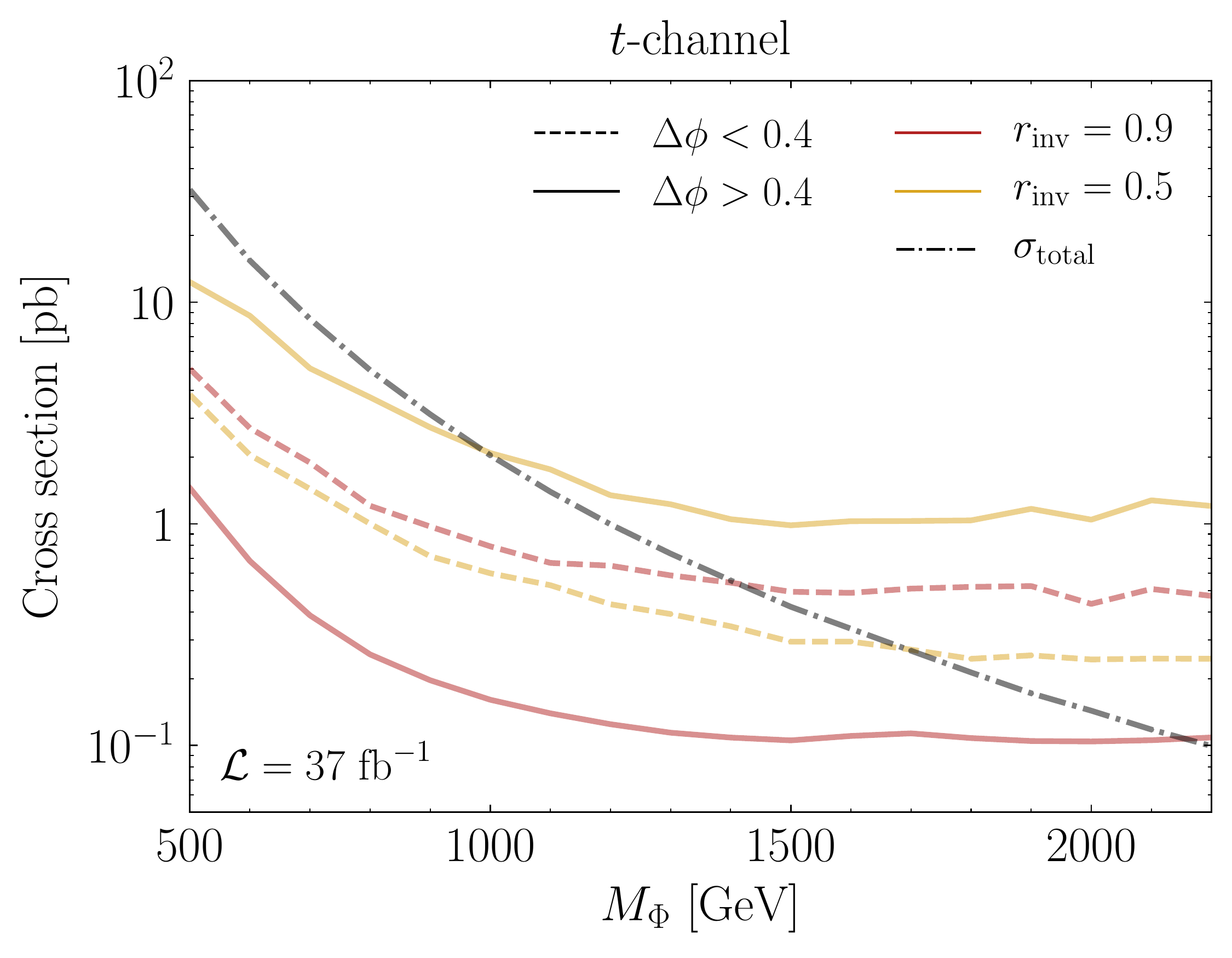} ~  \hspace{-0.8cm}
\caption{(Left) Projected sensitivity on the operator scale or $\Phi$ mass for the $t$-channel model with $\lambda = 1$.  The solid green line shows the canonical $\Delta\phi > 0.4$ cut in addition to the selection described in the text, while the dashed green line corresponds to the $\Delta\phi < 0.4$ cut. Note that for $r_\text{inv} \rightarrow 0$, a search strategy that does not have a minimum $\MET$ requirement should be investigated. The mapping onto the contact operator limit is $\Lambda = 2\,M_\Phi/\lambda$.  (Right)~The 95\% exclusion limits on the production cross section as a function of $M_{\Phi}$ for $r_{\inv}=0.9$ (red) or $0.5$ (yellow) corresponding to $\Delta\phi > 0.4$ (solid) and $\Delta\phi < 0.4$ (dashed). The total production cross section is shown as the dot-dashed black line.}
\label{fig:tchannel_limits}
\end{figure}

The left panel of Fig.~\ref{fig:tchannel_limits} shows the projected sensitivity bounds on the bi-fundamental mass, as a function of $r_\text{inv}$.  For $r_\text{inv} \lesssim 0.8$, the search with $\Delta \phi < 0.4$ is more powerful, but $\Delta \phi > 0.4$ does better at higher invisible fractions, as expected.  We also compare the results to  the expected reach for the contact operator limit.  At first glance, it would appear that the contact operator approach yields additional sensitivity, even though new channels are present for the full UV complete model.  However, this is spurious as the contact operator is not a good approximation for the mass scales relevant at the LHC.  In particular, the apparent improvement in the contact operator limit is an artifact of the larger tail in the $m_{\overline{\chi}\chi}$ distribution illustrated in Fig.~\ref{fig:tchannel_production}.  The right panel of Fig.~\ref{fig:tchannel_limits} shows the corresponding 95\% exclusion limit on the production cross section, as a function of $M_\Phi$.  We see explicitly that the $\Delta \phi < 0.4$ cut gives improved sensitivity when $r_\text{inv} =0.5$, but that the reverse is true when $r_\text{inv} = 0.9$. For this UV completion, the $\rho_d$ generally will not be displaced until $M_\Phi$ is larger than $\mathcal{O}(10\, {\rm TeV})$ for $\lambda = 1$ (as can be inferred from Eq.~(\ref{eq:displaced}) which is relevant in this model as well), which is well outside our expected sensitivity.   Additionally, the QCD pair-production of $\Phi$ is present for arbitrarily small values of $\lambda$.  As a result, we do not include a displaced region in Fig.~\ref{fig:tchannel_limits}.

There is potential room for improvement beyond the search presented here.  For example, a more sophisticated strategy could be devised to target small $r_\text{inv}$.   There is the additional complication that the dark shower tends to wash out the anticipated gains in sensitivity resulting from the additional production modes.  It may be that less inclusive variables, such as $M_{T2}$~\cite{Lester:1999tx} or its variants, could yield improved reach in certain regions of parameter space. We leave these investigations to future work.   

This completes our discussion of the collider projections for semi-visible jets.  The next section demonstrates that the direct detection of the $\eta_d$ is highly suppressed.

\section{Complementarity with Direct Detection Experiments}
\label{sec:directdetection}

Collider searches for DM in the contact operator limit ($\overline{q}\, q \rightarrow \overline{\chi}\, \chi$) are interesting in large part due to their complementarity with direct detection searches ($q \,\chi \rightarrow q\, \chi$).   A comparison of the limits derived using both experimental approaches has been explored in detail for the case of mono-$X$ signatures~\cite{Abdallah:2014hon, Abdallah:2015ter, Abercrombie:2015wmb}.  When the DM is composite, the comparison is complicated by the fact that $q^2_\text{LHC} \gg \Lambda_d^2 \gg q^2_\text{DD}$, where $q_\text{LHC(DD)}^2$ is the squared momentum transfer at the LHC(direct detection experiment).  In other words, the DM degrees of freedom are dark quarks at LHC energies, but become dark mesons at the scales probed by direct detection experiments.  The rest of this section provides some non-perturbative arguments to estimate the size of the direct detection rates for the strongly interacting models of interest here.  We will show that the direct detection rates are highly suppressed and fall below the neutrino background.  This section implicitly assumes that the $\eta_d$ comprises all of the DM.

Our goal is to compute the scattering of the composite DM particle, $\eta_d$, off a Standard Model nucleus for the vector contact operator given in Eq.~\eqref{eq:contact}.  It is worth noting that this was among the portals suggested in the first paper on direct detection, and was excluded long ago for non-composite DM interacting via the Standard Model $Z$ boson~\cite{Goodman:1984dc}.  For the composite DM candidate studied here, additional factors of momentum suppress the rate and make the model safe from direct detection.  

From Eq.~\eqref{eq:contact}, the direct detection scattering rate depends on the matrix element of a vector current involving the $\eta_d$.  Let the initial(final) momentum of the $\eta_d$ be $k(k')$ such that the total momentum is $P^\mu = (k' + k)^\mu$ and the momentum transfer to the nucleus is $q^\mu = (k-k')^\mu$.  By Lorentz invariance, the matrix element of interest requires the presence of an object that carries a vector index; $P^\mu$ and $q^\mu$ are the only vectors that are available.  Hence, the composite matrix element must take the form\footnote{In principle, these operators are  functions of $q^2/m_{\eta_d}^2$, \emph{i.e.}, the suppression scale is the physical mass of $\eta_d$.  However, because our parameter space takes $M_d \sim \Lambda_d \sim m_{\eta_d}$, we choose to use $\Lambda_d$ as a proxy for all relevant scales.  This is effectively just a change in the normalization of $F_1^d$ and $G^d$.}  
\begin{equation}
J^\mu \sim \overline{\chi}_a \gamma^\mu \chi_a \quad \longrightarrow \quad \eta_d^* \left[ \frac{P^\mu}{\Lambda_d} \, F_1^d(q^2) + \frac{q^\mu}{\Lambda_d} \, G^d(q^2) \right] \eta_d \, ,
\label{eq:DMmatrixelement}
\end{equation}
where $F_1^d(q^2)$ and $G^d(q^2)$ are DM form factors.  Note that the form factors only depend on $q^2$, which can be related to $P^2$ using $m_{\eta_d}^2$.  We use the standard notation that $F_1$ is the electric form factor; if $\eta_d$ had carried spin, there would be the possibility of an $F_2(q^2)$ magnetic form factor and its contribution to the current would be proportional to the spin vector.  The requirement that Eq.~\eqref{eq:DMmatrixelement} vanish by current conservation $\partial_\mu J^\mu = 0$ is directly related to the stability of the DM and imposes that $G^d(q^2) = 0$.  
 
 In the limit of small momentum transfer ($q^2 \ll \Lambda_d^2$), the remaining form factor can be expanded to first order as 
  \begin{equation}
 F_1^d(q^2) = F_1^d(0) + \frac{q^2}{\Lambda_d^2} \frac{\partial F_1^d (q^2)}{\partial q^2} \Bigg|_{q^2=0} \, .
 \label{eq:Fdq2}
 \end{equation} 
The first term in Eq.~\eqref{eq:Fdq2} is proportional to the charge of the $\eta_d$ under the new $U(1)'$ symmetry and consequently vanishes.  To see this explicitly, we integrate the $\mu=0$ component of the matrix element, which yields the conserved charge, $Q_\eta$: 
\begin{equation}
Q_\eta =   \int \text{d}^3x  \,\,  \psi_{\eta_d}^*(k') \, \frac{P^0}{\Lambda_d} \, \psi_{\eta_d}(k) F_1^d(q^2)  \sim F_1^d(0) \,,
\end{equation}  
since the integral over the wavefunctions $\psi_{\eta_d}$ yields $q^2=0$ by orthogonality.  As we argued in Sec.~\ref{sec:dynamics}, the DM is neutral with respect to this current, which immediately implies that $F_1^d(0) = 0$. 

Importantly, the higher-order contributions to $F_1^d(q^2)$ are non-zero.  Physically, as $q^2$ increases from zero,  the structure of the meson begins to reveal itself, as in deep inelastic scattering.  Because the partons are charged under the symmetry of interest, this leads to a non-zero contribution to $F_1^d(q^2)$.  In Standard Model physics, this is usually couched in terms of a non-zero ``charge radius," so we will use the same language here.  For example, in the case of the Standard Model neutron, $F_1^n(q^2) = q^2 \times (-R_n^2/6 + \kappa/(4 m_n^2)) + \mathcal{O}(q^4)$~\cite{Sachs:1962zzc, Bawin:1999ks}, where $R_n$ is the charge radius of the neutron, $\kappa$ is the dimensionless magnetic moment of the neutron, and $m_n$ is the mass of the neutron.  Because the dark meson $\eta_d$ is a scalar, the magnetic moment is zero and the only contribution to $F_1^{\eta_d}(q^2)$ at $\mathcal{O}(q^2)$ is proportional to the square of the charge radius.  For concreteness, we will assume that $R_d \sim 1/\Lambda_d$, which is reasonable up to order-one numbers since this is the only scale of relevance for the dark meson (under our assumption that $M_d \sim \Lambda_d$).  

The net result of these arguments is that the cross section is suppressed by  four powers of the momentum exchange.  Using these parametrics, we can make a rough estimate of the spin-independent direct detection cross section per nucleon: 
\begin{align}
\sigma_\text{DD} \sim \mathcal{O}(1) \times 10^{-52} \text{ cm}^2 \left(\frac{q^2/\Lambda_d^2}{10^{-6}}\right)^2 \left(\frac{R_d/\Lambda_d}{1}\right)^4 \left(\frac{1\text{ TeV}}{\Lambda}\right)^4\,.
\label{eq:DDEstimate}
\end{align}
Noting that the neutrino background begins to dominate at cross sections of $10^{-45}$ to $10^{-49}$ for a DM mass of 10 and 100~GeV respectively~\cite{Billard:2013qya}, this is a very challenging signal to observe at a direct detection experiment.

The result in Eq.~\eqref{eq:DDEstimate} clearly applies for the $s$-channel UV completion.  Unsurprisingly, the situation for the $t$-channel case is very similar.  In the heavy-mediator limit, the DM-quark effective interaction can be written in a useful form by applying the Fierz identities:
\begin{equation}
\mathcal{L}_\text{eff} = \frac{\lambda_{ac}^\dagger \lambda_{cb}}{8\, M_\Phi^2} \, \left[\overline{\chi}_a \gamma_\mu \left(1-\gamma^5\right) \chi_b\right] \, \left[\overline{q} \,\gamma^\mu \left(1+ \gamma^5\right) q\right]\,.
\end{equation}
The DM matrix element for the vector current is the same as in Eq.~\eqref{eq:DMmatrixelement}; there are no axial-vector contributions because there are no combinations of $\eta_d$ that yield the correct Lorentz and parity structure as the quark-level operator $\overline{\chi}\, \gamma^\mu \gamma^5\,\chi$.  Because the vector operator is the only one that contributes, this means that the direct detection estimate in Eq.~\eqref{eq:DDEstimate} applies in this case as well. 

As we have seen, the direct detection signals for these composite DM models are highly suppressed for the operators considered in this work, which suggests that the LHC provides a unique opportunity for discovery.  It is worth emphasizing, however, that the arguments in this section rely on the assumption that the DM is a scalar and ignore the possibility of inelastic transitions between the DM to a nearby state in the dark hadronic spectrum.  While the latter can provide a potential detection window, the detection rate depends on the mass splittings of the lightest states~\cite{Lisanti:2009vy}.  Because we remain agnostic to the details of the dark spectrum, we do not consider this possibility here.

\section{Conclusions}
\label{sec:conclusions}
This paper proposes a comprehensive discovery program for dark sector parton showers at the LHC.  Such signatures arise in a broad range of theories, but an inclusive search program can be designed by using a simplified parametrization of the dark sector and portal physics.  The LHC observables depend primarily on four parameters that divide into: 
\begin{itemize}
\item \emph{Dark Sector Parameters}:  The dark sector strong coupling constant ($\alpha_d$), the dark hadron mass scale ($M_d$), and the ratio of invisible to visible hadrons that are produced in the parton shower ($r_\inv$).
\item \emph{Portal Parameter}:  The operator scale ($\Lambda$) associated with the portal interaction.
\end{itemize}
In the spirit of mono-$X$ searches, we consider the contact operator limit, and then UV complete this portal with either an $s$- or $t$-channel mediator.  Targeted  search strategies can improve the sensitivity reach to the resolved operators, at the expense of being less model independent.  

We focused specifically on the scenario where the visible states produced in the dark parton shower are light quarks, and the visible hadronic shower is aligned with a collimated spray of DM particles, forming ``semi-visible jets."  In this case, the missing energy typically points in the same direction as one of the jets in the event, resulting in low signal efficiency under standard preselection cuts for jets + $\MET$ searches, which require $\Delta \phi > 0.4$.  We show that reversing this requirement to $\Delta \phi < 0.4$ significantly improves the signal reach for a wide range of $r_\inv$ for both the contact operator and its UV completions.  We demonstrate these gains by optimizing search strategies over simple cuts in jet number, $\MET$, and $H_T$.  While it has been demonstrated that these cuts are sufficient to cover the variety of phase space that can be realized by Simplified Models with weakly coupled DM~\cite{Cohen:2016nzv}, it is entirely possible that more detailed searches would improve the sensitivity to semi-visible signals.  For example, variables such as razor~\cite{Rogan:2010kb} or $\alpha_T$~\cite{Randall:2008rw, Chatrchyan:2013mys} might provide additional handles.  Furthermore, developing a search that directly targets the small $r_\text{inv}$ region would be interesting to investigate.  Strategies that use the substructure of the semi-visible jets could lead to further improvements, although one must be careful to not rely on detailed features of the dark hadronization given the large uncertainty implicit in modeling these non-perturbative effects.  

Semi-visible jets populate the control region typically utilized by ATLAS and CMS in standard jets + $\MET$ studies.  Therefore, care needs to be taken to establish a data-driven background strategy for these new types of searches.  To avoid complications in the projections made in this paper, we cut aggressively on the missing energy to eliminate the QCD background in the $\Delta \phi < 0.4$ region for the contact operator and $t$-channel searches, and relied on a bump-hunt for the $s$-channel search.  A more sophisticated determination of the background uncertainties could potentially relax the missing energy cuts used here and improve the signal reach.  One possibility\footnote{We thank S.~Thomas for this suggestion.} is to use a high-statistics photon+jets sample to determine the missing energy contribution from QCD---specifically, the photon energy could be measured and used to constrain the energy of the jets in the event.  This sample could then be used to characterize the QCD background in the signal region with $\Delta \phi < 0.4$, where an isolated photon veto would ensure orthogonality with the control region.  A detailed experimental study is needed to establish the viability of this method. 

This paper focused on the spectacular under-explored collider signatures that result from a strongly interacting hidden sector.  We studied the vector contact operator and two of its UV completions, but a variety of other operators are possible~\cite{Fan:2010gt, Hill:2011be, Fitzpatrick:2012ix, Hill:2014yka, Hill:2014yxa, Bishara:2016hek} and should be considered.  Additionally, we focused on the case where the visible decay products in the shower are light Standard Model quarks.  This is one of the most challenging possibilities because of the potentially large QCD backgrounds.  Other decay modes---say, to leptons or $b$-quarks---are not only feasible, but may provide additional handles to improve signal discrimination.  The analysis strategy presented here can easily be  generalized to these scenarios.  For each of these variations, it would be interesting to consider the complementarity of the LHC searches with direct detection experiments.  While the vector contact operator leads to suppressed direct detection rates, prospects may improve for other operators.  In addition, astrophysical probes, which we have not discussed here, may also shed light on these non-minimal sectors, either through  cascades produced in annihilation events~\cite{Freytsis:2016dgf, Elor:2015bho} or self-interactions~\cite{Tulin:2017ara}.   
 
As we have demonstrated, the LHC can play a unique and critical role in the discovery of hidden dark sectors.  The framework laid out in this paper provides an exciting opportunity to extend the current DM program at the LHC to these new model frontiers.   

\section*{Acknowledgments}
We are especially grateful to O. Mattelaer, whose prompt and helpful feedback regarding \texttt{MadGraph} was critical for the completion of this work;  the patches he provided enabled us to use the bias and gridpack modules together and to successfully generate the $t$-channel events.  We also thank A.~De~Cosa, N.~Desai, B.~Nachman, M.~Pierini, and S.~Thomas for useful conversations.  TC is supported by an LHC Theory Initiative Postdoctoral Fellowship, under the National Science Foundation grant PHY-0969510.  ML is supported by the DOE under contract DESC0007968, the Alfred P.~Sloan Foundation, and the Cottrell Scholar program through the Research Corporation for Science Advancement. HL is supported by the DOE under contract DE-AC02-05CH11231 and NSF grant PHY-1316783.  TC and ML are grateful to the CKC-CERN for their hospitality during the completion of this project.

\appendix
\section{Approaching the Contact Operator Limit}
\label{sec:ApproachContactLimit}
In this appendix, we study how the contact operator limit is approached for the $s/t$-channel UV completions. In the large $M_{Z'}$ ($M_{\Phi}$) limit, the $\overline \chi \,\chi$ production can be described by the effective contact operator by integrating out the $Z'$ ($\Phi$). In the $Z'$ case, this gives
\begin{align}
  \mathcal{O}^{Z'}_{\rm contact} =   \frac{1}{\Lambda_{Z'}^2}\,\big(\overline q_i \,\gamma^\mu \,q_i \big)\big (\overline \chi_a \,\gamma_\mu \,\chi_a\big) \, 
\quad\text{with}\quad \Lambda_{Z'} = \frac{M_{Z'}}{\sqrt{g_\chi\, g_q}}\,,
\end{align}
where we have taken $c=1$ as defined in Eq.~(\ref{eq:contact}).  For the $t$-channel case, the spin structure is different since only $q_R$ couples to the dark sector. The contact operator in this case is
\begin{align}
  \mathcal{O}^{\Phi}_{\rm contact} =   \frac{2}{\Lambda_{\Phi}^2}\,\big(\overline q_i\, \gamma^\mu P_R \,q_i \big)\big (\overline \chi_a \,\gamma_\mu P_L\,\chi_a\big) \, 
\quad \text{with} \quad \Lambda_{\Phi} = \frac{2\,M_{\Phi}}{\lambda}\,,
\end{align}
where $P_{L,R}$ are the projection operators for the corresponding helicity component. The extra factor of 2 is present so that when $\Lambda_{Z'}=\Lambda_{\Phi}$, the total production cross sections for the two cases are equal. Because the protons are not polarized at the LHC, the helicity structures do not lead to differences in the distributions of interest here. 

\begin{figure}[h!] 
\hspace{-0.9 cm} 
\vspace{-20pt}
 \includegraphics[width=0.84\textwidth]{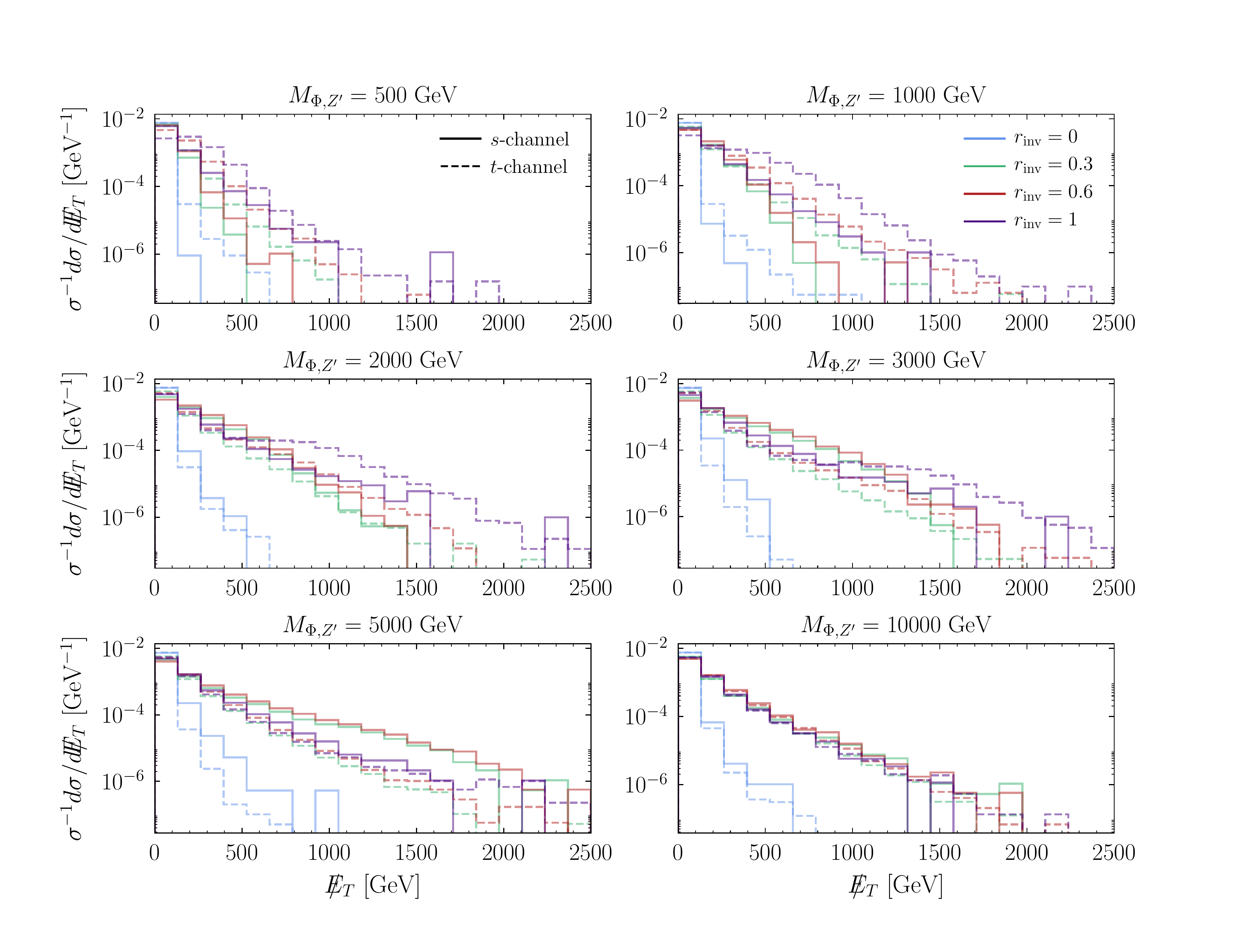} \hspace{-0.29 cm}
\caption{$\MET$ distributions for the $s$- and $t$-channel models for a range of mediator masses and $r_\text{inv}$.}
\label{fig:contactlimit_met}
\end{figure}

\begin{figure}[h!] 
\hspace{-0.9 cm} 
\vspace{-20pt}
 \includegraphics[width=0.84\textwidth]{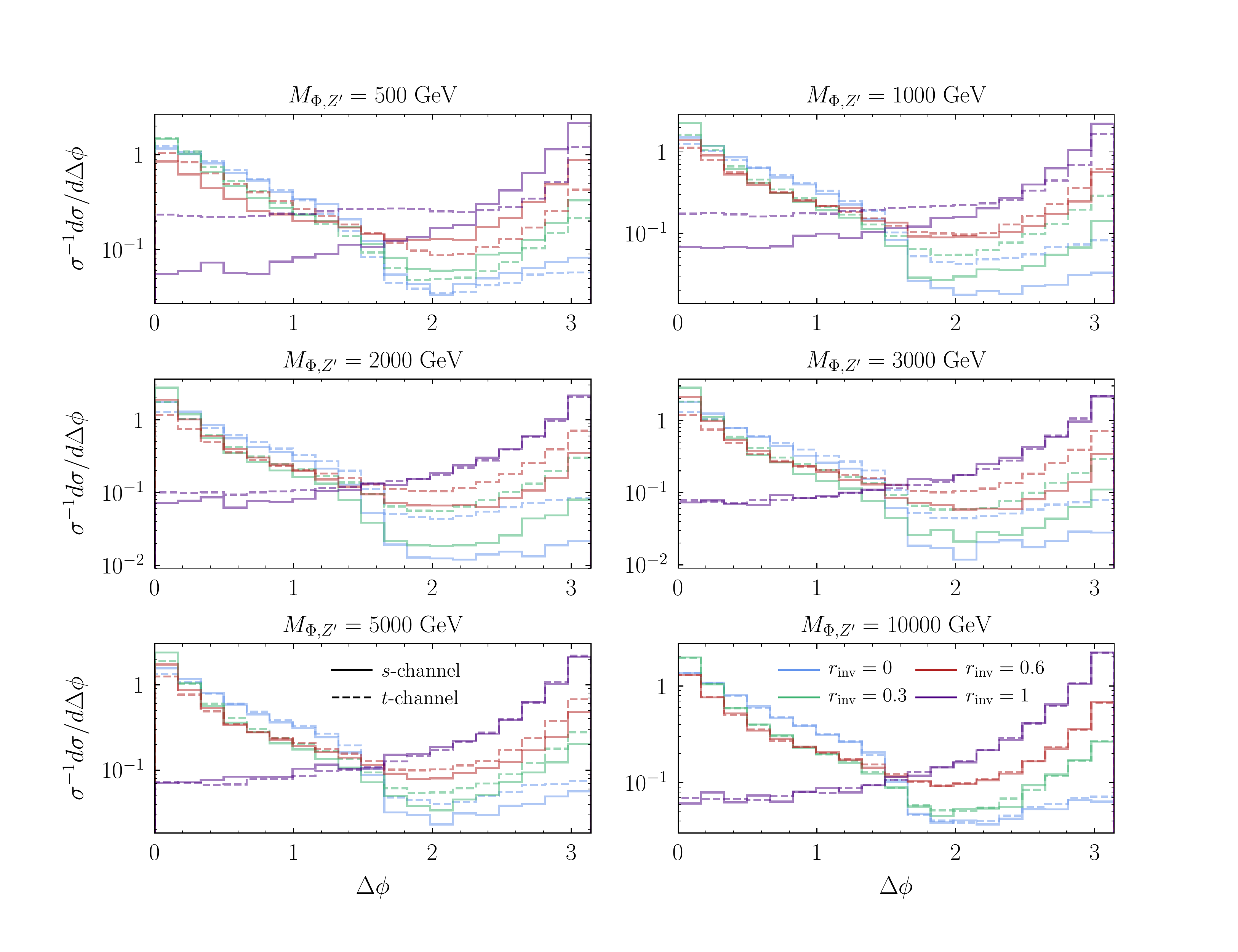} \hspace{-0.29 cm}
\caption{$\Delta\phi$ distributions for the $s$- and $t$-channel models for a range of mediator masses and $r_\text{inv}$.}
\label{fig:contactlimit_dphi}
\end{figure}

To illustrate how quickly the contact operator limit is approached, Fig.~\ref{fig:contactlimit_met} and \ref{fig:contactlimit_dphi} show the normalized $\MET$ and $\Delta \phi$ distributions for different values of the mediator mass and $r_\text{inv}$.  For the $s$-channel model, we take $g_q = 0.1$ and $g_\chi = 1$, and for the $t$-channel model, we take $\lambda = 1$.  For low masses, there are significant differences between the two cases.  This is due to different production channels along with modifications to the $m_{\chi \bar{\chi}}$ distributions. As the masses increase to $\mathcal{O}(10 \; {\rm TeV})$, the distributions converge to the universal contact operator limit.  We take $M_{Z'} = 100 \text{ TeV}$ for the contact operator event generation.

\newpage

\hbox{ }\vspace{-30pt}

\hbox{ }\vspace{-35pt}

\section{Search Insensitivity to Dark Shower Parameters}
\label{sec:VaryParameters}
\begin{figure}[h!] 
 \includegraphics[width=0.45\textwidth]{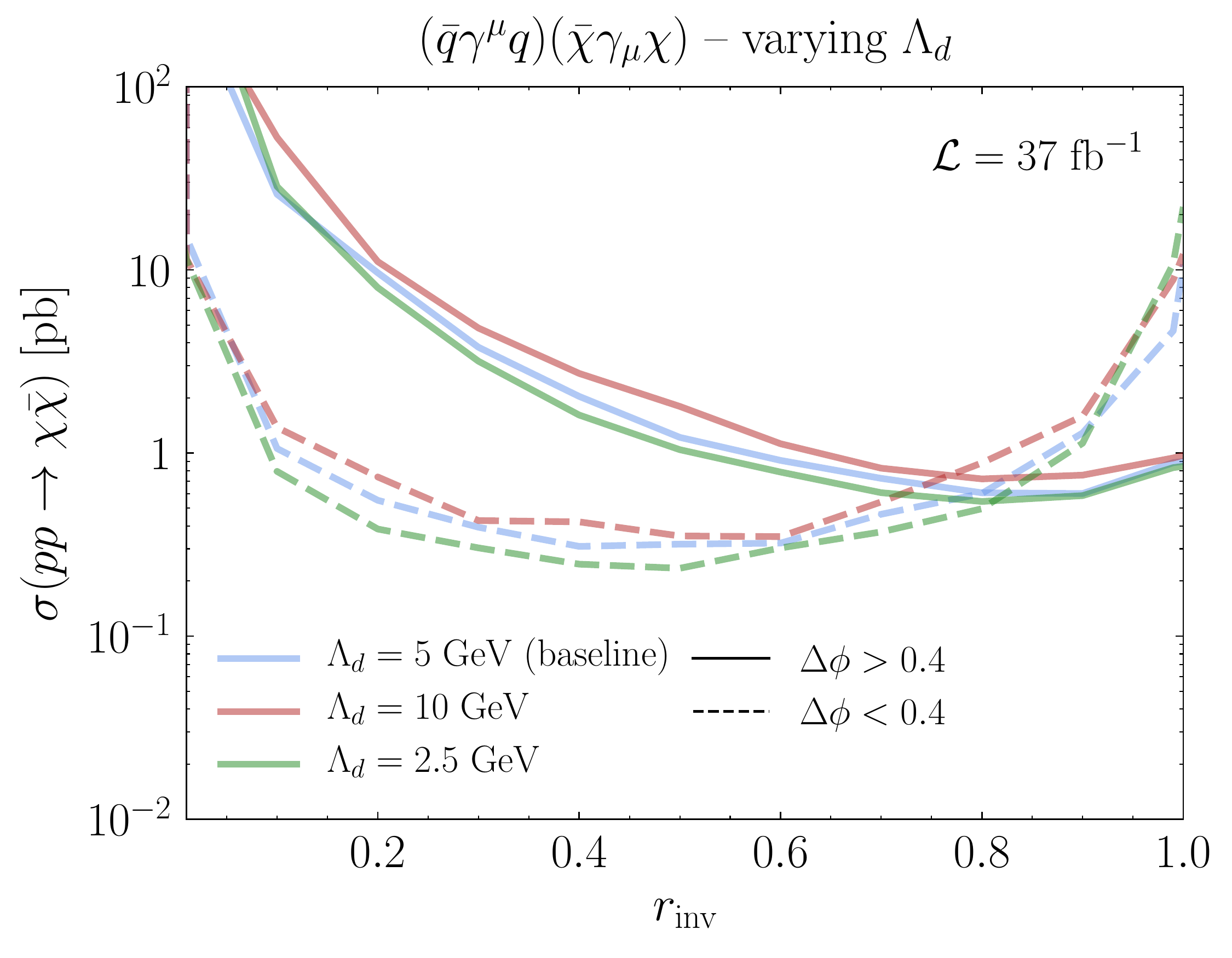}
 \includegraphics[width=0.45\textwidth]{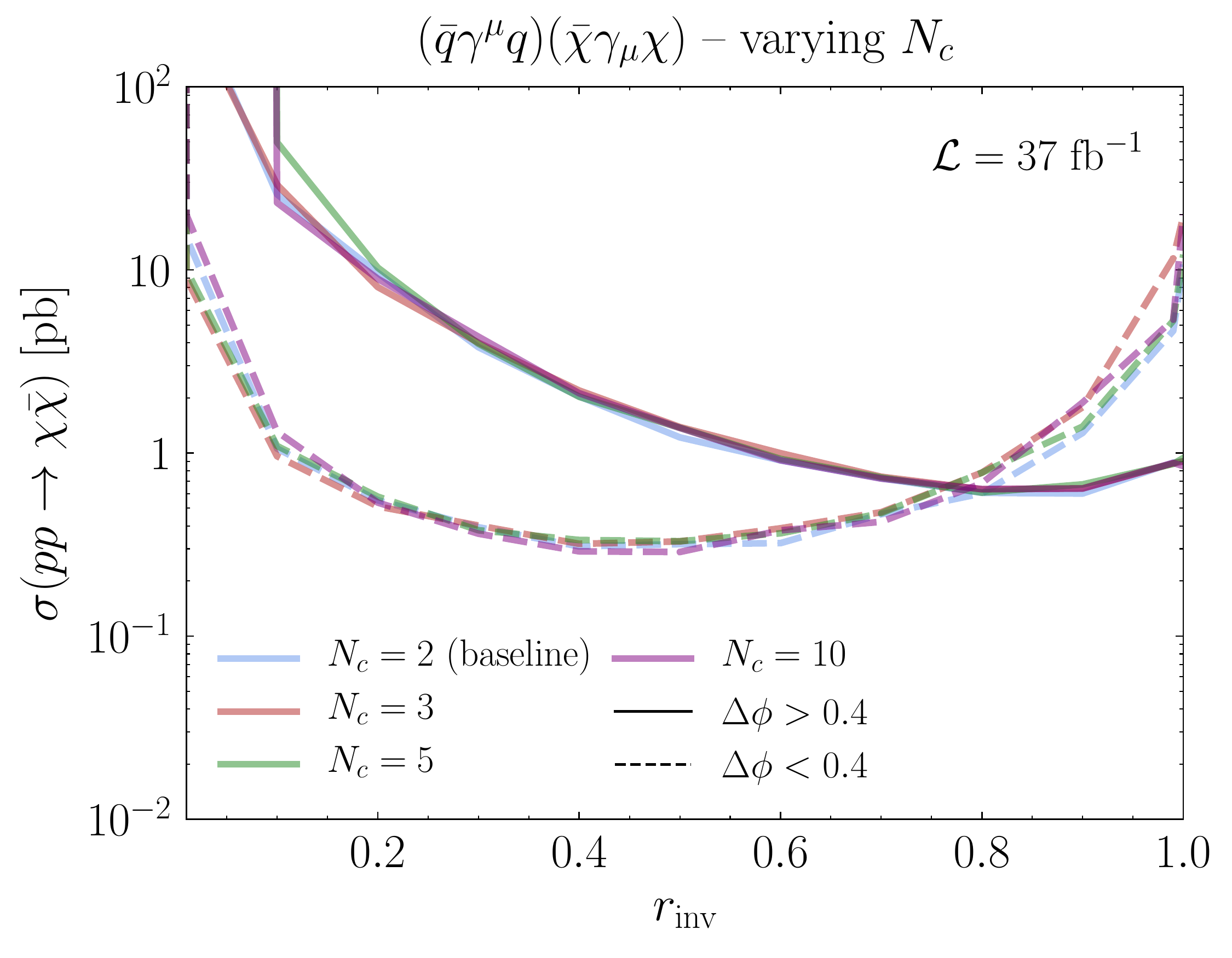}
  \includegraphics[width=0.45\textwidth]{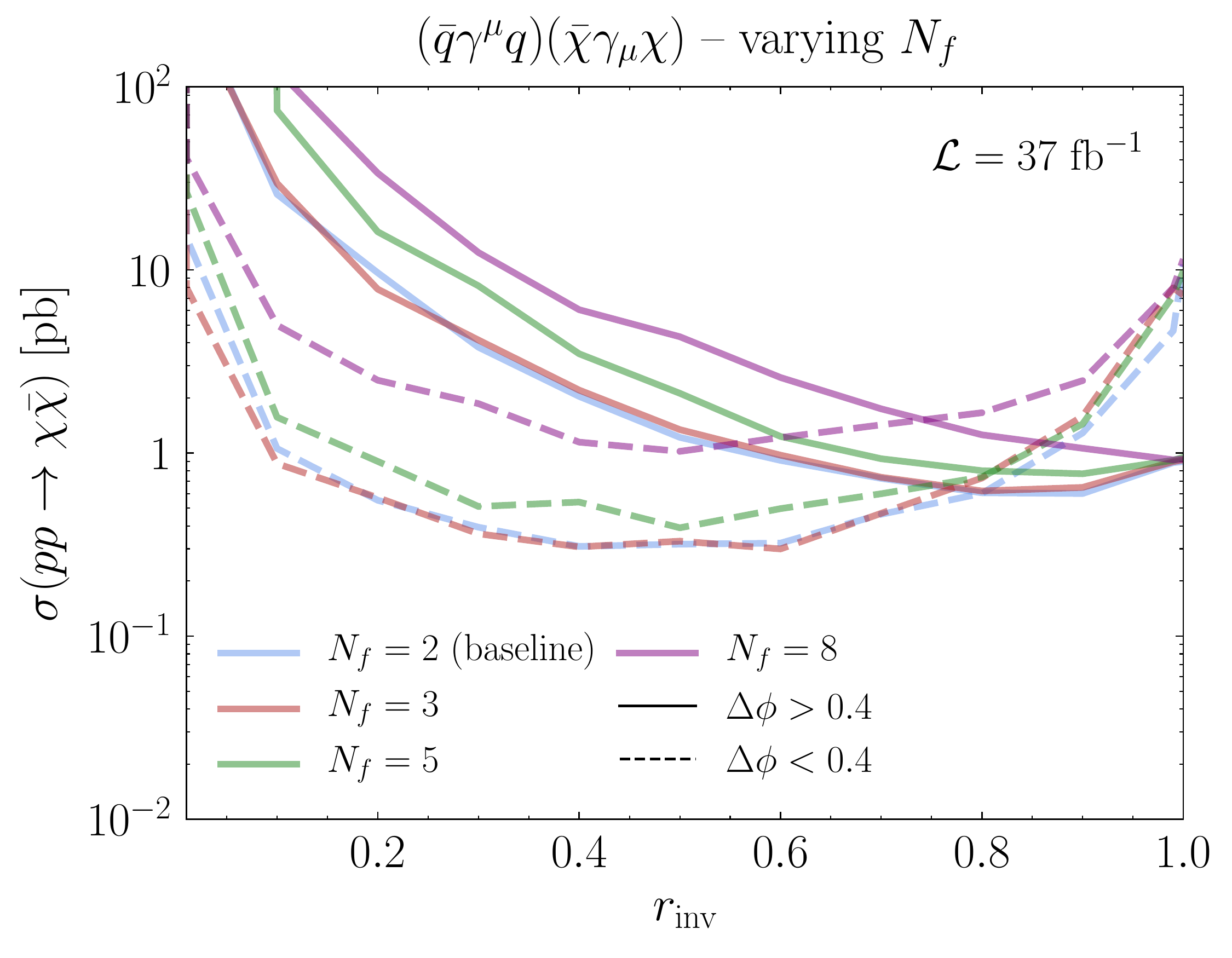} 
  \includegraphics[width=0.45\textwidth]{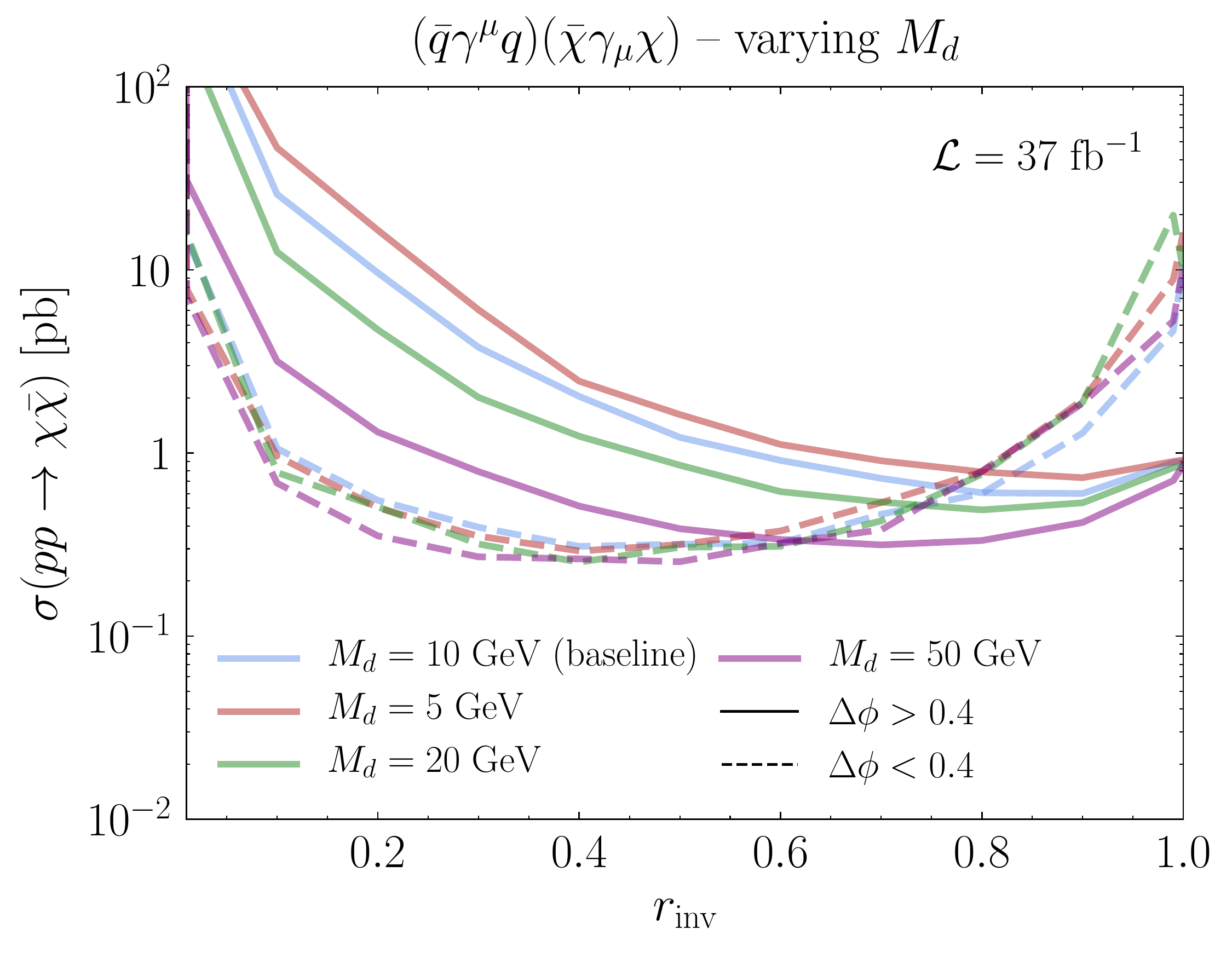} 

\caption{This figure shows that there is little variation in the limits on the cross section, assuming the contact operator approximation, when changing the detailed properties of the dark sector.  
}
\label{fig:param_var}
\end{figure}

In this Appendix, we provide a concrete illustration that the searches studied here are insensitive to the detailed choices made for the dark sector parameters, and thus are inclusive.   In particular, we vary the following parameters: the dark confinement scale $\Lambda_d = 2.5, 5, 10 \text{ GeV}$; the number of dark colors $N_c = 2, 3, 5, 10$; the number of dark flavors $N_f = 2, 3, 5, 8$; and the mass of the dark quark $M_d = 5,  10, 20, 50 \text{ GeV}$ in the simulation.  For each variation, we process the resulting events through the simulation pipeline.  Note that we are ignoring any subtleties related to the lifetime and flavor content of the decay products, \emph{i.e.}, we promptly decay all dark mesons to light flavor quarks.  The results are shown in Fig.~\ref{fig:param_var}, where we see that the limits are essentially unchanged as we scan the dark shower parameter space.  The largest variation in the limits is due to varying $N_f$, which is a result of the change in the running of the coupling.  As $N_f \rightarrow 11$ (for $N_c = 2$), the one-loop $\beta$-function goes to zero.  
Varying $M_d$ affects the limits for the monojet-style search with $\Delta \phi > 0.4$, but leaves the semi-visible search essentially unchanged.  

\end{spacing}
\clearpage

\begin{spacing}{1.1}
\bibliographystyle{utphys}
\bibliography{SemiVisibleJetsSimplified}

\providecommand{\href}[2]{#2}\begingroup\raggedright\begin{thebibliography}{100}

\bibitem{Goldberg:1986nk}
H.~Goldberg and L.~J. Hall, ``{A New Candidate for Dark Matter},''
\href{http://dx.doi.org/10.1016/0370-2693(86)90731-8}{{\em Phys. Lett.} {\bf
  B174} (1986)  151}.

\bibitem{Strassler:2006im}
M.~J. Strassler and K.~M. Zurek, ``{Echoes of a hidden valley at hadron
  colliders},'' \href{http://dx.doi.org/10.1016/j.physletb.2007.06.055}{{\em
  Phys. Lett.} {\bf B651} (2007)  374--379},
\href{http://arxiv.org/abs/hep-ph/0604261}{{\tt arXiv:hep-ph/0604261
  [hep-ph]}}.

\bibitem{Pospelov:2007mp}
M.~Pospelov, A.~Ritz, and M.~B. Voloshin, ``{Secluded WIMP Dark Matter},''
  \href{http://dx.doi.org/10.1016/j.physletb.2008.02.052}{{\em Phys. Lett.}
  {\bf B662} (2008)  53--61},
\href{http://arxiv.org/abs/0711.4866}{{\tt arXiv:0711.4866 [hep-ph]}}.

\bibitem{Feng:2008ya}
J.~L. Feng and J.~Kumar, ``{The WIMPless Miracle: Dark-Matter Particles without
  Weak-Scale Masses or Weak Interactions},''
  \href{http://dx.doi.org/10.1103/PhysRevLett.101.231301}{{\em Phys. Rev.
  Lett.} {\bf 101} (2008)  231301},
\href{http://arxiv.org/abs/0803.4196}{{\tt arXiv:0803.4196 [hep-ph]}}.

\bibitem{ArkaniHamed:2008qn}
N.~Arkani-Hamed, D.~P. Finkbeiner, T.~R. Slatyer, and N.~Weiner, ``{A Theory of
  Dark Matter},'' \href{http://dx.doi.org/10.1103/PhysRevD.79.015014}{{\em
  Phys. Rev.} {\bf D79} (2009)  015014},
\href{http://arxiv.org/abs/0810.0713}{{\tt arXiv:0810.0713 [hep-ph]}}.

\bibitem{Hochberg:2014kqa}
Y.~Hochberg, E.~Kuflik, H.~Murayama, T.~Volansky, and J.~G. Wacker, ``{Model
  for Thermal Relic Dark Matter of Strongly Interacting Massive Particles},''
  \href{http://dx.doi.org/10.1103/PhysRevLett.115.021301}{{\em Phys. Rev.
  Lett.} {\bf 115} (2015) no.~2, 021301},
\href{http://arxiv.org/abs/1411.3727}{{\tt arXiv:1411.3727 [hep-ph]}}.

\bibitem{Buen-Abad:2015ova}
M.~A. Buen-Abad, G.~Marques-Tavares, and M.~Schmaltz, ``{Non-Abelian dark
  matter and dark radiation},''
  \href{http://dx.doi.org/10.1103/PhysRevD.92.023531}{{\em Phys. Rev.} {\bf
  D92} (2015) no.~2, 023531},
\href{http://arxiv.org/abs/1505.03542}{{\tt arXiv:1505.03542 [hep-ph]}}.

\bibitem{Nussinov:1985xr}
S.~Nussinov, ``{Technocosmology: Could a Technibaryon Excess Provide a
  `Natural' Missing Mass Candidate?},''
\href{http://dx.doi.org/10.1016/0370-2693(85)90689-6}{{\em Phys. Lett.} {\bf
  165B} (1985)  55--58}.

\bibitem{Barr:1990ca}
S.~M. Barr, R.~S. Chivukula, and E.~Farhi, ``{Electroweak Fermion Number
  Violation and the Production of Stable Particles in the Early Universe},''
\href{http://dx.doi.org/10.1016/0370-2693(90)91661-T}{{\em Phys. Lett.} {\bf
  B241} (1990)  387--391}.

\bibitem{Khlopov:2005ew}
M.~{\relax Yu}. Khlopov, ``{Composite dark matter from 4th generation},''
  \href{http://dx.doi.org/10.1134/S0021364006010012}{{\em Pisma Zh. Eksp. Teor.
  Fiz.} {\bf 83} (2006)  3--6},
\href{http://arxiv.org/abs/astro-ph/0511796}{{\tt arXiv:astro-ph/0511796
  [astro-ph]}}.

\bibitem{Gudnason:2006ug}
S.~B. Gudnason, C.~Kouvaris, and F.~Sannino, ``{Towards working technicolor:
  Effective theories and dark matter},''
  \href{http://dx.doi.org/10.1103/PhysRevD.73.115003}{{\em Phys. Rev.} {\bf
  D73} (2006)  115003},
\href{http://arxiv.org/abs/hep-ph/0603014}{{\tt arXiv:hep-ph/0603014
  [hep-ph]}}.

\bibitem{Gudnason:2006yj}
S.~B. Gudnason, C.~Kouvaris, and F.~Sannino, ``{Dark Matter from new
  Technicolor Theories},''
  \href{http://dx.doi.org/10.1103/PhysRevD.74.095008}{{\em Phys. Rev.} {\bf
  D74} (2006)  095008},
\href{http://arxiv.org/abs/hep-ph/0608055}{{\tt arXiv:hep-ph/0608055
  [hep-ph]}}.

\bibitem{Khlopov:2008ty}
M.~{\relax Yu}. Khlopov and C.~Kouvaris, ``{Composite dark matter from a model
  with composite Higgs boson},''
  \href{http://dx.doi.org/10.1103/PhysRevD.78.065040}{{\em Phys. Rev.} {\bf
  D78} (2008)  065040},
\href{http://arxiv.org/abs/0806.1191}{{\tt arXiv:0806.1191 [astro-ph]}}.

\bibitem{Ryttov:2008xe}
T.~A. Ryttov and F.~Sannino, ``{Ultra Minimal Technicolor and its Dark Matter
  TIMP},'' \href{http://dx.doi.org/10.1103/PhysRevD.78.115010}{{\em Phys. Rev.}
  {\bf D78} (2008)  115010},
\href{http://arxiv.org/abs/0809.0713}{{\tt arXiv:0809.0713 [hep-ph]}}.

\bibitem{Foadi:2008qv}
R.~Foadi, M.~T. Frandsen, and F.~Sannino, ``{Technicolor Dark Matter},''
  \href{http://dx.doi.org/10.1103/PhysRevD.80.037702}{{\em Phys. Rev.} {\bf
  D80} (2009)  037702},
\href{http://arxiv.org/abs/0812.3406}{{\tt arXiv:0812.3406 [hep-ph]}}.

\bibitem{Alves:2009nf}
D.~S.~M. Alves, S.~R. Behbahani, P.~Schuster, and J.~G. Wacker, ``{Composite
  Inelastic Dark Matter},''
  \href{http://dx.doi.org/10.1016/j.physletb.2010.08.006}{{\em Phys. Lett.}
  {\bf B692} (2010)  323--326},
\href{http://arxiv.org/abs/0903.3945}{{\tt arXiv:0903.3945 [hep-ph]}}.

\bibitem{Mardon:2009gw}
J.~Mardon, Y.~Nomura, and J.~Thaler, ``{Cosmic Signals from the Hidden
  Sector},'' \href{http://dx.doi.org/10.1103/PhysRevD.80.035013}{{\em Phys.
  Rev.} {\bf D80} (2009)  035013},
\href{http://arxiv.org/abs/0905.3749}{{\tt arXiv:0905.3749 [hep-ph]}}.

\bibitem{Kribs:2009fy}
G.~D. Kribs, T.~S. Roy, J.~Terning, and K.~M. Zurek, ``{Quirky Composite Dark
  Matter},'' \href{http://dx.doi.org/10.1103/PhysRevD.81.095001}{{\em
  Phys.Rev.} {\bf D81} (2010)  095001},
\href{http://arxiv.org/abs/0909.2034}{{\tt arXiv:0909.2034 [hep-ph]}}.

\bibitem{Frandsen:2009mi}
M.~T. Frandsen and F.~Sannino, ``{iTIMP: isotriplet Technicolor Interacting
  Massive Particle as Dark Matter},''
  \href{http://dx.doi.org/10.1103/PhysRevD.81.097704}{{\em Phys. Rev.} {\bf
  D81} (2010)  097704},
\href{http://arxiv.org/abs/0911.1570}{{\tt arXiv:0911.1570 [hep-ph]}}.

\bibitem{Lisanti:2009am}
M.~Lisanti and J.~G. Wacker, ``{Parity Violation in Composite Inelastic Dark
  Matter Models},'' \href{http://dx.doi.org/10.1103/PhysRevD.82.055023}{{\em
  Phys. Rev.} {\bf D82} (2010)  055023},
\href{http://arxiv.org/abs/0911.4483}{{\tt arXiv:0911.4483 [hep-ph]}}.

\bibitem{Belyaev:2010kp}
A.~Belyaev, M.~T. Frandsen, S.~Sarkar, and F.~Sannino, ``{Mixed dark matter
  from technicolor},'' \href{http://dx.doi.org/10.1103/PhysRevD.83.015007}{{\em
  Phys. Rev.} {\bf D83} (2011)  015007},
\href{http://arxiv.org/abs/1007.4839}{{\tt arXiv:1007.4839 [hep-ph]}}.

\bibitem{Alves:2010dd}
D.~Spier Moreira~Alves, S.~R. Behbahani, P.~Schuster, and J.~G. Wacker, ``{The
  Cosmology of Composite Inelastic Dark Matter},''
  \href{http://dx.doi.org/10.1007/JHEP06(2010)113}{{\em JHEP} {\bf 06} (2010)
  113},
\href{http://arxiv.org/abs/1003.4729}{{\tt arXiv:1003.4729 [hep-ph]}}.

\bibitem{Lewis:2011zb}
R.~Lewis, C.~Pica, and F.~Sannino, ``{Light Asymmetric Dark Matter on the
  Lattice: SU(2) Technicolor with Two Fundamental Flavors},''
  \href{http://dx.doi.org/10.1103/PhysRevD.85.014504}{{\em Phys. Rev.} {\bf
  D85} (2012)  014504},
\href{http://arxiv.org/abs/1109.3513}{{\tt arXiv:1109.3513 [hep-ph]}}.

\bibitem{Buckley:2012ky}
M.~R. Buckley and E.~T. Neil, ``{Thermal dark matter from a confining
  sector},'' \href{http://dx.doi.org/10.1103/PhysRevD.87.043510}{{\em Phys.
  Rev.} {\bf D87} (2013) no.~4, 043510},
\href{http://arxiv.org/abs/1209.6054}{{\tt arXiv:1209.6054 [hep-ph]}}.

\bibitem{Hietanen:2012qd}
A.~Hietanen, C.~Pica, F.~Sannino, and U.~I. Sondergaard, ``{Isotriplet Dark
  Matter on the Lattice: SO(4)-gauge theory with two Vector Wilson fermions},''
  {\em PoS} {\bf LATTICE2012} (2012)  065,
\href{http://arxiv.org/abs/1211.0142}{{\tt arXiv:1211.0142 [hep-lat]}}.

\bibitem{Blinov:2012hq}
N.~Blinov, D.~E. Morrissey, K.~Sigurdson, and S.~Tulin, ``{Dark Matter
  Antibaryons from a Supersymmetric Hidden Sector},''
  \href{http://dx.doi.org/10.1103/PhysRevD.86.095021}{{\em Phys. Rev.} {\bf
  D86} (2012)  095021},
\href{http://arxiv.org/abs/1206.3304}{{\tt arXiv:1206.3304 [hep-ph]}}.

\bibitem{Cline:2013zca}
J.~M. Cline, Z.~Liu, G.~Moore, and W.~Xue, ``{Composite strongly interacting
  dark matter},'' \href{http://dx.doi.org/10.1103/PhysRevD.90.015023}{{\em
  Phys. Rev.} {\bf D90} (2014) no.~1, 015023},
\href{http://arxiv.org/abs/1312.3325}{{\tt arXiv:1312.3325 [hep-ph]}}.

\bibitem{Bai:2013xga}
Y.~Bai and P.~Schwaller, ``{Scale of dark QCD},''
  \href{http://dx.doi.org/10.1103/PhysRevD.89.063522}{{\em Phys. Rev.} {\bf
  D89} (2014) no.~6, 063522},
\href{http://arxiv.org/abs/1306.4676}{{\tt arXiv:1306.4676 [hep-ph]}}.

\bibitem{Appelquist:2014jch}
{\bf Lattice Strong Dynamics (LSD)} Collaboration, T.~Appelquist {\em et al.},
  ``{Composite bosonic baryon dark matter on the lattice: SU(4) baryon spectrum
  and the effective Higgs interaction},''
  \href{http://dx.doi.org/10.1103/PhysRevD.89.094508}{{\em Phys. Rev.} {\bf
  D89} (2014) no.~9, 094508},
\href{http://arxiv.org/abs/1402.6656}{{\tt arXiv:1402.6656 [hep-lat]}}.

\bibitem{Detmold:2014qqa}
W.~Detmold, M.~McCullough, and A.~Pochinsky, ``{Dark Nuclei I: Cosmology and
  Indirect Detection},''
  \href{http://dx.doi.org/10.1103/PhysRevD.90.115013}{{\em Phys. Rev.} {\bf
  D90} (2014) no.~11, 115013},
\href{http://arxiv.org/abs/1406.2276}{{\tt arXiv:1406.2276 [hep-ph]}}.

\bibitem{Detmold:2014kba}
W.~Detmold, M.~McCullough, and A.~Pochinsky, ``{Dark nuclei. II. Nuclear
  spectroscopy in two-color QCD},''
  \href{http://dx.doi.org/10.1103/PhysRevD.90.114506}{{\em Phys. Rev.} {\bf
  D90} (2014) no.~11, 114506},
\href{http://arxiv.org/abs/1406.4116}{{\tt arXiv:1406.4116 [hep-lat]}}.

\bibitem{Hochberg:2014dra}
Y.~Hochberg, E.~Kuflik, T.~Volansky, and J.~G. Wacker, ``{Mechanism for Thermal
  Relic Dark Matter of Strongly Interacting Massive Particles},''
  \href{http://dx.doi.org/10.1103/PhysRevLett.113.171301}{{\em Phys. Rev.
  Lett.} {\bf 113} (2014)  171301},
\href{http://arxiv.org/abs/1402.5143}{{\tt arXiv:1402.5143 [hep-ph]}}.

\bibitem{Appelquist:2015zfa}
T.~Appelquist {\em et al.}, ``{Detecting Stealth Dark Matter Directly through
  Electromagnetic Polarizability},''
  \href{http://dx.doi.org/10.1103/PhysRevLett.115.171803}{{\em Phys. Rev.
  Lett.} {\bf 115} (2015) no.~17, 171803},
\href{http://arxiv.org/abs/1503.04205}{{\tt arXiv:1503.04205 [hep-ph]}}.

\bibitem{Appelquist:2015yfa}
T.~Appelquist {\em et al.}, ``{Stealth Dark Matter: Dark scalar baryons through
  the Higgs portal},'' \href{http://dx.doi.org/10.1103/PhysRevD.92.075030}{{\em
  Phys. Rev.} {\bf D92} (2015) no.~7, 075030},
\href{http://arxiv.org/abs/1503.04203}{{\tt arXiv:1503.04203 [hep-ph]}}.

\bibitem{Han:2007ae}
T.~Han, Z.~Si, K.~M. Zurek, and M.~J. Strassler, ``{Phenomenology of hidden
  valleys at hadron colliders},''
  \href{http://dx.doi.org/10.1088/1126-6708/2008/07/008}{{\em JHEP} {\bf 0807}
  (2008)  008},
\href{http://arxiv.org/abs/0712.2041}{{\tt arXiv:0712.2041 [hep-ph]}}.

\bibitem{Strassler:2006ri}
M.~J. Strassler and K.~M. Zurek, ``{Discovering the Higgs through
  highly-displaced vertices},''
  \href{http://dx.doi.org/10.1016/j.physletb.2008.02.008}{{\em Phys.Lett.} {\bf
  B661} (2008)  263--267},
\href{http://arxiv.org/abs/hep-ph/0605193}{{\tt arXiv:hep-ph/0605193
  [hep-ph]}}.

\bibitem{Harnik:2008ax}
R.~Harnik and T.~Wizansky, ``{Signals of New Physics in the Underlying
  Event},'' \href{http://dx.doi.org/10.1103/PhysRevD.80.075015}{{\em Phys.
  Rev.} {\bf D80} (2009)  075015},
\href{http://arxiv.org/abs/0810.3948}{{\tt arXiv:0810.3948 [hep-ph]}}.

\bibitem{Falkowski:2010cm}
A.~Falkowski, J.~T. Ruderman, T.~Volansky, and J.~Zupan, ``{Hidden Higgs
  Decaying to Lepton Jets},''
  \href{http://dx.doi.org/10.1007/JHEP05(2010)077}{{\em JHEP} {\bf 05} (2010)
  077},
\href{http://arxiv.org/abs/1002.2952}{{\tt arXiv:1002.2952 [hep-ph]}}.

\bibitem{Falkowski:2010gv}
A.~Falkowski, J.~T. Ruderman, T.~Volansky, and J.~Zupan, ``{Discovering Higgs
  Decays to Lepton Jets at Hadron Colliders},''
  \href{http://dx.doi.org/10.1103/PhysRevLett.105.241801}{{\em Phys. Rev.
  Lett.} {\bf 105} (2010)  241801},
\href{http://arxiv.org/abs/1007.3496}{{\tt arXiv:1007.3496 [hep-ph]}}.

\bibitem{Co:2015pka}
R.~T. Co, F.~D'Eramo, L.~J. Hall, and D.~Pappadopulo, ``{Freeze-In Dark Matter
  with Displaced Signatures at Colliders},''
  \href{http://dx.doi.org/10.1088/1475-7516/2015/12/024}{{\em JCAP} {\bf 1512}
  (2015) no.~12, 024},
\href{http://arxiv.org/abs/1506.07532}{{\tt arXiv:1506.07532 [hep-ph]}}.

\bibitem{Izaguirre:2015zva}
E.~Izaguirre, G.~Krnjaic, and B.~Shuve, ``{Discovering Inelastic Thermal-Relic
  Dark Matter at Colliders},''
  \href{http://dx.doi.org/10.1103/PhysRevD.93.063523}{{\em Phys. Rev.} {\bf
  D93} (2016) no.~6, 063523},
\href{http://arxiv.org/abs/1508.03050}{{\tt arXiv:1508.03050 [hep-ph]}}.

\bibitem{Garcia:2015toa}
I.~Garcia~Garcia, R.~Lasenby, and J.~March-Russell, ``{Twin Higgs Asymmetric
  Dark Matter},'' \href{http://dx.doi.org/10.1103/PhysRevLett.115.121801}{{\em
  Phys. Rev. Lett.} {\bf 115} (2015) no.~12, 121801},
\href{http://arxiv.org/abs/1505.07410}{{\tt arXiv:1505.07410 [hep-ph]}}.

\bibitem{Zhang:2016sll}
M.~Zhang, M.~Kim, H.-S. Lee, and M.~Park, ``{Probing the chirality of dark
  matter at colliders with dark photon showering},''
\href{http://arxiv.org/abs/1612.02850}{{\tt arXiv:1612.02850 [hep-ph]}}.

\bibitem{Freytsis:2016dgf}
M.~Freytsis, S.~Knapen, D.~J. Robinson, and Y.~Tsai, ``{Gamma-rays from Dark
  Showers with Twin Higgs Models},''
  \href{http://dx.doi.org/10.1007/JHEP05(2016)018}{{\em JHEP} {\bf 05} (2016)
  018},
\href{http://arxiv.org/abs/1601.07556}{{\tt arXiv:1601.07556 [hep-ph]}}.

\bibitem{Daci:2015hca}
N.~Daci, I.~De~Bruyn, S.~Lowette, M.~H.~G. Tytgat, and B.~Zaldivar,
  ``{Simplified SIMPs and the LHC},''
  \href{http://dx.doi.org/10.1007/JHEP11(2015)108}{{\em JHEP} {\bf 11} (2015)
  108},
\href{http://arxiv.org/abs/1503.05505}{{\tt arXiv:1503.05505 [hep-ph]}}.

\bibitem{Hochberg:2015vrg}
Y.~Hochberg, E.~Kuflik, and H.~Murayama, ``{SIMP Spectroscopy},''
  \href{http://dx.doi.org/10.1007/JHEP05(2016)090}{{\em JHEP} {\bf 05} (2016)
  090},
\href{http://arxiv.org/abs/1512.07917}{{\tt arXiv:1512.07917 [hep-ph]}}.

\bibitem{Schwaller:2015gea}
P.~Schwaller, D.~Stolarski, and A.~Weiler, ``{Emerging Jets},''
\href{http://arxiv.org/abs/1502.05409}{{\tt arXiv:1502.05409 [hep-ph]}}.

\bibitem{Knapen:2016hky}
S.~Knapen, S.~Pagan~Griso, M.~Papucci, and D.~J. Robinson, ``{Triggering Soft
  Bombs at the LHC},''
\href{http://arxiv.org/abs/1612.00850}{{\tt arXiv:1612.00850 [hep-ph]}}.

\bibitem{Curtin:2015fna}
D.~Curtin and C.~B. Verhaaren, ``{Discovering Uncolored Naturalness in Exotic
  Higgs Decays},'' \href{http://dx.doi.org/10.1007/JHEP12(2015)072}{{\em JHEP}
  {\bf 12} (2015)  072},
\href{http://arxiv.org/abs/1506.06141}{{\tt arXiv:1506.06141 [hep-ph]}}.

\bibitem{Cohen:2015toa}
T.~Cohen, M.~Lisanti, and H.~K. Lou, ``{Semivisible Jets: Dark Matter
  Undercover at the LHC},''
  \href{http://dx.doi.org/10.1103/PhysRevLett.115.171804}{{\em Phys. Rev.
  Lett.} {\bf 115} (2015) no.~17, 171804},
\href{http://arxiv.org/abs/1503.00009}{{\tt arXiv:1503.00009 [hep-ph]}}.

\bibitem{Alves:2011wf}
{\bf LHC New Physics Working Group} Collaboration, D.~Alves {\em et al.},
  ``{Simplified Models for LHC New Physics Searches},''
  \href{http://dx.doi.org/10.1088/0954-3899/39/10/105005}{{\em J.Phys.} {\bf
  G39} (2012)  105005},
\href{http://arxiv.org/abs/1105.2838}{{\tt arXiv:1105.2838 [hep-ph]}}.

\bibitem{Abdallah:2014hon}
J.~Abdallah {\em et al.}, ``{Simplified Models for Dark Matter and Missing
  Energy Searches at the LHC},''
\href{http://arxiv.org/abs/1409.2893}{{\tt arXiv:1409.2893 [hep-ph]}}.

\bibitem{Abdallah:2015ter}
J.~Abdallah {\em et al.}, ``{Simplified Models for Dark Matter Searches at the
  LHC},'' \href{http://dx.doi.org/10.1016/j.dark.2015.08.001}{{\em Phys. Dark
  Univ.} {\bf 9-10} (2015)  8--23},
\href{http://arxiv.org/abs/1506.03116}{{\tt arXiv:1506.03116 [hep-ph]}}.

\bibitem{Abercrombie:2015wmb}
D.~Abercrombie {\em et al.}, ``{Dark Matter Benchmark Models for Early LHC
  Run-2 Searches: Report of the ATLAS/CMS Dark Matter Forum},''
\href{http://arxiv.org/abs/1507.00966}{{\tt arXiv:1507.00966 [hep-ex]}}.

\bibitem{Kahlhoefer:2017dnp}
F.~Kahlhoefer, ``{Review of LHC Dark Matter Searches},''
  \href{http://dx.doi.org/10.1142/S0217751X1730006X}{{\em Int. J. Mod. Phys.}
  {\bf A32} (2017) no.~13, 1730006},
\href{http://arxiv.org/abs/1702.02430}{{\tt arXiv:1702.02430 [hep-ph]}}.

\bibitem{Hietanen:2013fya}
A.~Hietanen, R.~Lewis, C.~Pica, and F.~Sannino, ``{Composite Goldstone Dark
  Matter: Experimental Predictions from the Lattice},''
  \href{http://dx.doi.org/10.1007/JHEP12(2014)130}{{\em JHEP} {\bf 12} (2014)
  130},
\href{http://arxiv.org/abs/1308.4130}{{\tt arXiv:1308.4130 [hep-ph]}}.

\bibitem{Hochberg:2017khi}
Y.~Hochberg, E.~Kuflik, and H.~Murayama, ``{Dark Spectroscopy},''
\href{http://arxiv.org/abs/1706.05008}{{\tt arXiv:1706.05008 [hep-ph]}}.

\bibitem{Andersson:1983ia}
B.~Andersson, G.~Gustafson, G.~Ingelman, and T.~Sjostrand, ``{Parton
  Fragmentation and String Dynamics},''
\href{http://dx.doi.org/10.1016/0370-1573(83)90080-7}{{\em Phys.Rept.} {\bf 97}
  (1983)  31--145}.

\bibitem{Goodman:2010yf}
J.~Goodman, M.~Ibe, A.~Rajaraman, W.~Shepherd, T.~M. Tait, {\em et al.},
  ``{Constraints on Light Majorana dark Matter from Colliders},''
  \href{http://dx.doi.org/10.1016/j.physletb.2010.11.009}{{\em Phys.Lett.} {\bf
  B695} (2011)  185--188},
\href{http://arxiv.org/abs/1005.1286}{{\tt arXiv:1005.1286 [hep-ph]}}.

\bibitem{Beltran:2010ww}
M.~Beltran, D.~Hooper, E.~W. Kolb, Z.~A.~C. Krusberg, and T.~M.~P. Tait,
  ``{Maverick dark matter at colliders},''
  \href{http://dx.doi.org/10.1007/JHEP09(2010)037}{{\em JHEP} {\bf 09} (2010)
  037},
\href{http://arxiv.org/abs/1002.4137}{{\tt arXiv:1002.4137 [hep-ph]}}.

\bibitem{Fox:2011pm}
P.~J. Fox, R.~Harnik, J.~Kopp, and Y.~Tsai, ``{Missing Energy Signatures of
  Dark Matter at the LHC},''
  \href{http://dx.doi.org/10.1103/PhysRevD.85.056011}{{\em Phys.Rev.} {\bf D85}
  (2012)  056011},
\href{http://arxiv.org/abs/1109.4398}{{\tt arXiv:1109.4398 [hep-ph]}}.

\bibitem{Frandsen:2012rk}
M.~T. Frandsen, F.~Kahlhoefer, A.~Preston, S.~Sarkar, and K.~Schmidt-Hoberg,
  ``{LHC and Tevatron Bounds on the Dark Matter Direct Detection Cross-Section
  for Vector Mediators},''
  \href{http://dx.doi.org/10.1007/JHEP07(2012)123}{{\em JHEP} {\bf 07} (2012)
  123},
\href{http://arxiv.org/abs/1204.3839}{{\tt arXiv:1204.3839 [hep-ph]}}.

\bibitem{An:2012va}
H.~An, X.~Ji, and L.-T. Wang, ``{Light Dark Matter and $Z'$ Dark Force at
  Colliders},'' \href{http://dx.doi.org/10.1007/JHEP07(2012)182}{{\em JHEP}
  {\bf 07} (2012)  182},
\href{http://arxiv.org/abs/1202.2894}{{\tt arXiv:1202.2894 [hep-ph]}}.

\bibitem{An:2012ue}
H.~An, R.~Huo, and L.-T. Wang, ``{Searching for Low Mass Dark Portal at the
  LHC},'' \href{http://dx.doi.org/10.1016/j.dark.2013.03.002}{{\em Phys. Dark
  Univ.} {\bf 2} (2013)  50--57},
\href{http://arxiv.org/abs/1212.2221}{{\tt arXiv:1212.2221 [hep-ph]}}.

\bibitem{An:2013xka}
H.~An, L.-T. Wang, and H.~Zhang, ``{Dark matter with $t$-channel mediator: a
  simple step beyond contact interaction},''
  \href{http://dx.doi.org/10.1103/PhysRevD.89.115014}{{\em Phys. Rev.} {\bf
  D89} (2014) no.~11, 115014},
\href{http://arxiv.org/abs/1308.0592}{{\tt arXiv:1308.0592 [hep-ph]}}.

\bibitem{Buchmueller:2013dya}
O.~Buchmueller, M.~J. Dolan, and C.~McCabe, ``{Beyond Effective Field Theory
  for Dark Matter Searches at the LHC},''
  \href{http://dx.doi.org/10.1007/JHEP01(2014)025}{{\em JHEP} {\bf 01} (2014)
  025},
\href{http://arxiv.org/abs/1308.6799}{{\tt arXiv:1308.6799 [hep-ph]}}.

\bibitem{Chang:2013oia}
S.~Chang, R.~Edezhath, J.~Hutchinson, and M.~Luty, ``{Effective WIMPs},''
  \href{http://dx.doi.org/10.1103/PhysRevD.89.015011}{{\em Phys. Rev.} {\bf
  D89} (2014) no.~1, 015011},
\href{http://arxiv.org/abs/1307.8120}{{\tt arXiv:1307.8120 [hep-ph]}}.

\bibitem{Bai:2013iqa}
Y.~Bai and J.~Berger, ``{Fermion Portal Dark Matter},''
  \href{http://dx.doi.org/10.1007/JHEP11(2013)171}{{\em JHEP} {\bf 11} (2013)
  171},
\href{http://arxiv.org/abs/1308.0612}{{\tt arXiv:1308.0612 [hep-ph]}}.

\bibitem{Dreiner:2013vla}
H.~Dreiner, D.~Schmeier, and J.~Tattersall, ``{Contact Interactions Probe
  Effective Dark Matter Models at the LHC},''
  \href{http://dx.doi.org/10.1209/0295-5075/102/51001}{{\em Europhys. Lett.}
  {\bf 102} (2013) no.~5, 51001},
\href{http://arxiv.org/abs/1303.3348}{{\tt arXiv:1303.3348 [hep-ph]}}.

\bibitem{DiFranzo:2013vra}
A.~DiFranzo, K.~I. Nagao, A.~Rajaraman, and T.~M.~P. Tait, ``{Simplified Models
  for Dark Matter Interacting with Quarks},''
  \href{http://dx.doi.org/10.1007/JHEP11(2013)014,
  10.1007/JHEP01(2014)162}{{\em JHEP} {\bf 11} (2013)  014},
  \href{http://arxiv.org/abs/1308.2679}{{\tt arXiv:1308.2679 [hep-ph]}}.
[Erratum: JHEP01,162(2014)].

\bibitem{Papucci:2014iwa}
M.~Papucci, A.~Vichi, and K.~M. Zurek, ``{Monojet versus the rest of the world
  I: t-channel models},'' \href{http://dx.doi.org/10.1007/JHEP11(2014)024}{{\em
  JHEP} {\bf 11} (2014)  024},
\href{http://arxiv.org/abs/1402.2285}{{\tt arXiv:1402.2285 [hep-ph]}}.

\bibitem{Buchmueller:2014yoa}
O.~Buchmueller, M.~J. Dolan, S.~A. Malik, and C.~McCabe, ``{Characterising dark
  matter searches at colliders and direct detection experiments: Vector
  mediators},'' \href{http://dx.doi.org/10.1007/JHEP01(2015)037}{{\em JHEP}
  {\bf 01} (2015)  037},
\href{http://arxiv.org/abs/1407.8257}{{\tt arXiv:1407.8257 [hep-ph]}}.

\bibitem{Hamaguchi:2014pja}
K.~Hamaguchi, S.~P. Liew, T.~Moroi, and Y.~Yamamoto, ``{Isospin-Violating Dark
  Matter with Colored Mediators},''
  \href{http://dx.doi.org/10.1007/JHEP05(2014)086}{{\em JHEP} {\bf 05} (2014)
  086},
\href{http://arxiv.org/abs/1403.0324}{{\tt arXiv:1403.0324 [hep-ph]}}.

\bibitem{Garny:2014waa}
M.~Garny, A.~Ibarra, S.~Rydbeck, and S.~Vogl, ``{Majorana Dark Matter with a
  Coloured Mediator: Collider vs Direct and Indirect Searches},''
  \href{http://dx.doi.org/10.1007/JHEP06(2014)169}{{\em JHEP} {\bf 06} (2014)
  169},
\href{http://arxiv.org/abs/1403.4634}{{\tt arXiv:1403.4634 [hep-ph]}}.

\bibitem{Harris:2014hga}
P.~Harris, V.~V. Khoze, M.~Spannowsky, and C.~Williams, ``{Constraining Dark
  Sectors at Colliders: Beyond the Effective Theory Approach},''
  \href{http://dx.doi.org/10.1103/PhysRevD.91.055009}{{\em Phys. Rev.} {\bf
  D91} (2015)  055009},
\href{http://arxiv.org/abs/1411.0535}{{\tt arXiv:1411.0535 [hep-ph]}}.

\bibitem{Jacques:2015zha}
T.~Jacques and K.~Nordstr{\"o}m, ``{Mapping monojet constraints onto Simplified
  Dark Matter Models},'' \href{http://dx.doi.org/10.1007/JHEP06(2015)142}{{\em
  JHEP} {\bf 06} (2015)  142},
\href{http://arxiv.org/abs/1502.05721}{{\tt arXiv:1502.05721 [hep-ph]}}.

\bibitem{Liew:2016oon}
S.~P. Liew, M.~Papucci, A.~Vichi, and K.~M. Zurek, ``{Mono-X Versus Direct
  Searches: Simplified Models for Dark Matter at the LHC},''
  \href{http://dx.doi.org/10.1007/JHEP06(2017)082}{{\em JHEP} {\bf 06} (2017)
  082},
\href{http://arxiv.org/abs/1612.00219}{{\tt arXiv:1612.00219 [hep-ph]}}.

\bibitem{Englert:2016joy}
C.~Englert, M.~McCullough, and M.~Spannowsky, ``{S-Channel Dark Matter
  Simplified Models and Unitarity},''
  \href{http://dx.doi.org/10.1016/j.dark.2016.09.002}{{\em Phys. Dark Univ.}
  {\bf 14} (2016)  48--56},
\href{http://arxiv.org/abs/1604.07975}{{\tt arXiv:1604.07975 [hep-ph]}}.

\bibitem{Alwall:2014hca}
J.~Alwall, R.~Frederix, S.~Frixione, V.~Hirschi, F.~Maltoni, {\em et al.},
  ``{The automated computation of tree-level and next-to-leading order
  differential cross sections, and their matching to parton shower
  simulations},'' \href{http://dx.doi.org/10.1007/JHEP07(2014)079}{{\em JHEP}
  {\bf 1407} (2014)  079},
\href{http://arxiv.org/abs/1405.0301}{{\tt arXiv:1405.0301 [hep-ph]}}.

\bibitem{Ball:2013hta}
{\bf NNPDF} Collaboration, R.~D. Ball, V.~Bertone, S.~Carrazza, L.~Del~Debbio,
  S.~Forte, A.~Guffanti, N.~P. Hartland, and J.~Rojo, ``{Parton distributions
  with QED corrections},''
  \href{http://dx.doi.org/10.1016/j.nuclphysb.2013.10.010}{{\em Nucl. Phys.}
  {\bf B877} (2013)  290--320},
\href{http://arxiv.org/abs/1308.0598}{{\tt arXiv:1308.0598 [hep-ph]}}.

\bibitem{Sjostrand:2007gs}
T.~Sjostrand, S.~Mrenna, and P.~Z. Skands, ``{A Brief Introduction to PYTHIA
  8.1},'' \href{http://dx.doi.org/10.1016/j.cpc.2008.01.036}{{\em
  Comput.Phys.Commun.} {\bf 178} (2008)  852--867},
\href{http://arxiv.org/abs/0710.3820}{{\tt arXiv:0710.3820 [hep-ph]}}.

\bibitem{Carloni:2011kk}
L.~Carloni, J.~Rathsman, and T.~Sjostrand, ``{Discerning Secluded Sector gauge
  structures},'' \href{http://dx.doi.org/10.1007/JHEP04(2011)091}{{\em JHEP}
  {\bf 1104} (2011)  091},
\href{http://arxiv.org/abs/1102.3795}{{\tt arXiv:1102.3795 [hep-ph]}}.

\bibitem{Carloni:2010tw}
L.~Carloni and T.~Sjostrand, ``{Visible Effects of Invisible Hidden Valley
  Radiation},'' \href{http://dx.doi.org/10.1007/JHEP09(2010)105}{{\em JHEP}
  {\bf 1009} (2010)  105},
\href{http://arxiv.org/abs/1006.2911}{{\tt arXiv:1006.2911 [hep-ph]}}.

\bibitem{deFavereau:2013fsa}
{\bf DELPHES 3} Collaboration, J.~de~Favereau {\em et al.}, ``{DELPHES 3, A
  modular framework for fast simulation of a generic collider experiment},''
  \href{http://dx.doi.org/10.1007/JHEP02(2014)057}{{\em JHEP} {\bf 1402} (2014)
   057},
\href{http://arxiv.org/abs/1307.6346}{{\tt arXiv:1307.6346 [hep-ex]}}.

\bibitem{Cacciari:2008gp}
M.~Cacciari, G.~P. Salam, and G.~Soyez, ``{The Anti-k(t) jet clustering
  algorithm},'' \href{http://dx.doi.org/10.1088/1126-6708/2008/04/063}{{\em
  JHEP} {\bf 0804} (2008)  063},
\href{http://arxiv.org/abs/0802.1189}{{\tt arXiv:0802.1189 [hep-ph]}}.

\bibitem{Nachman:2014kla}
B.~Nachman, P.~Nef, A.~Schwartzman, M.~Swiatlowski, and C.~Wanotayaroj, ``{Jets
  from Jets: Re-clustering as a tool for large radius jet reconstruction and
  grooming at the LHC},'' \href{http://dx.doi.org/10.1007/JHEP02(2015)075}{{\em
  JHEP} {\bf 02} (2015)  075},
\href{http://arxiv.org/abs/1407.2922}{{\tt arXiv:1407.2922 [hep-ph]}}.

\bibitem{Alwall:2007fs}
J.~Alwall {\em et al.}, ``{Comparative study of various algorithms for the
  merging of parton showers and matrix elements in hadronic collisions},''
  \href{http://dx.doi.org/10.1140/epjc/s10052-007-0490-5}{{\em Eur. Phys. J.}
  {\bf C53} (2008)  473--500},
\href{http://arxiv.org/abs/0706.2569}{{\tt arXiv:0706.2569 [hep-ph]}}.

\bibitem{Aaboud:2016tnv}
{\bf ATLAS} Collaboration, M.~Aaboud {\em et al.}, ``{Search for new phenomena
  in final states with an energetic jet and large missing transverse momentum
  in $pp$ collisions at $\sqrt{s}=13$  TeV using the ATLAS detector},''
  \href{http://dx.doi.org/10.1103/PhysRevD.94.032005}{{\em Phys. Rev.} {\bf
  D94} (2016) no.~3, 032005},
\href{http://arxiv.org/abs/1604.07773}{{\tt arXiv:1604.07773 [hep-ex]}}.

\bibitem{Aaboud:2017yvp}
{\bf ATLAS} Collaboration, M.~Aaboud {\em et al.}, ``{Search for new phenomena
  in dijet events using 37 fb$^{-1}$ of $pp$ collision data collected at
  $\sqrt{s}=$13 TeV with the ATLAS detector},''
\href{http://arxiv.org/abs/1703.09127}{{\tt arXiv:1703.09127 [hep-ex]}}.

\bibitem{Cowan:2010js}
G.~Cowan, K.~Cranmer, E.~Gross, and O.~Vitells, ``{Asymptotic formulae for
  likelihood-based tests of new physics},''
  \href{http://dx.doi.org/10.1140/epjc/s10052-011-1554-0,
  10.1140/epjc/s10052-013-2501-z}{{\em Eur. Phys. J.} {\bf C71} (2011)  1554},
  \href{http://arxiv.org/abs/1007.1727}{{\tt arXiv:1007.1727
  [physics.data-an]}}.
[Erratum: Eur. Phys. J.C73,2501(2013)].

\bibitem{Englert:2016knz}
C.~Englert, K.~Nordstram, and M.~Spannowsky, ``{Towards resolving
  strongly-interacting dark sectors at colliders},''
  \href{http://dx.doi.org/10.1103/PhysRevD.94.055028}{{\em Phys. Rev.} {\bf
  D94} (2016) no.~5, 055028},
\href{http://arxiv.org/abs/1606.05359}{{\tt arXiv:1606.05359 [hep-ph]}}.

\bibitem{Ismail:2017ulg}
A.~Ismail, A.~Katz, and D.~Racco, ``{On Dark Matter Interactions with the
  Standard Model through an Anomalous $Z'$},''
\href{http://arxiv.org/abs/1707.00709}{{\tt arXiv:1707.00709 [hep-ph]}}.

\bibitem{Neubert:2015fka}
M.~Neubert, J.~Wang, and C.~Zhang, ``{Higher-Order QCD Predictions for Dark
  Matter Production in Mono-$Z$ Searches at the LHC},''
  \href{http://dx.doi.org/10.1007/JHEP02(2016)082}{{\em JHEP} {\bf 02} (2016)
  082},
\href{http://arxiv.org/abs/1509.05785}{{\tt arXiv:1509.05785 [hep-ph]}}.

\bibitem{Mattelaer:2015haa}
O.~Mattelaer and E.~Vryonidou, ``{Dark matter production through loop-induced
  processes at the LHC: the s-channel mediator case},''
  \href{http://dx.doi.org/10.1140/epjc/s10052-015-3665-5}{{\em Eur. Phys. J.}
  {\bf C75} (2015) no.~9, 436},
\href{http://arxiv.org/abs/1508.00564}{{\tt arXiv:1508.00564 [hep-ph]}}.

\bibitem{Backovic:2015soa}
M.~Backovi{\'c}, M.~Kr{\"a}mer, F.~Maltoni, A.~Martini, K.~Mawatari, and
  M.~Pellen, ``{Higher-order QCD predictions for dark matter production at the
  LHC in simplified models with s-channel mediators},''
  \href{http://dx.doi.org/10.1140/epjc/s10052-015-3700-6}{{\em Eur. Phys. J.}
  {\bf C75} (2015) no.~10, 482},
\href{http://arxiv.org/abs/1508.05327}{{\tt arXiv:1508.05327 [hep-ph]}}.

\bibitem{Alloul:2013bka}
A.~Alloul, N.~D. Christensen, C.~Degrande, C.~Duhr, and B.~Fuks, ``{FeynRules
  2.0 - A complete toolbox for tree-level phenomenology},''
  \href{http://dx.doi.org/10.1016/j.cpc.2014.04.012}{{\em Comput. Phys.
  Commun.} {\bf 185} (2014)  2250--2300},
\href{http://arxiv.org/abs/1310.1921}{{\tt arXiv:1310.1921 [hep-ph]}}.

\bibitem{Bentvelsen:1998ug}
S.~Bentvelsen and I.~Meyer, ``{The Cambridge jet algorithm: Features and
  applications},'' \href{http://dx.doi.org/10.1007/s100520050232}{{\em
  Eur.Phys.J.} {\bf C4} (1998)  623--629},
\href{http://arxiv.org/abs/hep-ph/9803322}{{\tt arXiv:hep-ph/9803322
  [hep-ph]}}.

\bibitem{Aad:2015zva}
{\bf ATLAS Collaboration} Collaboration, G.~Aad {\em et al.}, ``{Search for new
  phenomena in final states with an energetic jet and large missing transverse
  momentum in pp collisions at $\sqrt{s}=8$ TeV with the ATLAS detector},''
\href{http://arxiv.org/abs/1502.01518}{{\tt arXiv:1502.01518 [hep-ex]}}.

\bibitem{Lester:1999tx}
C.~G. Lester and D.~J. Summers, ``{Measuring masses of semiinvisibly decaying
  particles pair produced at hadron colliders},''
  \href{http://dx.doi.org/10.1016/S0370-2693(99)00945-4}{{\em Phys. Lett.} {\bf
  B463} (1999)  99--103},
\href{http://arxiv.org/abs/hep-ph/9906349}{{\tt arXiv:hep-ph/9906349
  [hep-ph]}}.

\bibitem{Goodman:1984dc}
M.~W. Goodman and E.~Witten, ``{Detectability of Certain Dark Matter
  Candidates},''
\href{http://dx.doi.org/10.1103/PhysRevD.31.3059}{{\em Phys. Rev.} {\bf D31}
  (1985)  3059}.

\bibitem{Sachs:1962zzc}
R.~G. Sachs, ``{High-Energy Behavior of Nucleon Electromagnetic Form
  Factors},''
\href{http://dx.doi.org/10.1103/PhysRev.126.2256}{{\em Phys. Rev.} {\bf 126}
  (1962)  2256--2260}.

\bibitem{Bawin:1999ks}
M.~Bawin and S.~A. Coon, ``{Neutron charge radius and the Dirac equation},''
  \href{http://dx.doi.org/10.1103/PhysRevC.60.025207}{{\em Phys. Rev.} {\bf
  C60} (1999)  025207},
\href{http://arxiv.org/abs/nucl-th/9906014}{{\tt arXiv:nucl-th/9906014
  [nucl-th]}}.

\bibitem{Billard:2013qya}
J.~Billard, L.~Strigari, and E.~Figueroa-Feliciano, ``{Implication of neutrino
  backgrounds on the reach of next generation dark matter direct detection
  experiments},'' \href{http://dx.doi.org/10.1103/PhysRevD.89.023524}{{\em
  Phys. Rev.} {\bf D89} (2014) no.~2, 023524},
\href{http://arxiv.org/abs/1307.5458}{{\tt arXiv:1307.5458 [hep-ph]}}.

\bibitem{Lisanti:2009vy}
M.~Lisanti and J.~G. Wacker, ``{Disentangling Dark Matter Dynamics with
  Directional Detection},''
  \href{http://dx.doi.org/10.1103/PhysRevD.81.096005}{{\em Phys. Rev.} {\bf
  D81} (2010)  096005},
\href{http://arxiv.org/abs/0911.1997}{{\tt arXiv:0911.1997 [hep-ph]}}.

\bibitem{Cohen:2016nzv}
T.~Cohen, M.~J. Dolan, S.~El~Hedri, J.~Hirschauer, N.~Tran, and A.~Whitbeck,
  ``{Dissecting Jets and Missing Energy Searches Using $n$-body Extended
  Simplified Models},'' \href{http://dx.doi.org/10.1007/JHEP08(2016)038}{{\em
  JHEP} {\bf 08} (2016)  038},
\href{http://arxiv.org/abs/1605.01416}{{\tt arXiv:1605.01416 [hep-ph]}}.

\bibitem{Rogan:2010kb}
C.~Rogan, ``{Kinematical variables towards new dynamics at the LHC},''
\href{http://arxiv.org/abs/1006.2727}{{\tt arXiv:1006.2727 [hep-ph]}}.

\bibitem{Randall:2008rw}
L.~Randall and D.~Tucker-Smith, ``{Dijet Searches for Supersymmetry at the
  LHC},'' \href{http://dx.doi.org/10.1103/PhysRevLett.101.221803}{{\em Phys.
  Rev. Lett.} {\bf 101} (2008)  221803},
\href{http://arxiv.org/abs/0806.1049}{{\tt arXiv:0806.1049 [hep-ph]}}.

\bibitem{Chatrchyan:2013mys}
{\bf CMS} Collaboration, S.~Chatrchyan {\em et al.}, ``{Search for
  supersymmetry in hadronic final states with missing transverse energy using
  the variables $\alpha_T$ and b-quark multiplicity in pp collisions at $\sqrt
  s=8$ TeV},'' \href{http://dx.doi.org/10.1140/epjc/s10052-013-2568-6}{{\em
  Eur. Phys. J.} {\bf C73} (2013) no.~9, 2568},
\href{http://arxiv.org/abs/1303.2985}{{\tt arXiv:1303.2985 [hep-ex]}}.

\bibitem{Fan:2010gt}
J.~Fan, M.~Reece, and L.-T. Wang, ``{Non-relativistic effective theory of dark
  matter direct detection},''
  \href{http://dx.doi.org/10.1088/1475-7516/2010/11/042}{{\em JCAP} {\bf 1011}
  (2010)  042},
\href{http://arxiv.org/abs/1008.1591}{{\tt arXiv:1008.1591 [hep-ph]}}.

\bibitem{Hill:2011be}
R.~J. Hill and M.~P. Solon, ``{Universal behavior in the scattering of heavy,
  weakly interacting dark matter on nuclear targets},''
  \href{http://dx.doi.org/10.1016/j.physletb.2012.01.013}{{\em Phys. Lett.}
  {\bf B707} (2012)  539--545},
\href{http://arxiv.org/abs/1111.0016}{{\tt arXiv:1111.0016 [hep-ph]}}.

\bibitem{Fitzpatrick:2012ix}
A.~L. Fitzpatrick, W.~Haxton, E.~Katz, N.~Lubbers, and Y.~Xu, ``{The Effective
  Field Theory of Dark Matter Direct Detection},''
  \href{http://dx.doi.org/10.1088/1475-7516/2013/02/004}{{\em JCAP} {\bf 1302}
  (2013)  004},
\href{http://arxiv.org/abs/1203.3542}{{\tt arXiv:1203.3542 [hep-ph]}}.

\bibitem{Hill:2014yka}
R.~J. Hill and M.~P. Solon, ``{Standard Model anatomy of WIMP dark matter
  direct detection I: weak-scale matching},''
  \href{http://dx.doi.org/10.1103/PhysRevD.91.043504}{{\em Phys. Rev.} {\bf
  D91} (2015)  043504},
\href{http://arxiv.org/abs/1401.3339}{{\tt arXiv:1401.3339 [hep-ph]}}.

\bibitem{Hill:2014yxa}
R.~J. Hill and M.~P. Solon, ``{Standard Model anatomy of WIMP dark matter
  direct detection II: QCD analysis and hadronic matrix elements},''
  \href{http://dx.doi.org/10.1103/PhysRevD.91.043505}{{\em Phys. Rev.} {\bf
  D91} (2015)  043505},
\href{http://arxiv.org/abs/1409.8290}{{\tt arXiv:1409.8290 [hep-ph]}}.

\bibitem{Bishara:2016hek}
F.~Bishara, J.~Brod, B.~Grinstein, and J.~Zupan, ``{Chiral Effective Theory of
  Dark Matter Direct Detection},''
  \href{http://dx.doi.org/10.1088/1475-7516/2017/02/009}{{\em JCAP} {\bf 1702}
  (2017) no.~02, 009},
\href{http://arxiv.org/abs/1611.00368}{{\tt arXiv:1611.00368 [hep-ph]}}.

\bibitem{Elor:2015bho}
G.~Elor, N.~L. Rodd, T.~R. Slatyer, and W.~Xue, ``{Model-Independent Indirect
  Detection Constraints on Hidden Sector Dark Matter},''
  \href{http://dx.doi.org/10.1088/1475-7516/2016/06/024}{{\em JCAP} {\bf 1606}
  (2016) no.~06, 024},
\href{http://arxiv.org/abs/1511.08787}{{\tt arXiv:1511.08787 [hep-ph]}}.

\bibitem{Tulin:2017ara}
S.~Tulin and H.-B. Yu, ``{Dark Matter Self-interactions and Small Scale
  Structure},''
\href{http://arxiv.org/abs/1705.02358}{{\tt arXiv:1705.02358 [hep-ph]}}.

\end{thebibliography}\endgroup
\end{spacing}
\end{document}